\documentclass[11pt,a4paper]{article}
\pdfoutput=1
\usepackage{jheppub}

\usepackage{amssymb,amsmath,amstext,amsfonts,empheq,mdframed,xcolor}
\usepackage{url}
\usepackage{bbold}
\usepackage{bm}
\usepackage{graphicx}
\usepackage{xcolor}
\usepackage{overpic}
\usepackage{subcaption}
\usepackage{braket}
\usepackage{soul}
\definecolor{lgray}{gray}{0.90}
\newcommand*\graybox[1]{ \colorbox{lgray}{\hspace{1em}#1\hspace{1em}}}

\newcommand{\be}{\begin{equation}}
\newcommand{\ee}{\end{equation}}
\newcommand{\bea}{\begin{eqnarray}}
\newcommand{\eea}{\end{eqnarray}}
\newcommand{\nn}{\nonumber}
\newcommand{\bal}{\begin{aligned}}
\newcommand{\eal}{\end{aligned}}

\newcommand{\Mp}{M_{\rm Pl}}

\newcommand{\Omegainf}{\Omega_{\textrm{GW}}^{{\mathrm{inf}}}}
\newcommand{\Omegarad}{\Omega_{\textrm{GW}}^{\mathrm{rad}}}
\newcommand{\OGW}{\Omega_\textrm{GW}}

\newcommand{\etaperp}{\eta_\perp}

\newcommand{\kf}{k_{\textrm{out}}}

\newcommand{\bog}{Bogoliubov}
\newcommand{\de}{\delta}

\newcommand{\muu}{\mu}

\newcommand{\x}{\kappa}

\newcommand{\kstar}{k_*}

\newcommand{\bk}{\boldsymbol{k}}
\newcommand{\bp}{\boldsymbol{p}}

\newcommand{\bx}{\boldsymbol{x}}
\definecolor{alizarin}{rgb}{0.82, 0.1, 0.26}

\newcommand{\X}{X}

\newcommand{\deltaeta}{\etaperp \delta}

\newcommand{\g}{g}

\newcommand{\di}{{\rm d}}

\newcommand{\indX}{X}

\newcommand{\p}{\mathcal{P}}

\newcommand{\ra}{\rho}

\newcommand{\tauf}{\tau_{\rm out}}
\newcommand{\zf}{z_{\mathrm{out}}}
\newcommand{\keq}{k_{\mathrm{eq}}}

\newcommand{\gam}{\gamma}
\newcommand{\cG}{\mathcal{G}}
\title{\centering \huge  
Primordial gravitational waves\\ from excited states}
\author[a,b,c]{Jacopo Fumagalli,}
\author[d]{Gonzalo~A. Palma,}
\author[a]{S\'{e}bastien Renaux-Petel,}
\author[e,f]{Spyros Sypsas,}
\author[a]{Lukas~T. Witkowski}
\author[d]{and Cristobal Zenteno}

\affiliation[a]{Institut d'Astrophysique de Paris, GReCO, UMR 7095 du CNRS et de Sorbonne Universit\'{e}, 98bis
boulevard Arago, 75014 Paris, France}
\affiliation[b]{Instituto de F\'{ı}sica Te\'{o}rica UAM/CSIC, Calle Nicol\'as Cabrera 13-15, Cantoblanco E-28049 Madrid, Spain}
\affiliation[c]{
Departamento de F\'{ı}sica Te\'{o}rica, Universidad Aut\'{o}noma de Madrid (UAM), Campus de Cantoblanco, E-28049 Madrid, Spain
}
\affiliation[d]{Grupo de Cosmolog\'ia y Astrof\'isica Te\'orica, Departamento de F\'{i}sica, FCFM,\\ \mbox{Universidad de Chile}, Blanco Encalada 2008, Santiago, Chile}
\affiliation[e]{Department of Physics, Faculty of Science, Chulalongkorn University, Phayathai Rd., Bangkok 10330, Thailand}
\affiliation[f]{National Astronomical Research Institute of Thailand, Don Kaeo, Mae Rim, Chiang Mai 50180, Thailand}

\emailAdd{jacopo.fumagalli@uam.es}
\emailAdd{gpalmaquilod@ing.uchile.cl}
\emailAdd{renaux@iap.fr}
\emailAdd{s.sypsas@gmail.com}
\emailAdd{lukas.witkowski@iap.fr}
\emailAdd{cristobal.zenteno@ing.uchile.cl}

\abstract{We show that a scalar excited state with large occupation numbers during inflation leads to an enhancement of tensor modes and a characteristic pattern of order-one oscillations in the associated stochastic gravitational wave background (SGWB) sourced during inflation.
An effective excited state, i.e.~a departure from the Bunch-Davies vacuum, can emerge dynamically as the result of a transient non-adiabatic evolution,
e.g.~a sharp feature along the inflationary history.
We provide an explicit example in a multifield context where the sharp feature triggering the excited state is identified with a strong turn in the inflationary trajectory. \textit{En passant}, we 
derive a universal expression for the tensor power spectrum sourced at second order by an arbitrary number of scalar degrees of freedom during inflation,
crucially taking into account the nontrivial structure of the Hilbert space in multifield setups. The SGWB sourced during inflation can overcome the standard scalar-induced SGWB sourced at horizon re-entry of the fluctuations after inflation, while being less constrained by perturbativity and backreaction bounds. In addition, one may entertain the possibility of detecting both since they peak at different frequencies exhibiting oscillations with distinct periods. 
}

\begin{document}
\hfill{\flushright {IFT-UAM/CSIC-21-140}}
\maketitle

\newpage

\section{Introduction}
\label{sec:intro}

Since LIGO's ``first light"~\cite{Abbott:2016blz}, an astounding picture has been rapidly unfolding: our universe is filled with gravitational waves (GWs) sourced by black hole mergers. In fact, black holes are just one among numerous sources expected to contribute to the so-called stochastic gravitational wave background (SGWB); a bath of gravitational waves with frequencies spanning at least 20 orders of magnitude. Another important source is cosmic inflation~\cite{Guth:1980zm,Starobinsky:1980te,Linde:1981mu,Albrecht:1982wi,Mukhanov:1981xt}, the period of exponential expansion that preceded the hot Big-Bang phase of our universe, responsible for the cosmic microwave background (CMB) anisotropies and the large-scale structure (LSS). Indeed, one of inflation's most remarkable predictions is the existence of a background of primordial GWs~\cite{Starobinsky:1979ty, Rubakov:1982df,Fabbri:1983us, Abbott:1984fp} with wavelengths ranging from planetary scales up to $10^4$~Mpc (the largest observable scale). This corresponds to a range of frequencies between $10^{-17}$ and $10^{2}$~Hz. Current CMB observatories are indirectly searching for primordial GWs with frequencies $10^{-17}-10^{-13}$~Hz (in the form of B-mode polarisation), whereas the $10^{-9}-10^{-1}$~Hz band will be accessible to future surveys such as LISA~\cite{LISA:web}, SKA~\cite{SKA:web} and IPTA~\cite{IPTA:web}.

Within the inflationary paradigm, these anisotropies are traced back to quantum fluctuations around a quasi de Sitter background. 
In single-field, slow-roll inflation, the latter are predicted to evolve adiabatically, leaving the universe filled with inhomogeneities characterised by almost scale-invariant power spectra. This can be understood as the result of the smooth evolution of the background, disfavouring any particular time-slice as wavelengths are continuously stretched from sub- to super-horizon scales due to the exponential expansion of space. Scale invariance (i.e.~spectra with constant amplitudes) implies that a detection of B-modes in the CMB polarisation would allow us to deduce the entire GW spectrum from its low-frequency amplitude. To date, CMB observations~\cite{BICEP:2021xfz} have constrained the power spectrum of primordial GWs to be far below the signal sensitivity available to the next generation of detectors (one order of magnitude for SKA, three orders of magnitude for LISA) leaving essentially no possibility for a direct detection of the SGWB's primordial component. 

However, the simplest models of inflation might not constitute an accurate description of the origin of primordial perturbations throughout the entire band of observable frequencies. CMB surveys cover but a narrow window of scales, rendering the extrapolation of the observed scale invariance of scalar perturbations to higher frequencies a potentially premature assumption. Actually, from a theoretical perspective, given the ultraviolet sensitivity of inflation and the associated difficulties to realise a prolonged stage thereof (see e.g.~\cite{Baumann:2014nda}), it is natural to entertain the possibility that inflation has occurred in successive periods with possibly vastly different properties. As a matter of fact, there exists circumstantial evidence pointing towards this possibility \cite{Clesse:2017bsw,Garcia-Bellido:2020pwq,Franciolini:2021tla}: the black holes observed by LIGO/Virgo might be primordial black holes~\cite{Novikov-pbh,Hawking:1971ei} resulting from the collapse of extreme over-densities on small scales.
We thus have phenomenological, theoretical and observational motivations to consider 
non-standard realisations of inflation in order to describe the origin of cosmological perturbations, including primordial gravitational waves. 
\begin{figure}[t]
\centering
\includegraphics[width=0.75\textwidth]{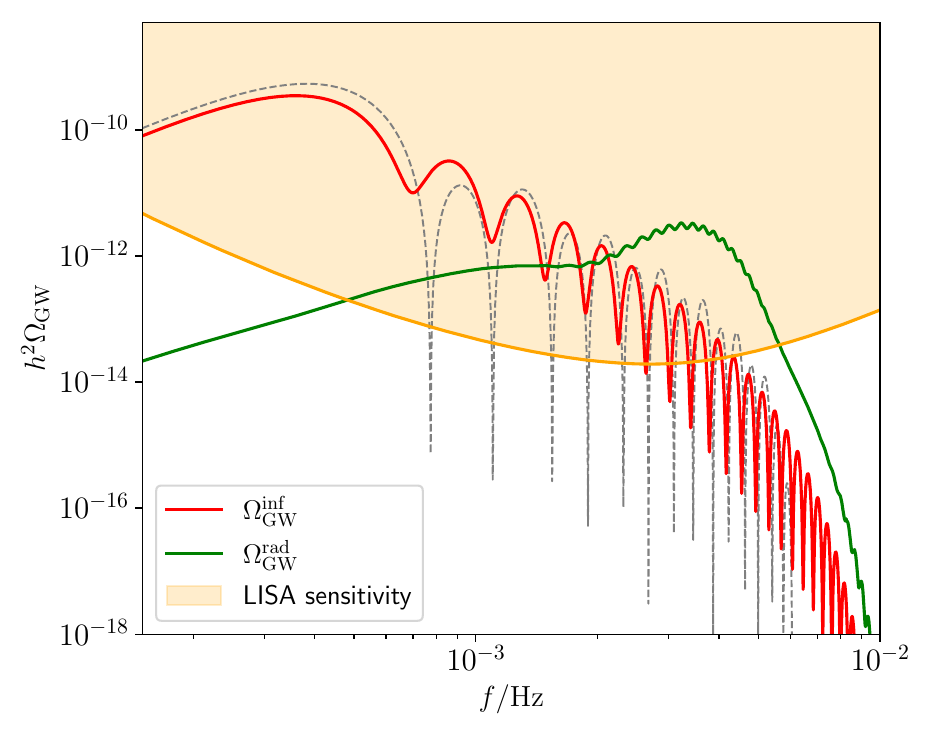}
\caption{\textit{
In red, the characteristic inflationary gravitational wave energy density from the emergence of an excited state during inflation. This spectrum is the subject of the current work where it is computed explicitly. In green, the scalar induced post-inflationary gravitational wave spectrum resulting from the presence of the same excited state triggered by a sharp feature during inflation. The shape of this contribution has been computed in \cite{Fumagalli:2020nvq}.
The grey dashed line represents the analytical template computed in Sec.~\ref{sec:GW-ExSt}, and explicitly given in Eq.~\eqref{template}, which is independent of the specific model generating the excited state.
}}
\label{fig:Full28}
\end{figure}

The purpose of this article is to uncover the characteristic frequency profile imprinted in the inflationary-era SGWB spectrum stemming from an excited state that is dynamically generated during inflation at wavelengths well below those relevant for the CMB and LSS. 
Even if one considers standard Bunch-Davies initial conditions in the remote past, an effective excited state can appear as a general outcome of a momentary departure from the adiabatic evolution of the scalar fluctuations, which isolates a preferred set of scales during inflation, leading to (possibly) sizable, scale-dependent enhancements of the power spectrum of scalar perturbations. As we will show, the latter can cause a significant production of primordial GWs at localised frequencies, both during and after inflation. With the appropriate conditions, excited states during inflation can produce observable signals accessible for instance to LISA or SKA, turning them, along with the next-generation of CMB observatories, into main contributors to the nascent field of multi-messenger primordial cosmology~\cite{Adshead:2020bji,Unal:2020mts,Malhotra:2020ket,Ricciardone:2021kel,Braglia:2021fxn,Dimastrogiovanni:2021mfs}.

Excited states can be generated in many scenarios of the primordial universe, a non-exhaustive (and non mutually exclusive) 
list of phenomena and references including: brief departures from slow-roll in single-field inflation, caused e.g by a step in the inflaton potential or a time-dependent effective parameter~\cite{Starobinsky:1992ts,Kaloper:2003nv,Ashoorioon:2006wc,Bean:2008na,Ashoorioon:2017toq,Ashoorioon:2018uey,Ballesteros:2018wlw,Ballesteros:2021fsp,Tasinato:2020vdk,Dalianis:2021iig,Inomata:2021tpx}; particle production due to time-dependent masses and resonance phenomena~\cite{Chung:1999ve,Barnaby:2009dd,Cook:2011hg,Carney:2012pk}; sharp turns in the multifield landscape~\cite{Achucarro:2010da,Palma:2020ejf,Fumagalli:2020adf,Fumagalli:2020nvq,Braglia:2020taf,Iacconi:2021ltm}; multiple-stage inflation~\cite{Polarski:1992dq,Adams:1997de,Pi:2017gih,Pi:2019ihn,DAmico:2020euu,DAmico:2021vka}, etc. Note that in all of these examples, the excited state emerges dynamically at some moment during the evolution, which is crucially different from initialising the system in a non Bunch-Davies state (as in e.g.~\cite{Ragavendra:2020vud}).
Independently of any particular microphysics, which plays an interesting but secondary role in what follows, the formalism and the results presented in this work have a broad range of applicability; they are derived under the single assumption of an excited state with large occupation numbers. We have kept the formalism as general as possible, taking into account the possible sourcing of GWs by multiple sources, which, in general, can be non-trivially quantum-mechanically correlated. Upon doing so, we will see that there is an interesting effect special to multifield dynamics, which amounts to a significant enhancement of the spectrum, even in two-field inflation, as a consequence of the quantum mixing among the fields.

As we will show, the presence of an excited state of scalar perturbations leads to a significant enhancement of the tensor modes generated during inflation compared to a sourcing by fields in their vacuum state. The intuitive physical reason is clear: an excited state means the presence of particles during inflation; these particles carry some energy beyond the minimal one of vacuum quantum fluctuations, and this energy --- more accurately the transverse traceless part of the corresponding energy-momentum tensor --- sources GWs. Something worth highlighting is that this component of the SGWB is governed by the particle content and dynamics of inflation. It thus offers a probe of the physical processes during inflation that is complementary to the scalar-induced SGWB generated after inflation, which is sourced as the wavelengths of the primordial fluctuations re-enter the horizon after inflation and hence is sensitive only to the statistics of the fluctuations at the end of inflation
(see~\cite{Domenech:2021ztg} for a recent review).

Some of us have recently studied the scalar-induced GW background generated after inflation. The corresponding primordial curvature power spectrum displays large oscillations on small scales, characteristic of sharp features during inflation leading to substantial particle production~\cite{Fumagalli:2020nvq}. The present paper thus complements this study by characterising the unavoidable inflationary-era SGWB that is also generated in this context. 
As we shall see, 
these two complementary signals of GWs appear in the SGWB in the form of two characteristic bumps in $\OGW$ (the density of primordial GWs), located at different frequencies ---see Fig.~\ref{fig:Full28}.\footnote{Throughout the paper, the LISA sensitivity curve that we add for illustrative purposes corresponds to the power law integrated sensitivity of~\cite{Thrane:2013oya} for a threshold signal-to-noise ratio $\textrm{SNR}_\textrm{th}=1$ and an effective total observation time $T_\textrm{obs}=3 \textrm{ years}$.} This can be understood as the result of the emergence of two distinct comoving scales marking the positions of these maxima: the horizon size at the time of particle production, and the wavelength of the maximally enhanced mode of the primordial scalar power spectrum. Moreover, these GW spectra are modulated by oscillations with (distinct) frequencies, determined by the aforementioned scales, and whose relative amplitudes are different in the two cases: oscillations in the inflationary-era SGWB are of order one, while they are of order $10 \%$ for the ones generated after inflation.
The resulting frequency profile of the total SGWB thus displays a very rich structure, offering a smoking gun signature of nontrivial inflationary dynamics far away from the CMB window.
\subsection{Layout}
We begin in Sec.~\ref{sec:prelims} with a review of generic aspects of the SGWB. In Sec.~\ref{sec:multi}, we derive a universal expression for the tensor power spectrum sourced at second order by an arbitrary number of scalar degrees of freedom during inflation, highlighting previously overlooked effects due to the quantum mixing thereof. We analyse the detailed structure of the spectrum in Sec.~\ref{sec:GW-ExSt} providing analytical estimates of all the spectral characteristics. In Sec.~\ref{sec:obs}, we discuss the interplay between theoretical constraints related to backreaction and perturbative control and prospects for detection in future GWs observatories, while Sec.~\ref{sec:conc} contains our conclusions. Finally, the (approximate) equivalence of the spectra computed with the retarded Green's function and with the in-in formalism is shown in App.~\ref{app:in-in}, further elaborating on points discussed in Sec.~\ref{sec:multi}; appendix~\ref{Appapprox} contains details relevant for the analytic results obtained in Sec.~\ref{sec:GW-ExSt}.

\section{Stochastic gravitational wave background basics}
\label{sec:prelims}
Let us start by defining the density parameter of gravitational waves, $\OGW$, which will constitute the main observable quantity to be dealt with throughout this work. We are interested in studying the generation and evolution of tensor perturbations in a Friedmann–Lema\^itre–Robertson–Walker (FLRW) background. This may be done by perturbing the FLRW metric as
\be
\di s^2 = a^2(\tau) \left[ - \di\tau^2 +  (e^{2\zeta}\delta_{ij} + h_{ij}) \di x^i \di x^j \right],
\ee
where $\tau$ denotes conformal time. Here, $h_{ij}(\bx,\tau)$ is the tensor perturbation which is traceless ($\delta^{ij} h_{ij} = 0$) and transverse ($\partial^i h_{ij} = 0$), while $\zeta$ is the curvature perturbation in this (comoving) gauge. 
It will be convenient to express $h_{ij}(\bx,\tau)$ in terms of Fourier modes $\tilde h_{ij} (\bk , \tau)$:
\be
h_{ij}(\bx,\tau) = \int \frac{\di^3 k}{(2\pi)^3} \tilde h_{ij} (\bk , \tau) e^{i \bk \cdot \bx} .
\ee
The transverse and traceless conditions then translate to $k^i \tilde h_{ij} = \delta^{ij} \tilde h_{ij}= 0$, allowing us to further decompose $\tilde h_{ij} (\bk , \tau)$ in terms of spin-$2$ polarisation tensors as:
\be
\tilde h_{ij}(\bk,\tau) =\sum_{r= +, \times}  e^r_{ij}(\bk)h^r_{\bk}(\tau) ,
\ee
with the two polarisation tensors $e^+_{ij}(\bk)$ and $e^\times_{ij}(\bk)$ taken to be real.\footnote{We use $e_{ij}^+(\bk) =\frac{1}{\sqrt{2}}(\hat{m}_i\hat{m}_j-\hat{n}_i\hat{n}_j)$ and $e_{ij}^\times(\bk) =\frac{1}{\sqrt{2}}( \hat{m}_i\hat{n}_j+\hat{n}_i\hat{m}_j)$, where $\{\hat{m}_i,\hat{n}_i\}$ are the unit vectors orthogonal to $\bk$. In this way,
$e_{ij}^r(\bk) = e_{ij}^r(-\bk)$ and $e_{ij}^r e_{ij}^ s = \delta^{rs}$.} 

Consistent with the fact that the source of GWs considered in this work comprises scalar degrees of freedom, in what follows, we consider a statistically homogeneous, isotropic and unpolarised SGWB, and we write the two-point function of $h^r_{\bk}$ as 
\be\label{defpower}
\left \langle h^r_{\bk} (\tau) h^s_{\bk '} (\tau) \right \rangle =  (2\pi)^3 \delta^{rs} \delta(\bk + \bk ')\frac{2\pi^2}{k^3} \frac{1}{2}\mathcal{P}_t (k,\tau), 
\ee
where $\mathcal{P}_t (k,\tau)$ is the (total) dimensionless power spectrum, and where the brackets denote the statistical average, equivalent in our context to a quantum expectation value. 

When tensor modes are well inside the horizon and truly behave as free gravitational waves, with $(h^{r}_{\bm{k}})' = k h^r_{\bm{k}}$, one can define their energy density as \cite{Maggiore:2007ulw,Caprini:2018mtu}
\be \label{defrho}
\rho_{\rm GW} (\tau) =   \frac{\Mp^2}{4 a^2}  \left\langle  h_{ij}' h_{ij}' 
\right\rangle ,
\ee
where we have used units such that $\Mp^2 = (8\pi G)^{-1}$, while $'\equiv\di/\di\tau$. 
This expression allows one to obtain the density parameter $\OGW = \rho_{\rm GW} / \rho_{\rm cr}$ (where $\rho _{\mathrm{cr}}= 3\Mp^2 H^2$) as
\be \label{Omegas}
\OGW (\tau) =  \int \di \ln(k)  \OGW (k,\tau) , \qquad  \OGW (k , \tau) =  \frac{k^2}{12 a^2 H^2 }   \mathcal P_t(k, \tau) ,
\ee
with $\OGW (k,\tau)$ the density parameter per comoving logarithmic scale. This expression gives an explicit relation between $\OGW$ and the power spectrum evaluated at a time where the modes of interest are sub-horizon.\footnote{In this context, it is customary to average ${\cal P}_t$ over many periods of the GWs, a procedure often denoted with a bar that we do not write for simplicity.}

In order to determine $\p_t(k,\tau)$, we will consider scenarios where scalar fluctuations are enhanced during inflation, making them a relevant source of tensor perturbations $h_{ij}$. To proceed, we must consider Einstein's equations linearised with respect to $h_{ij}$ in an FLRW background. This gives the following equation of motion for the Fourier mode $\tilde h_{ij}$:
\be
\tilde h_{ij} '' (\bm{k}, \tau) + 2 \mathcal H \tilde h_{ij} ' (\bm{k}, \tau) + k^2 \tilde h_{ij} (\bm{k}, \tau) = \tilde S_{ij} (\bm{k}, \tau) ,
\ee
where $\tilde S_{ij} (\bm{k}, \tau)$ represents the Fourier mode of the transverse and traceless component of the energy-momentum tensor acting as a source term. Just as we did with the tensor perturbation, we may decompose this source term as $\tilde S_{ij} (\bm{k}, \tau) = \sum_{r= +, \times}  e^r_{ij}(\bk) S^r_{\bk}(\tau)$, giving an equation of motion for the polarisation modes $h_{\bm{k}}^{r}$:
\begin{align}
	h_{\bm{k}}^{\lambda} {}'' (\tau) + 2 \mathcal H h_{\bm{k}}^{\lambda} {}' (\tau) + k^2 h_{\bm{k}}^{\lambda} (\tau) = S_{\bm{k}}^{\lambda} (\tau)\,,
	\label{eq:eom-h}
\end{align}
where ${\cal H}=a'/a$ is the co-moving Hubble parameter. 

The source $S_{\bm{k}}^{\lambda} (\tau)$ depends on the matter content of the universe at different stages of its evolution. During inflation, $S_{\bm{k}}^{\lambda} (\tau)$ receives contributions from the curvature perturbation $\zeta$ and (if present) other degrees of freedom. This gives rise to an inflationary component of the tensor field that we label $h_{ij}^{\mathrm{inf}}$. On the other hand, after inflation $S_{\bm{k}}^{\lambda} (\tau)$ receives contributions from the super-horizon fluctuations $\zeta$, produced during inflation, as they re-enter the horizon \cite{Acquaviva:2002ud, Mollerach:2003nq, Ananda:2006af, Baumann:2007zm}. The result of this second source is a post-inflationary component of the tensor field that we label $h_{ij}^{\mathrm{rad}}$. 

Thus, $h_{ij}$ can be written as the sum of two terms:\footnote{Since it is subdominant and anyway uncorrelated to the component discussed here, we neglect the homogeneous solution during inflation, i.e.~the usual vacuum tensor modes generated during inflation.}
\be\label{hdeco}
h_{ij}(\bx,\tau) = h_{ij}^{\mathrm{inf}}(\bx,\tau) + h_{ij}^{\mathrm{rad}}(\bx,\tau).
\ee 
Inserting this decomposition into Eqs.~\eqref{defpower}-\eqref{Omegas} we can schematically (and with obvious notation) write $\OGW$ as follows:
\be \label{omdeco}
\OGW = \Omegarad + \Omegainf + \OGW^{\mathrm{mix}}. 
\ee
As we will see in a specific example, for excited states, the size of the two contributions $\Omegarad$ and $\Omegainf$ depends on different quantities and for most of the parameter space one contribution can easily overcome the other. For this reason, 
we do not consider the mixed term here.
$\Omegarad$ induced by sharp features creating excited states at a given time along the inflationary history has been computed in~\cite{Fumagalli:2020nvq} and the main result will be briefly reviewed in Sec.~\ref{sec:Omega-rad-review}. The purpose of the current work is then to compute, in all generality, the pattern arising in $\Omegainf$, i.e.~the contribution from tensor modes sourced \textit{during} inflation, due to the presence of excited scalar fluctuations.  
Usually the inflation-generated contribution is sub-dominant compared to the one sourced during the post-inflationary era. The reason is that, as we will see in more detail later, while in the latter case the source term in Eq.~\eqref{eq:eom-h} is schematically given by $S_{\bk}\propto \zeta^2 \propto \p_{\zeta}$, in the former case, considering the adiabatic perturbation $Q_{\zeta} \propto \sqrt{\epsilon} \zeta $, it reads $S_{\bk}\propto Q_{\zeta}^2 \propto \epsilon \p_{\zeta}$; hence, naively $\Omegainf\sim\epsilon^2\Omegarad$. However, as we will show, the situation can drastically change if the temporal behaviour of the scalar modes becomes non-standard before horizon crossing and/or additional entropic degrees of freedom become relevant in the source term (see also~\cite{Biagetti:2013kwa}).

\subsection{Stochastic background today}\label{today}
We are interested in the stochastic background of GWs as measured in the present cosmic era ($\tau_0$). For a given mode in Fourier space, the frequency of GWs today is given by
\be\label{freq0}
f = \frac{k c}{2\pi a_0} = 1.5\times10^{-15} k\,\mathrm{Mpc\,\, Hz}.
\ee
With the scale crossing the Hubble radius at matter-radiation equality being $k_{\mathrm{eq}} \simeq 1.3\times 10^{-2}\, \mathrm{Mpc}^{-1}$, all
modes with frequencies $f \gtrsim 10^{-17}\,$Hz have re-entered the horizon during radiation domination, unless a non-standard thermal history is considered between the end of inflation and the radiation-dominated era. 

To compute the post-inflationary induced GWs today, it is sufficient to note that the source term is active when the corresponding mode re-enters the horizon and soon decays (as $\propto \tau^{-2}$ during radiation) leaving a free propagating GW with an energy density $\rho_{\mathrm{GW}}\propto 1/a^4$. Thus, if we consider $\Omegarad$ at $\tau_{\mathrm{p}}$, a time after horizon crossing for a given mode such that the source has become negligible, we have 
\be\label{radtod}
\Omegarad (k,\tau_0) = \Omegarad(k,\tau_{\mathrm{p}}) \frac{\rho_{\mathrm{cr}}(\tau_{\mathrm{p}})}{\rho_{\mathrm{cr}}(\tau_0)}\left(\frac{a_{\mathrm{p}}}{a_0}\right)^4.
\ee
Deep inside the radiation era $\rho_{\mathrm{cr}}(\tau_{\mathrm{p}})\simeq \rho_r(\tau_{\mathrm{p}})$ and
\be\label{cg}
\left(\frac{a_{\mathrm{p}}}{a_0}\right)^4 = c_g \frac{\rho_r(\tau_0)}{\rho_{r}(\tau_{\mathrm{p}})}, \quad \mathrm{with}\qquad c_g= \frac{g_{*}(T_{\mathrm{p}})}{g_{*}(T_0)} {\bigg( \frac{g_{S}(T_0)}{g_{S}(T_{\mathrm{p}})} \bigg)}^{4/3},
\ee
where $g_S$ and $g_*$ are respectively the effective number of entropic and relativistic degrees of freedom as a function of the temperature $T$.
Thus, Eq.~\eqref{radtod} becomes 
 \be\label{radtod2}
\Omegarad(k,\tau_0) = c_g \Omega_{r, 0}\Omegarad(k,\tau_{\mathrm{p}}),
\ee
with $\Omega_{r, 0}$ the energy density fraction in radiation today.\\[-0.3cm]

The contribution to the stochastic background sourced during inflation is more conveniently expressed in terms of the tensor power spectrum at the end of inflation. In fact, $\p_t^{\textrm{inf}}$ at a given time after inflation can be written by means of a transfer function that takes into account the evolution of the tensor modes throughout the cosmic history, i.e., $\p^{\textrm{inf}}_t(k,\tau)=T^2(k,\tau)\p_t(k,\tau_{\mathrm{end}})$. By inserting this expression into Eq.~\eqref{Omegas} (with the decomposition~\eqref{omdeco} in mind), we obtain \cite{Turner:1993vb,Caprini:2018mtu}\footnote{Reference \cite{Turner:1993vb} assumes the same time dependence for all modes deep inside matter domination (an assumption that is correct once the average over the oscillatory terms is considered) and finds the $k$ dependent coefficient in the transfer function by interpolating the full numerical result, i.e.
\be\label{transferint}
T^2(k,\tau_0) =  \left(\frac{3 j_1(k\tau_0)}{k\tau_0}\right)^2\left(1+ 1.34 \frac{k}{\keq}+ 2.5\left(\frac{k}{\keq}\right)^2\right).
\ee
In the review \cite{Caprini:2018mtu}, an analytical result for $T^2(k,\tau_0)$ has been computed by considering modes that enter the horizon during radiation and matching their behaviour at a given time $\tau_*$ (the time when the pure radiation and matter solutions for the scale factor cross) related in a precise way to the time of matter-radiation equality. This analytical result is in good agreement with the full numerical one and has no substantial difference with respect to the interpolation in Eq.~\eqref{transferint}. We thus use the analytical result obtained in this way to estimate the prefactor in Eq.~\eqref{redshift2}.
}
\be\label{betterestimate2}
\Omegainf = \frac{3}{128}c_g\Omega_{\mathrm{r},0 }\left(\frac{1}{2}\left(\frac{\keq}{k}\right)^2 + \frac{16}{9}\right)\cdot {\cal P}_t(k,\tau_{\mathrm{end}}).
\ee
For consistency, we have multiplied the transfer function one can find in \cite{Caprini:2018mtu} (obtained under the assumption that $\rho_r$ has always red-shifted as $1/a^4$) by $c_g$; the factor that takes into account the different number of relativistic degrees of freedom when the modes of interest re-enter the horizon.\\[-0.3cm]

The first Standard Model degree of freedom to become non-relativistic is the the top quark that annihilates at about $T\simeq m_t/6 \simeq 30 \, \mathrm{GeV}$. By recalling that the frequency of a GW produced at horizon crossing (during radiation) can be directly related to the temperature of the universe at that time\footnote{By using the conservation of entropy $g_S(T)T^3 a^3  =\mathrm{const}$ and $\rho_{\mathrm{rad}}= \frac{\pi^2}{30}g_*(T)T^4$ we can rewrite Eq.~\eqref{freq0} as
\be
f(T_p) = 2.5 \cdot 10^{-8} \, \mathrm{Hz}\,\left[\frac{g_*(T_p)}{100}\right]^{1/6}\frac{T_p}{\mathrm{GeV}},
\ee
where we used that for $T\gtrsim \mathrm{Mev}$, $g_S\simeq g_*$.
}, one deduces that $f(T\gtrsim 30 \,\mathrm{GeV}) \gtrsim 8\cdot 10^{-7}\,\mathrm{Hz}$. Thus, if one is interested in frequencies relevant for GWs observatories like, for instance, LISA ($10^{-5}\,\mathrm{Hz}\lesssim f \lesssim 10^{-1}\,\mathrm{Hz} $) and LVK ($1\,\mathrm{Hz}\lesssim f \lesssim 10^{4}\,\mathrm{Hz} $)\footnote{See \cite{Liu:2015psa} for a study of cosmic phase transitions with PTA ($10^{-9}\,\mathrm{Hz}\lesssim f \lesssim 10^{-6}\,\mathrm{Hz} $).}, all Standard Model degrees of freedom can be safely treated as relativistic at the time of production. Therefore, $g_*(T_p)\simeq g_S(T_p) \simeq 106.75$, which, together with the present-era values $g_S(T_0)\simeq 3.91$ and $g_*(T_0) = 2$, leads to $c_g \simeq 0.4$. Furthermore, as per common practice, in order to avoid propagation of uncertainties on the measurements of the Hubble parameter, we will consider the quantity $h^2\OGW$ with $H_0 = h\cdot 100\, \mathrm{Km /(s\cdot Mpc)}$. \\[-0.3cm]

Summarising, it is convenient to write the two contributions to $h^2\OGW$ as 
\be\label{redshif1}
h^2\Omegarad(k,\tau_0) = r_r \Omegarad(k,\tau_p),\quad r_r \equiv h^2c_g\Omega_{\mathrm{r,0}} \simeq 1.6\cdot 10 ^{-5},
\ee
\be\label{redshift2}
h^2\Omegainf(k,\tau_0) = r_i \p_t(k,\tau_{\mathrm{end}}),\quad r_i \equiv  h^2 \cdot  0.0416 \cdot c_g\Omega_{\mathrm{r,0}} ,
\ee
where the explicit expression for
$\Omegarad(k,\tau_p)$ will be specified later in Eq.~\eqref{eq:OmegaGW-i} (we will re-consider the ``red-shifting" factors $r_i,r_r$ when discussing observability in Sec.~\ref{sec:obs}). The following sections will focus on computing $\p_t(k,\tau_{\mathrm{end}})\equiv \p_t(k)$ in presence of an excited state. For notational convenience we will also omit the $\tau_0$ argument and write $\Omega_{\mathrm{GW}}(k)\equiv \Omega_{\mathrm{GW}}(k,\tau_0) $.

\section{Multisourced primordial gravitational waves}
\label{sec:multi}

\subsection{Multifield quantisation}

We will be interested in GWs sourced by ${\cal N}$ scalar fluctuations to second order in perturbation theory. 
To this end, let us denote the corresponding gauge-invariant quantum operators (at linear order) by $\hat{Q}_{\indX}(\bk,\tau)$ (more on their normalisation below), where $X$ runs from $1$ to ${\cal N}$, and expand them in a basis of canonical creation/annihilation operators as
\begin{align}\label{generic}
\hat{Q}_{\indX}(\bk,\tau)=\sum_{i=1}^{\cal N} Q_{\indX i}(k,\tau) \hat{a}_i(\bk)+\textrm{h.c.}(-\bk)\,, \end{align}
where
\begin{align}
\left[\hat{a}_i(\bk),\hat{a}_j^\dagger(\bk')\right]=(2\pi)^3 \delta^{ij}\delta^{(3)}(\bk-\bk^\prime).\label{quantiz}
\end{align}
A crucial aspect of a multi-species system is that the ladder operator basis consists of ${\cal N}$ vectors labeled by the indices $i,j$ (see e.g.~\cite{Salopek:1988qh,GrootNibbelink:2001qt,Tsujikawa:2002qx,Weinberg:2008zzc}). This is the result of properly taking into account the interaction of ${\cal N}$ scalar degrees of freedom during inflation: the Hilbert space of the system is the tensor product of the individual Hilbert spaces, which is itself isomorphic to the ${\cal N}$ complex-dimensional vector-space spanned by ladder operators associated with one degree of freedom each.

As usual, creation/annihilation operators and their corresponding vacuum state are defined by specifying a set of mode functions. Deep enough on sub-Hubble scales, one can always consider suitably defined fluctuations $\hat{Q}_{\indX}(\bk,\tau)$
that are decoupled and hence quantum-mechanically independent, and throughout this paper, we impose thereon the Bunch-Davies (BD) vacuum, i.e.~the mode functions are chosen so that when $|k\tau|\gg 1$ their corresponding vacuum state is the Minkowski one (the unique one minimising the Hamiltonian at early times). This is equivalent to imposing the following initial conditions:
\be\label{BD}
Q_{Xi}(k,\tau) = \frac{1}{a(\tau)\sqrt{2k}}e^{-ik\tau}\delta_{X i},\qquad \tau\rightarrow -\infty,
\ee
with $\delta_{X i}$ the Kronecker delta, i.e.~one can always choose the ${\cal N}$ elements of the ladder basis to be ``aligned'' with the ${\cal N}$ initially independent fluctuating degrees of freedom (see e.g.~Sec 3.2 of~\cite{Pinol:2020cdp}). In addition, Eq.~\eqref{BD} assumes that each field has been canonically normalised, i.e.~$S = \frac12 \int \di t \di^3 \bx \sum_X \dot{Q}_X^2+\ldots\;$.

Then, a specific model determines a system of $\mathcal{N}$ coupled differential equations satisfied by the operators $\hat{Q}_X$. In particular, the associated ${\cal N}^2$ mode functions $Q_{X i}$ in~\eqref{generic} correspond, for each field index $X$, to $\mathcal{N}$ independent solutions (labelled by the index $i$) of the $\mathcal{N}$-field system at hand.
To explicitly find the mode functions $Q_{X i}$, one simply solves the corresponding equations of motion with the $\mathcal{N}$ different sets of initial conditions; each one given by fixing the index $i$ in Eq.~\eqref{BD} and let $X$ vary.

In Sec.~\ref{sec:GW-ExSt}, we will focus on GW sourced by excited states, which imply a specific form of $Q_{\indX i}(k,\tau)$. For the moment though, let us keep the discussion as generic as possible and derive the second-order scalar induced inflationary tensor power spectrum in a general multifield context.\\[-0.3cm]

Due to the SVT decomposition, tensors can only be sourced (to lowest order in perturbation theory) by the transverse, traceless component of the energy-momentum tensor furnished by the kinetic terms of whatever scalar fields comprise the model at hand. Therefore, in the notation of Eq.~\eqref{generic}, the scalar source in Eq.~\eqref{eq:eom-h} 
will be given by
\begin{align}
\hat{S}_{\bm{k}}^{\lambda} (\tau)=
\frac{2}{\Mp^2}
\sum_{\indX}
\int \frac{\mathrm{d}^3 \bp}{(2 \pi)^{3}} e^\lambda(\bk,\bp)  
\hat{Q}_{\indX}(\bp,\tau) \hat{Q}_{\indX} (\bk - \bp,\tau), 
\label{eq:source}
\end{align}
where
\be\label{pola}
e^{+,\times}(\bk,\bp) \equiv p^i p^j e_{ij}^{+,\times}(\bk)=\frac{p^2}{\sqrt{2}} \sin^2 \theta \Big(\cos(2 \phi),\sin(2 \phi) \Big),
\ee
with $\theta$ the  angle  between  the  wavevectors $\bk$ (of the induced GWs) and $\bp$ (of the source), while $\phi$ is the azimuthal angle of $\bp$.

Note again that Eq.~\eqref{eq:source} is not restricted to models with canonical scalar fields; for instance, given a non-trivial field-space metric one can always diagonalise the kinetic term of the fluctuations by projecting them on a set of vielbeins (see for instance Sec.~\ref{Explicit}). Although our general formalism is independent of this, in such scenarios, it may be convenient to choose these vielbeins such that one of the fields $Q_X$ corresponds to the instantaneous adiabatic fluctuation, which we will do in concrete examples. That is, we will distinguish the first of the ${\cal N}$ fluctuating degrees of freedom by identifying it with $Q_\zeta \equiv \Mp\sqrt{2\epsilon}\zeta$, with the other fields corresponding to instantaneous entropic fluctuations $\psi_1,\psi_2, \ldots \psi_{{\cal N}-1}$. Equivalently, one may use interchangeably the notations $\indX=\{1,2, \ldots, {\cal N} \}$ or $\indX=\{\zeta,\psi_1, \ldots, \psi_{{\cal N}-1} \}$.\\[-0.3cm]

Viewed as an operatorial statement, the formal solution of the tensor equation of motion~\eqref{eq:eom-h} can then be expressed as 
\begin{align} \label{S-to-I}
\hat{h}^\lambda_{\bk}(\tau) = \hat{h}^\lambda_{\bk;0}(\tau) + \int^{\tau} \mathrm{d}\tau_1\; g_k(\tau,\tau_1)  \hat{S}_{\bm{k}}^{\lambda} (\tau_1).
\end{align}
The Green's function
can be written in terms of the vacuum mode functions. In de Sitter (dS) this becomes\footnote{More precisely, the Green's function associated with the canonical (Mukhanov-Sasaki) variable $v_k(\tau) =a(\tau)h_k(\tau)$ can be expressed as $g^v_k(\tau,\tau')= i[v^0_k(\tau)v^{0*}_k(\tau')-v^{0*}_k(\tau)v^0_k(\tau')]$, where $v_k^0(\tau)$ is the solution of the corresponding homogeneous equation, satisfying the Wronskian condition $v_k^0(\tau)\partial_{\tau}v_k^{0*}(\tau)-\partial_\tau v_k^0(\tau)v_k^{0*}(\tau)=i$, i.e.~$v_k^0(\tau)=-\frac{i}{\sqrt{2k^3}}\frac{e^{-i k \tau}}{ \tau}\left(1+ik\tau \right)$.} 
\be \label{eq:green-function}
g_k(\tau,\tau') = i \frac{\zeta(k\tau) \zeta^*(k\tau') - \zeta^*(k\tau) \zeta(k\tau')}{2\tau'^2 k^3} \Theta(\tau-\tau'),
\ee
with 
\be \label{zpm}
\zeta(k\tau) =  e^{-i k\tau}(1+i k\tau), 
\ee
the standard dS mode functions with Bunch-Davies asymptotics (where for convenience, we have not included the $H/\sqrt{2k^3}$ factor), which are the same for tensor and scalar modes (hence the use of $\zeta$). Furthermore, the vacuum contribution (homogeneous solution) $\hat{h}^\lambda_{\bk;0}(\tau)$ is uncorrelated
to the source term (it comes with its own quanta):
\be \label{3pt-vac-source}
\left\langle \hat{h}^\lambda_{\bk;0}  \hat{S}_{\bm{k}}^{\lambda} \right\rangle = 0.
\ee
Before computing the power spectrum, let us comment on the use of the Green's function in this context, a subject that has been discussed in~\cite{Weinberg:2005vy,Musso:2006pt,Seery:2008qj,Senatore:2009cf,Adshead:2009cb,Baumgart:2020oby}.
\subsection{Field evolution in the interaction-picture}
\label{ssec:int-inin}
Equation~\eqref{S-to-I} can be viewed as the first term in the expansion of the field operator in a series over the interaction-picture free fields:
\be \label{h-in-in}
\hat{h}^\lambda_{\bk}(\tau) =  \left[ \bar{\rm T} e^{i \int^\tau_{-\infty_+} \!\! \di\tau_1 \; \hat{\cal H}_{\rm int}(\tau_1)} \right]  \hat{h}^\lambda_{\bk;0}(\tau)  \left[ {\rm T} e^{-i \int^\tau_{-\infty_-} \!\! \di\tau_1 \; \hat{\cal H}_{\rm int}(\tau_1)} \right],
\ee
where T denotes time- and $\bar{\rm T}$ anti time-ordering, while $\infty_\pm \equiv \infty (1 \pm i\epsilon)$ accounts for the contour deformation in the infinite past. In this language, $\hat{h}^\lambda_{\bk;0}(\tau)$ is the interaction-picture field, while $\hat{\cal H}_{\rm int}$ is the interaction picture Hamiltonian:
\be \label{Hint}
\hat{\cal H}_{\rm int}(\tau) = \frac{1}{(2\pi)^3} \int \di \bp\, a^2(\tau)\hat{h}_{\bm{p};0}(\tau) \hat{S}_{-\bm{p}}^{\lambda} (\tau),
\ee
with $\hat{S}$ given by Eq.~\eqref{eq:source}.

To verify this, we may expand the exponentials to first order in ${\cal H}_{\rm int}$, to obtain
\bea
\hat{h}^\lambda_{\bk}(\tau) =   \hat{h}^\lambda_{\bk;0}(\tau)  &+& i \int^\tau_{-\infty_+} \!\! \di\tau_1 \; \left[ \hat{\cal H}_{\rm int}(\tau_1),\hat{h}^\lambda_{\bk;0}(\tau) \right] 
\nn\\ &-&i \int^\tau_{-\infty_-} \!\! \di\tau_1 \;  \left[ \hat{h}^\lambda_{\bk;0}(\tau), \hat{\cal H}_{\rm int}(\tau_1)\right] +\ldots \, ,
\eea
which, upon inserting~\eqref{Hint}, reads
\bea
\hat{h}^\lambda_{\bk}(\tau) =   \hat{h}^\lambda_{\bk;0}(\tau)  &+& \frac{i}{(2\pi)^3}  \int^\tau_{-\infty_+}  \!\! \di\tau_1 \int \di \bp \; a^2(\tau_1) \left[ \hat{h}^\lambda_{\bp;0}(\tau_1),\hat{h}^\lambda_{\bk;0}(\tau) \right] \hat{S}_{-\bm{p}}^{\lambda} (\tau_1) \nn\\&+&  
\frac{i}{(2\pi)^3} \int^\tau_{-\infty_-} \!\! \di\tau_1 \int \di \bp \; a^2(\tau_1) \left[ \hat{h}^\lambda_{\bp;0}(\tau_1),\hat{h}^\lambda_{\bk;0}(\tau) \right] \hat{S}_{-\bm{p}}^{\lambda} (\tau_1)+\ldots \,. 
\eea
Next, let us isolate the infinite past by inserting an arbitrary time\footnote{In the next section, this arbitrary time will be related to the characteristic time of the ``feature'' creating the excited state.
\label{ft:tbar}}
$\bar\tau$~\cite{Senatore:2009cf}. After expanding the graviton in the canonical basis, 
\be 
\hat{h}_{\bk;0}(\tau)= \frac{H}{\sqrt{4k^3}} \left(\zeta(k\tau)\hat{a}(\bk) + \zeta^*(k\tau) \hat{a}^\dag(-\bk) \right),
\ee 
with $\zeta$ the dS mode function~\eqref{zpm} and $\hat{a},\hat{a}^{\dag}$ satisfying the commutation relations \eqref{quantiz}, we may use~\eqref{eq:green-function} to finally obtain
\bea \label{h-tauf}
\hat{h}^\lambda_{\bk}(\tau) &=& \hat{h}^\lambda_{\bk;0}(\tau)  + \int_{\bar\tau}^{\tau} \mathrm{d}\tau_1\; g_k(\tau,\tau_1)  \hat{S}_{\bm{k}}^{\lambda} (\tau_1)
\nn\\ &+& \frac12
\int^{\bar\tau}_{-\infty_+} \!\! \di\tau_1 \; g_k(\tau,\tau_1)  \hat{S}_{\bm{k}}^{\lambda} (\tau_1) +  \frac12
\int^{\bar\tau}_{-\infty_-} \!\! \di\tau_1 \; g_k(\tau,\tau_1)  \hat{S}_{\bm{k}}^{\lambda} (\tau_1) +\ldots \, . 
\eea
The first line of this equation coincides with Eq.~\eqref{S-to-I} (with the part of the integral between $\bar\tau$ and $\tau$; for $i\epsilon\to0$, the matching of the leading terms is exact). We may thus draw two conclusions: \emph{i}) the Green's function solution for the tensor field~\eqref{S-to-I} is an approximation of the nonlinear result~\eqref{h-in-in} ~\cite{Musso:2006pt}, and \emph{ii}) it also differs from the latter as far as the implementation of the $i\epsilon$ prescription is concerned~\cite{Adshead:2009cb,Senatore:2009cf}. We further elaborate on this discussion in App.~\ref{app:in-in}.

Both characteristics can be thought of as manifestations of the quantum nature of the inflationary tensor modes since both the contour deformation and the nonlinearity arise from a quantum-mechanical treatment which is appropriate at $\tau=-\infty$, where the BD initial condition is imposed. However, since we are considering an effective excited state emerging at some later time, the ``quantum'' character here translates into a statistical property of the random variable $h$: all the terms in the expansion~\eqref{h-in-in} express the non-Gaussian variable\footnote{Since the tensors are sourced to second order in the scalars, they are intrinsically non-Gaussian even if the latter are Gaussian. 
}
as a series over the Gaussian random fields $h_0$, $Q_0$ much like the familiar local ansatz~\cite{Komatsu:2001rj} for the scalar fluctuation (see also App.~A.1 of~\cite{Palma:2019lpt} for a related discussion).

\subsection{The tensor power spectrum}
\label{sec:general-result}
Let us for a moment (see the end of the section) neglect the last line of Eq.~\eqref{h-tauf}. Then %Having justified Eq.~\eqref{S-to-I}, 
the graviton two-point function can be simply written as  
\begin{align} 
\left\langle \hat{h}^\lambda_{\bk}(\tau) \hat{h}^\mu_{{\bk}'}(\tau) \right\rangle
&=
 \int^\tau \mathrm{d}\tau_1 \int^\tau \mathrm{d}\tau_2 \; g_k(\tau,\tau_1) g_k(\tau,\tau_2) \left\langle \hat{S}^\lambda_{{\bk}}(\tau_1) \hat{S}^\mu_{{\bk}'}(\tau_2) \right\rangle,
\label{eq:h-h}
\end{align}
where we have dropped the vacuum contribution $\hat{h}_{\bk;0}$, since the sourced GWs will be the dominant component in the scenarios under consideration here. Note that the operators involved in Eq.~\eqref{eq:h-h} are the interaction-picture fields, which, here, have Gaussian eigenvalues, allowing us to proceed via Wick's theorem. Scalar non-Gaussianity boosted by the excited state, see e.g.~{\cite{Chen:2006nt,Holman:2007na,Meerburg:2009ys,Agarwal:2012mq,Ganc:2011dy,Flauger:2013hra,Aravind:2013lra},
will enter at two loops and beyond via insertions of (at least) the ever-present cubic vertices~\cite{Maldacena:2002vr}
$\hat{{\cal H}}^{(3)}_{\rm int} \supset \Mp^2 \epsilon^2  a^2 \,\hat{\zeta} \left( \hat{\zeta}^{'2}+ \left( \partial\hat{\zeta}\right)^2\right)$.

A back of the envelope estimation implies that the perturbativity/backreaction conditions discussed in Sec.~\ref{sec:obs} should automatically grant radiative stability. 

Performing the Wick contractions and ignoring the irrelevant disconnected contribution, one thus obtains
\bea
\left\langle \hat{S}^\lambda_{\bk}(\tau_1) \hat{S}^\mu_{\bk'}(\tau_2) \right\rangle &=&  \left(\frac{2}{\Mp^2} \right)^2  \int \mathrm{d}^3  \bp \; e^\lambda(\bk,\bp)\Big(e^\mu(-\bk,-\bp)+e^\mu(-\bk,\bp-\bk)\Big) \times \nonumber \\
&&\sum_{X,Y} P_{XY}(\tau_1,\tau_2;p) P_{XY}\left( \tau_1,\tau_2;|\bk-\bp| \right) \; \delta^{(3)}\left( \bk+\bk^\prime \right),
\label{eq:S-S}
\eea
where the scalar power spectra are given by
\begin{align}
   P_{XY}(\tau_1,\tau_2;k) = \sum_i  Q_{Xi}(k,\tau_1) Q^{*}_{Yi}(k,\tau_2)\,,
   \label{eq:P-without-renormlaisation}
\end{align}
with the mode functions defined in Eq.~\eqref{generic}.
Next, we may substitute the polarisation vectors \eqref{pola} noticing that $e^{+,\times}(\bk,\bp)=e^{+,\times}(-\bk,-\bp)=e^{+,\times}(-\bk,\bp-\bk)$. The integrals over $\phi$ then yield the polarisation Kronecker delta as $\pi \delta^{\lambda\mu}$.
Writing the scalar source in Eq.~\eqref{eq:h-h} explicitly and using the definition~\eqref{defpower},
we arrive at the main formula for the total power spectrum of tensor modes sourced by scalar degrees of freedom during inflation:
\begin{empheq}[box=\graybox]{align}
\label{Master}
{\cal P}_{t}(k,\tau) = \frac{k^3}{2\pi^4 \Mp^4} \sum_{i,j} & \int_0^{\infty} \di p \, p^6 \int_{0}^{\pi} \di \theta \,  \sin^5 \theta \; \times\nn \\ & \Bigg|\int^\tau \mathrm{d}\tau_1 \,   g_k(\tau,\tau_1) \sum_{X}Q_{Xi}(p,\tau_1) Q_{Xj}(|\mathbf{k}-\mathbf{p}|,\tau_1) \Bigg|^2 ,
\end{empheq}
where we recall that $g_k$ is given in \eqref{eq:green-function}. We thus see that by considering a multisource scenario, not only does there appear a summation over the distinct scalar degrees of freedom sourcing GWs but, due to the mixing, also over the quanta comprising each source. Notably, when all contributions are of the same order, this leads to an enhancement of the power spectrum proportional to ${\cal N}^4$. Finally, let us remind the reader that, as we show in App.~\ref{app:in-in}, this expression can be obtained via the  in-in formalism at one loop.\footnote{This has also been discussed in App.~C of~\cite{Barnaby:2011qe} in the context of axion inflation.}\\[-0.3cm]

Taken at face value, the integrations in Eq.~\eqref{Master} lead to divergences in the hard-momentum limit $p\to\infty$ and the infinite past limit $\tau_1\to-\infty$. 
However, one has to keep in mind that our focus is on scenarios where only a short range of modes, starting from a given time, are enhanced. Thus, for arbitrary large values of the internal momenta $(p,|\bk-\bp|)$, the mode functions follow a dS-like evolution and can be treated in the same way as in standard inflationary scenarios (without an excited state). The discussion regarding the exact finite part present in the literature (see~\cite{delRio:2018vrj} for instance) would not affect in anyway our results since
the former is orders of magnitude suppressed compared to the contribution coming from the enhanced modes computed in the next section.
Hence, for practical purposes, one can simply regularise the integral by introducing a finite cutoff in momentum space.\footnote{When reasonably chosen, results are independent of the cutoff. Numerically, our choice is such that momenta are included up to the last enhanced modes. Results are then robust upon changes spanning several orders of magnitude around this value. This means that, as expected, including the contribution of modes that are in Bunch-Davies throughout their entire evolution does not influence the final result. This is valid until one picks an unreasonably large cutoff. Then the integral starts growing (slowly) due to the standard UV divergence that, once properly renormalised, would leave a subdominant finite contribution.}

Regarding the time integral: in the next section we will argue that for the cases studied in this work, the time $\bar\tau$ that we used in Eq.~\eqref{h-tauf} acquires a physical meaning (see also footnote~\ref{ft:tbar}). In the $\tau_1<\bar\tau$ domain (which we call the ``in region" in the next section), the mode functions follow again the standard dS evolution, rendering this contribution to $\p_t$ subdominant compared to the one coming from $\tau_1>\bar\tau$. This preferred time thus serves here as another cutoff ``shielding" the infinite past. As discussed in Sec.~\ref{ssec:int-inin} and App.~\ref{app:in-in}, neglecting the $\tau_1<\bar\tau$ contribution renders the power spectrum computed via the retarded Green's function, approximately equal to the one-loop, in-in power spectrum.\\

For later convenience, let us also define the following two sets of dimensionless momenta
\begin{align} \label{xypq}
    x = \frac{p}{k},\quad y = \frac{|\bk-\bp|}{k},
\end{align}
and 
\begin{align}\label{sd}
    s = x+y,\quad d = |x-y|.
\end{align}
Using these, the geometrical factor in \eqref{Master} becomes 
\begin{align}\label{Geom}
\int_0^{\infty} \di p \; p^6 \int_{0}^{\pi} \di \theta  \; \sin^5 \theta &= k^7\int _0^{\infty} \di y \int_{|1-y|}^{1+y} \di x  \; xy \left(\frac{4x^2-(1+x^2-y^2) ^2}{4}\right)^2 \\
&= \frac{k^7}{4^3}\int _0^{1} \mathrm{d} d \int_{1}^{\infty} \di s \; \left(s^2-1 \right)^2 \left(s^2-d^2 \right) \left(d^2-1 \right)^2.
\end{align}

\section{Stochastic gravitational wave background from excited states}
\label{sec:GW-ExSt}

We are interested in studying the stochastic background of gravitational waves sourced during inflation associated with the appearance of excited states at a given time along the inflationary history.
Since interactions among multiple fields provide a natural playground for excited states to arise, we exemplify our claims in Sec.~\ref{Explicit} with a two-field model in which a short period of strongly non-geodesic motion along the inflationary trajectory lies at the origin of the excited state. Let us, however, emphasise that all our main results regarding the shape of the signal follow solely from the presence of an excited state and are equally valid when the mechanism triggering it occurs in different (e.g.~single-field) scenarios. 

As we will see, the precise origin (multifield/single-field) as well as the particularities of each model are encapsulated in the explicit functional form of the Bogoliubov coefficients, and play an interesting but secondary role. There exist though a characteristic that is specific to multisourced GWs: we have seen that our  master formula~\eqref{Master} is a nontrival 
generalisation of the single source result owing not only to the summation over the various types of fluctuations but also over the various types of quanta. In ``democratic'' situations like the one studied in Sec.~\ref{Explicit}, this will in turn enhance the tensor modes by orders of magnitude (depending on the number of fields) due to a combinatorial factor.

\subsection{Dynamically generated excited states}
\label{sec:dyn-ex-state}

The main results of this paper only depend on the existence of a dynamically generated excited state, whose definition is simple: although all modes are initialised in the Bunch-Davies vacuum, a non-trivial inflationary dynamics is such that after some time, some sub-Hubble $k$-modes are not in their ground state anymore. A paradigmatic and physically motivated large class of models in which this mechanism may be at play is the one of sharp features, and for concreteness, we formulate things in this language in the following, although our results have a broad range of applicability. By a sharp feature, we mean a sudden  change 
in some background parameter $f(N)$, where $N$ henceforth denotes the number of \textit{e}-folds.\footnote{More precisely, the duration of the feature has to be small enough so that a given scalar mode whose sub-Hubble dynamics is perturbed (enhanced) during the feature is still sub-Hubble at the end of it.}
This could be any background quantity controlling the dynamics of the perturbations during inflation. Our special focus is on sharp features that at the same time (exponentially) enhance the power spectrum for a limited range of scales.

In order to see that this naturally leads to an excited state with a large amount of particle production, we consider the evolution of the various degrees of freedom by splitting the time domain in three regions (see Fig.~\ref{SharpFeat}): $in$ for $N<N_{\rm out}-\delta$ (with $\delta$ the duration of the feature), where the mode functions are placed in the BD vacuum in the infinite past
and obey the standard dynamics on a slowly changing inflationary background that we approximate with a de Sitter epoch; $feature$ for the narrow region when the feature is active, i.e.~between $N_{\rm out}-\delta<N<N_{\rm out}$; and $out$ for $N>N_{\rm out}$, where the dynamics is back to standard (like in the in region) but now with different initial conditions set at $N_{\rm out}$ by matching to the feature-region solution. 
\begin{figure}[t]
\centering
\includegraphics[width=0.8\textwidth]{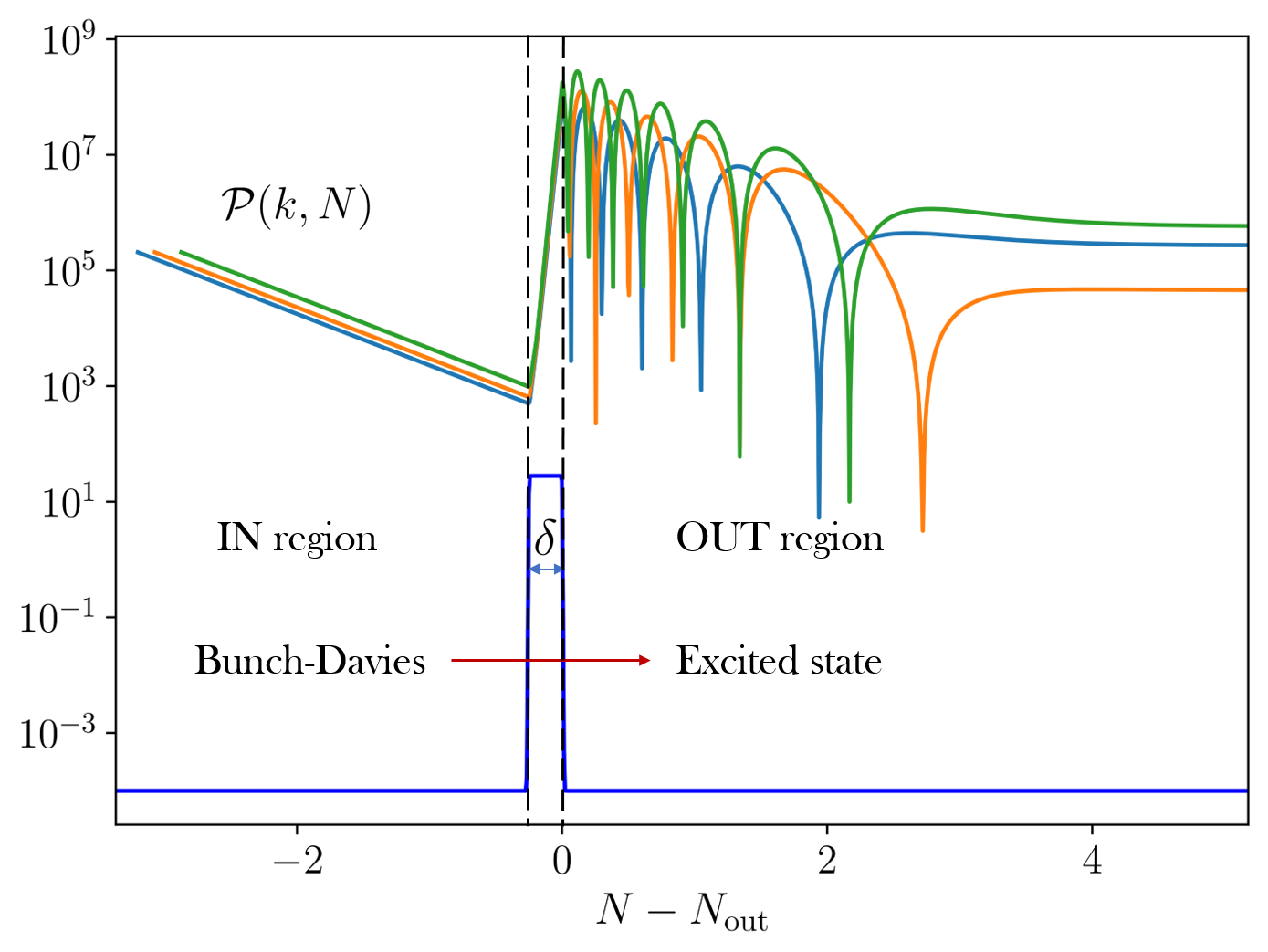} 
\caption{ 
\textit{Time evolution of the curvature perturbation power spectrum in presence of a sharp feature for three different $k$-modes (normalised to $\p_0$, i.e.~the single-field, slow-roll primordial power spectrum). In blue, an illustrative profile for the background parameter defining the feature. At the onset of the so-called \textit{out} region, the relevant modes are still deep inside the Hubble radius. By that time, the sharp feature has prepared the system in an excited state, so that the corresponding power spectrum oscillates with a $k$-dependent phase before freezing outside the horizon. Scalar induced GWs generated after inflation are only sensitive to the snapshot of the power spectrum of $\zeta$ at late times, i.e.~the primordial curvature power spectrum. Instead, GWs generated during inflation are sensitive not only to the various degrees of freedom entering in the energy-momentum tensor but also to their time evolution during inflation. The three curves have been computed explicitly by using the example of Sec.~\ref{Explicit} with parameters $(\etaperp,\delta)= (28,0.25) $.}}
\label{SharpFeat}
\end{figure}

The types of sharp features we are interested in are such that the maximally enhanced modes are deep inside the Hubble radius at the time of the feature. This naturally leads one to distinguish two relevant scales: $k_*$ marking the maximum of the scalar power spectrum, and $\kf$, the wavenumber of the mode that exits the horizon at $N_{\rm out}$. 
In order to quantify the hierarchy between them, let us introduce the parameter
\be \label{gamma}
\gam \equiv \frac{k_*}{\kf},
\ee
which will be useful when studying the enhancement of the GW energy density produced by excited states. Before discussing the behaviour of the mode functions appearing in the generic solution~\eqref{generic} in these three regions, let us re-write them as follows  
\begin{equation}\label{Qnormalized}
    Q_{Xi}(k,\tau) = \frac{H}{\sqrt{2 k^3}}\widetilde{Q}_{Xi}(k,\tau),
\end{equation}
where $H/{\sqrt{2k^3}}$ has been factored out for later convenience.

In the \textit{in} and \textit{out} regions, solutions take respectively the form (up to a global phase factor)
\begin{align}\label{Region1}
    \widetilde{Q}^{\mathrm{in}}_{Xi}(k,\tau) = \delta_{Xi} \zeta(k\tau),
\end{align}
and
\begin{align}\label{Region3}
    \widetilde{Q}^{\mathrm{out}}_{Xi}(k,\tau) = \alpha_{Xi}(k) \zeta(k\tau) +\beta_{Xi}(k) \zeta^*(k\tau).
\end{align} 
As before
\be \label{zbd}
\zeta(k\tau) =  e^{-i k\tau}(1+i k\tau), \quad\text{such that}\quad \zeta^*(k\tau)=\zeta(-k\tau) ,
\ee
is the standard de Sitter mode function. Note that we are considering scenarios in which the relevant enhanced modes are deep inside the Hubble radius at the onset of the \textit{out} region, which, as we will see, is the most relevant time for GW production since gradients suppress the source at subsequent times. There, it is thus a good simplifying approximation to use massless mode functions $\zeta(k\tau)$. It is straightforward in principle to include mass effects, but technically cumbersome with the appearance of Hankel functions throughout that would obscure the simple physics we want to describe.\\[-0.3cm]

The dS mode function and its complex conjugate provide two independent solutions to the corresponding equation of motion, so that $\widetilde{Q}^{\mathrm{out}}_{Xi}(k,\tau)$ is necessarily a linear combination of them, with coefficients $ \alpha_{Xi}(k)$ and $ \beta_{Xi}(k)$ called Bogoliubov coefficients. The latter are not arbitrary though, as they should be such that at any time, the $\hat{Q}_{X}({\bx})$ commute with one another, the same for their conjugate momenta 
$\hat{\pi}^{Q_{X}}({\bx})$, and that $\left[\hat{Q}_{X}({\bx}),\hat{\pi}^{Q_{Y}}({\bx}')\right]= i \delta_{XY}\delta(\bx-\bx')$ hold. This imposes the following set of relations (with an implicit sum over the repeated index $i$):
\begin{align}
\alpha_{Xi} \alpha_{Yi}^*-\beta_{Xi}^* \beta_{Yi}&=\delta_{XY}, \label{bog-constraint-1} \\
\alpha_{Xi} \beta_{Yi}^*-\beta_{Xi}^* \alpha_{Yi}&=0\,, \label{bog-constraint-2}
\end{align}
which are automatically satisfied by any unitary evolution from the \textit{in} to the \textit{out} region. In a single-field situation, they reduce to the well known relation $|\alpha|^2-|\beta|^2=1$, which generalises here, for any $X$, to
\bea \label{Bogo}
\sum_{i=1}^{\cal N} \left( \left| \alpha_{Xi}(k) \right|^2 - \left| \beta_{Xi}(k) \right|^2 \right) =1,
\eea
but we stress that the whole set of relations~\eqref{bog-constraint-1},~\eqref{bog-constraint-2} should hold.\footnote{The fact they are indeed satisfied in the example of Sec.~\ref{Explicit} provides a non-trivial check of the computations performed in~\cite{Palma:2020ejf,Fumagalli:2020nvq}.} We will explicitly verify them for any time in the example of Sec.~\ref{Explicit}.

In the \textit{in} region, the Bunch-Davies initial conditions \eqref{BD}, with $\alpha_{X i}(k) =\delta_{X i}$ and $\beta_{Xi}(k) = 0 $ for all $k$ modes, trivially satisfy \eqref{bog-constraint-1},~\eqref{bog-constraint-2}, with the mode functions in the $\tau\rightarrow -\infty$ limit behaving as the Minkowski positive frequency modes. Then, 
the vacuum associated with the operators $\hat{a}_i,\hat{a}_i^{\dagger}$ appearing in Eq.~\eqref{generic} corresponds to the lowest-energy state in the remote past, which we label as the \textit{in} vacuum.  
The latter also represents the time-independent state of the system (since, as it is customary, the dynamics is described in the Heisenberg picture). In the \textit{out} region, the dynamics is back to the standard one, with free fields propagating over a (quasi) dS background. Accordingly, the mode functions which behave as the positive frequency Minkowski modes in the remote past (here meaning, for each $k$ mode, $|k\tau|\gg 1$ although still $\tau>\tau_{\mathrm{out}}$) are analogous to the ones of the \textit{in} region given in Eq.~\eqref{Region1}.
Thus, one can write the fields operators in the \textit{out} region as
\be\label{bexp}
\hat{Q}_{\indX}(\bk,\tau>\tau_{\mathrm{out}})\equiv\hat{Q}^{\mathrm{out}}_{\indX}(\bk,\tau) = \frac{H}{\sqrt{2 k^3}}\sum_{j=1}^{\mathcal{N}}\left(\zeta(k\tau)\delta_{X j} \right)\hat{b}_j(\bk) + \mathrm{h.c.}(-\bk). 
\ee
where $\hat{b}_{i},\hat{b}_i^{\dagger}$ are a new set of creation/annihilation operators defining the vacuum (and the excited states) 
of the system in the \textit{out} region.
To find the relation between these new set of operators and the ones corresponding to the \textit{in} vacuum, one simply equates the expansion~\eqref{bexp} with the one in Eq.~\eqref{generic} after substituting the expressions~\eqref{Qnormalized}-\eqref{Region3} therein, giving:
\be
\hat{b}_X(\bk) =\sum^{\mathcal{N}}_{i=1} \left(\alpha_{X i} \hat{a}_i(\bk) + \beta^*_{X i} \hat{a}_i^{\dagger}(-\bk)\right)\,.
\ee
One can check that these operators satisfy the canonical commutation
relations~\eqref{quantiz} by virtue of Eqs.~\eqref{bog-constraint-1},~\eqref{bog-constraint-2}.
The mean number of particles for each species at the onset of the \textit{out} region is then given by the expectation value of the operators $\hat{N}_X=\hat{b}_X^{\dagger}\hat{b}_X$ in the state of the system i.e.~the \textit{in} vacuum), leading to
\be
n_X(k) =\sum_{i} |\beta_{X i}(k)|^2,
\ee
which is a generalisation of the standard result $n(k) = |\beta(k)|^2$ (see also e.g.~\cite{Nilles:2001fg}).

The occupation numbers $n_X(k)$ are exponentially greater than unity in the set-ups under investigation here. In fact, as we briefly review later, to have an enhancement of the power spectrum for a given range of modes due to the sharp feature, one needs $|\alpha(k)|\gg 1$ (removing indices for simplicity). This, together with the quantisation conditions \eqref{Bogo}, implies $|\beta(k)| \simeq |\alpha(k)| \gg 1$ for the relevant range of modes.
In addition, the matching of the mode functions at the onset of the \textit{out} region selects a $k$-dependent phase difference between the \bog \, coefficients (see for instance Sec.~2.3 of~\cite{Fumagalli:2020nvq} for a more detailed explanation):
\be\label{simpleBog}
\beta\simeq e^{2ik/\kf + i\varphi}\alpha,
\ee
with $\kf$ the scale corresponding to the time of the feature $N_{\mathrm{out}}$, i.e.~the mode $\kf$ exits the Hubble radius at $N_{\rm out}$, and $\varphi$ is a phase factor whose $k$-dependence is mild compared to the rapidly oscillating first term; we assume $\alpha$ real without loss of generality.

The exact time dependence of the mode functions in the region of the feature is model-dependent. However, since the features we are considering have the property to enhance the power spectrum of the scalar modes by several orders of magnitude during a short period of time, we parameterise this time dependence with an exponential enhancement, i.e.
\begin{align}\label{R2}
    \widetilde{Q}^{\mathrm{feat}}_{Xi}(k,\tau) = \tilde{Q}^{\mathrm{out}}_{Xi}(k,\tauf)e^{(N-N_{\rm out})g(k)}\simeq \tilde{Q}^{\mathrm{out}}_{Xi}(k,\tauf) \left(\frac{\tauf}{\tau}\right)^{\g(k)},
\end{align}
with $g(k)$ a model-dependent function of momentum.

Let us stress once more the sources of model dependence of the whole set-up under investigation, 
highlighting in parallel what is completely generic. 
Firstly, the precise functional form of the Bogoliubov coefficients in the \textit{out} region depends on the exact mechanism at hand. However, by inserting the parametrisation~\eqref{Region3} into Eq.~\eqref{Master} we derive generic formulae for the enhancement of the tensor modes that hold for arbitrary $k$-dependent coefficients. It is then reasonable, since one considers a narrowly peaked curvature power spectrum, to study the enhancement of tensor modes in the limit where these coefficients are peaked around a given scale $\kstar$. 
That will allow us to derive explicitly all the main features of the signal. However, as we stress in Sec. \ref{enhancement}, most of them are indeed independent from the shape of the \bog~coefficients. The second source of model dependence lies in the mode functions $\widetilde{Q}^{\mathrm{feat}}(k,\tau)$ in the region of the feature. As already stated, we parametrise the latter by means of Eq.~\eqref{R2}. In any case, the contribution to the tensor power spectrum coming from this region is subdominant (see discussion in Sec.~\ref{Region2}) and the main conclusion about the shape of $\OGW$ is robust against the exact form of this parametrisation.

\subsubsection*{Primordial power spectrum from a sharp feature}
Since the GWs sourced in the radiation era depend directly on the scalar power spectrum $\mathcal{P}_\zeta$, to facilitate the comparison with the inflationary-era sourced GWs, we close this section by briefly summarising the effect of a sharp feature on $\mathcal{P}_\zeta$ (see for instance ~\cite{Fumagalli:2020nvq} for more details).

For a sharp feature inducing an excited state, from the definition \eqref{eq:P-without-renormlaisation} and the solution \eqref{Region3} in the \textit{out} region, we have that
\be\label{Pzeta}
\p_{\zeta}  \equiv \frac{1}{2\epsilon \Mp^2}\p_{\zeta\zeta} =  
\p_0 \sum_i |\alpha_{\zeta i} + \beta_{\zeta i}|^2 \, ,
\ee
where the relevant quantities are evaluated at horizon crossing and $\p_0 = H ^2 / (8 \pi^2 \epsilon \Mp^2)$ is the single-field, slow-roll, dimensionless scalar power spectrum. As usual, dimensionless power spectra are defined as $\p_{XY} = \frac{k^3}{2\pi^2}P_{XY} $.

Suppressing the indices on the Bogoliubov coefficients for simplicity, one can expand the square in Eq.~\eqref{Pzeta} and write the power spectrum as 
\be
\label{eq:Pzeta-excited}
\p_{\zeta} \simeq \p_0 |\alpha|^2 \left( 1+\bigg|\frac{\beta}{\alpha}\bigg|^2+2 \bigg|\frac{\beta}{\alpha}\bigg| \cos\left(\frac{2 k}{\kf}\right)  \right) \, ,
\ee
where the appearance of the cosine is a direct consequence of the relation~\eqref{simpleBog}. For $|\alpha| \sim |\beta| \gg 1$, we have that ${|\beta|}/{|\alpha|} \simeq 1$ so that we can simplify this further to write 
 \be
 \label{eq:Pzeta-strong-feature}
\p_{\zeta} \simeq \frac{1}{2}\overline{\mathcal{P}}_\zeta \Bigg(1 + \cos \bigg( \frac{2 k}{\kf} \bigg) \Bigg) \, , \quad \textrm{with} \quad \overline{\mathcal{P}}_\zeta \equiv 2\cdot2 \p_0 |\alpha|^2 \, .
\ee
That is, for a sharp feature leading to a significant enhancement of fluctuations the scalar power spectrum takes the form of an enhanced envelope $\overline{\mathcal{P}}_\zeta$ modulated by sinusoidal oscillations in $k$ with unit amplitude. The frequency of this oscillation is $2/\kf$, i.e.~it is set by the scale of the feature.

For significant but not exponential particle production, i.e.~for $|\alpha| \sim |\beta| \sim 1$, the amplitude of oscillation is less than unity but still $\mathcal{O}(1)$. In that case, the power spectrum can still be boosted compared to its value at CMB scales as long as $\p_0$ is larger at the relevant scales, e.g.~if $\epsilon$ has a smaller value than when CMB modes cross the horizon.

\subsection{Post-inflationary generated GWs from excited states: brief review}
\label{sec:Omega-rad-review}
Excited states during inflation will also produce scalar-induced GWs in the post-inflationary era when the relevant fluctuations re-enter the horizon \cite{Acquaviva:2002ud, Mollerach:2003nq, Ananda:2006af, Baumann:2007zm}.\footnote{See also \cite{Domenech:2021and} for the impact of primordial dark matter isocurvature fluctuations on the scalar-induced SGWB generated after inflation.} The equation of motion for the tensor modes is again \eqref{eq:eom-h} but the source in this case depends to leading order on a four-point function of primordial curvature perturbations. If these are Gaussian, this can be written as a product of two instances of the scalar power spectrum, but in general there will also be a contribution proportional to the trispectrum, see e.g.~\cite{Garcia-Bellido:2017aan,Unal:2018yaa,Cai:2018dig,Atal:2021jyo,Adshead:2021hnm}. %Here we will ignore this latter more model-dependent contribution for simplicity, since in many explicit realisations of inflation it is expected to only give a small correction.
At CMB scales primordial fluctuations are highly Gaussian and one expects the trispectrum contribution to the GW spectrum to be negligible compared to the term involving the power spectrum. However, for the modes affected by the excited state this is not necessarily the case and the trispectrum contribution may become important. We leave this for future work.

The present-era fraction of energy density in GWs that were sourced in the post-inflationary era can then be written as~\cite{Ananda:2006af, Baumann:2007zm}:\footnote{Here we have rescaled the integration variables $(d,s)$ compared to their namesakes in the previous works \cite{Fumagalli:2020nvq,Fumagalli:2021cel} by a subset of the authors, to be consistent with the definition of $(d,s)$ in Eq.~\eqref{sd}.}
\begin{align}
\label{eq:OmegaGW-i}
    \Omegarad(k) = c_g \Omega_{\textrm{r},0} \int_0^1 \textrm{d} d \int_1^\infty \textrm{d} s \, \mathcal{T}_\textrm{rad}(d,s) \, \mathcal{P}_\zeta \bigg(\frac{k}{2}(s+d)\bigg) \mathcal{P}_\zeta \bigg(\frac{k}{2}(s-d)\bigg) \, .
\end{align}
The factor $c_g \Omega_{\textrm{r},0}$ relates the GW energy density fraction at the time of its sourcing in the radiation-dominated era to that of today --- c.f.~Sec.~\ref{today}.
The integration kernel $\mathcal{T}_\textrm{rad}(d,s)$ is computed in terms of time-integrals over the transfer functions and Green's function factors that relate the primordial curvature fluctuations to the source term in Eq.~\eqref{eq:eom-h}. It also includes a kinematic factor that arises from rewriting the integral over momenta in terms of the variables $(d,s)$ and an oscillation average has also been performed. The kernel depends on the equation of state of the universe when the relevant fluctuations re-enter the horizon. Assuming a standard thermal history of the universe, here we take this to be an era of radiation domination, in which case the relevant kernel is given in \cite{Espinosa:2018eve,Kohri:2018awv}. \footnote{The corresponding expressions for a different equation of state can be found in \cite{Inomata:2019zqy,Inomata:2019ivs,Domenech:2019quo,Domenech:2020kqm} and the resulting post-inflationary GW spectrum due to an excited state during inflation was analysed in \cite{Witkowski:2021raz}.}

One important property of $\mathcal{T}_\textrm{rad}(d,s)$ is the existence of a singularity for $s=\sqrt{3} $ signalling resonant amplification. As a result of this, for a finite-width peak in $\mathcal{P}_\zeta$ at $k=\kstar$ that is sufficiently narrow, the post-inflationary contribution to the GW spectrum will exhibit a narrow principal peak from resonant amplification at $k \simeq 2 \kstar / \sqrt{3}$. In addition, there is generically also a lower broad ``bump" around $k \approx \kstar / \sqrt{3}$. For broader $\mathcal{P}_\zeta$, the two peaks in $\Omegarad$ are increasingly ``blurred" into one by the double convolution in \eqref{eq:OmegaGW-i} so that one eventually finds a single broad peak in the GW spectrum around $k \sim \kstar$.

Let us now briefly review the effect of an excited state due to a sharp feature during inflation on the radiation-era GW spectrum~\cite{Fumagalli:2020nvq}. We consider the case where the scalar fluctuations are enhanced by the feature, i.e.~$|\alpha_{\zeta i}| \sim |\beta_{\zeta i}| \gg 1$. As described in Sec.~\ref{sec:dyn-ex-state}, this produces a peak in $\mathcal{P}_\zeta$ in virtue of $|\alpha_{\zeta i}| \gg 1$ that is further modulated by $\mathcal{O}(1)$ oscillations due to $|\beta_{\zeta i}| / |\alpha_{\zeta i}| \sim \mathcal{O}(1)$. The width of the peak $\Delta k$ is given by the range of scales affected by the feature. One strategy for understanding the corresponding GW spectrum is to treat the oscillation in $\mathcal{P}_\zeta$ as a series of individual peaks. These can in general be taken as narrow peaks in the sense described in the paragraph above, as every such spike only covers a fraction of $\Delta k$, which for a sharp feature already corresponds to a narrow interval in general. Every such peak  in $\mathcal{P}_\zeta$ then produces its own resonance peak in $\Omegarad$. There are also further resonance peaks from interactions between different spikes coming from the two factors of $\mathcal{P}_\zeta$ in Eq.~\eqref{eq:OmegaGW-i}~\cite{Cai:2019amo, Fumagalli:2020nvq}.

The sum of these contributions then gives rise to a spectral shape of $\Omegarad$ which for the scales of maximal enhancement is well-reproduced by the template \cite{Fumagalli:2020nvq}:
\begin{align}\label{Omegaradfeature}
\Omegarad (k) \underset{k \sim 2 \kstar / \sqrt{3}}{=} \overline{\Omega}_\textrm{GW}^{\textrm{ rad}} (k) \bigg[ 1 + \mathcal{A}_\textrm{lin} \cos \bigg(\frac{2 \sqrt{3} k}{\kf} + \phi \bigg) \bigg] \, .
\end{align} 
The sinusoidal modulation arises from the superposition of the various resonance bumps, with the maxima of the cosine coinciding with the maxima thereof. Note that the frequency of oscillation is increased by a factor $\sqrt{3}$ compared to that of the oscillation in $\mathcal{P}_\zeta$. As the resonance peaks have finite width, their multiple superposition has an effect of averaging out the modulation and suppressing the amplitude $\mathcal{A}_\textrm{lin}$. Even for the minimal case with just an $\mathcal{O}(1)$ number of oscillations within the interval $\Delta k$, one finds that at best $\mathcal{A}_\textrm{lin} \sim \mathcal{O}(20 \%)$, as can be seen from the green curve in Fig.~\ref{fig:Full28}. This decreases further as the frequency of the oscillation is increased. The smooth background spectrum $\overline{\Omega}_\textrm{GW}^{\textrm{ rad}} (k)$ can be shown to be given by the GW spectrum due to the smooth background of the scalar power spectrum $\overline{\mathcal{P}}_\zeta(k) \sim |\alpha_{\zeta i}(k)|^2$, which here is given by a peak of width $\Delta k$ with maximum at $k=\kstar$. If this is narrow, i.e.~$\Delta k / \kstar < 1$, the GW spectrum $\overline{\Omega}_\textrm{GW}^{\textrm{ rad}}$ takes the usual form of a broad lower bump at $k \sim \kstar / \sqrt{3}$ and a principal resonance peak at $k \simeq 2 \kstar / \sqrt{3}$, see again the green curve in Fig.~\ref{fig:Full28}.

\subsection{Different contributions to $\Omegainf$}
Let us now turn back our attention to the GWs sourced during inflation. By inserting Eq.~\eqref{Qnormalized} into Eq.~\eqref{Master} with the change of variables \eqref{sd} ---or \eqref{Geom}--- we can rewrite the tensor spectrum   
$\p_t(k)\equiv\lim_{k\tau\to 0}\p_t(k,\tau)$ as
\bea \label{Master2}
\p_t(k) &=&   \frac{H^4}{8\pi^4\Mp^4} \int _0^{\infty} \di y \int_{|1-y|}^{1+y} \di x  \;  \muu(x,y) \times 
\nn\\
&& \sum_{i,j} \Bigg| \sum_{X}  \int_{-\infty}^0 \di z' \; { \widetilde{Q}}_{X i}(kx,z'/k){ \widetilde{Q}}_{X j }(ky,z'/k) G(0,z') \Bigg|^2,
\eea
with the geometrical factor given by
\begin{align}
    \muu(x,y) = \frac{\left(4x^2-(1+x^2-y^2)^2\right)^2}{(4x y)^2} = \frac{(d^2-1)^2(s^2 -1)^2}{(s^2-d^2)^2}.
\end{align}
Furthermore,
\begin{align} \label{Gzz}
    \quad G(z,z') = k g_k (\tau,\tau')= i \frac{\zeta(z) \zeta^*(z') - \zeta^*(z) \zeta(z')}{2 z'^2 } \Theta(z-z');
\end{align}
in particular
\begin{align} \label{Gzz0}
    \quad G(0,z) = k g_k (0,\tau)= \frac{z\cos z - \sin z}{z^2} , 
\end{align}
where we have introduced the dimensionless variable $z$, defined as
\be
z \equiv k \tau.
\ee 

The time integral inside the modulus square in Eq.~\eqref{Master2} can be split by exploiting the piecewise construction of the solution, taking into account the different time dependence in each region. By introducing
\be
z_{\rm in }= e^{\delta} \zf, 
\ee
(with $\delta$ accounting for the small but finite duration of the feature), the total result can be organised as follows:
\bea \label{main-34}
\p_t(k) = \p_t^{\mathrm{ out}}(k)+\p_t^{\mathrm{ feat}}(k)+\p_t^{\mathrm{mix}}(k).
\eea
The constituent power spectra (let us henceforth suppress the integration limits), 
\bea\label{constituent}
\p_t^{\mathrm {x}}(k) =\frac{H^4}{8\pi^4\Mp^4}\int \di x \int \di y \; \muu(x,y) \Delta^{\mathrm{x}}_t,
\eea 
indexed by $\mathrm{x} =  \{\rm out/feat/mix\}$, are given by
\begin{align} 
\label{Dout}
\Delta^{\mathrm{out}}_t &= \sum_{ij} \left|  I_{ij}(xk, yk; \zf,0) \right|^2 \equiv \sum_{ij} \left|I_{ij}^\mathrm{out} \right|^2 ,
\\
\label{Dfeat}
\Delta^{\mathrm{feat}}_t &= \sum_{ij} |  I_{ij}(xk, yk; z_{\mathrm{in}},z_{\mathrm{out}})|^2\equiv\sum_{ij} \left|I_{ij}^\mathrm{feat}\right|^2 ,
\\
\label{Dmix}
\Delta^{\mathrm{mix}}_t &= 2\mathrm{Re} \left(\sum_{ij}  I_{ij}^{\mathrm{feat}}\left(I^{\mathrm{out}}_{ij}\right)^* \right) ,
\end{align}
with
\bea \label{time-int}
I_{ij}(xk,yk;a,b) \equiv \sum_{X}  \int_{a}^b \di z \; { \widetilde{Q}}_{X i}(xk,z/k){ \widetilde{Q}}_{X j }(yk,z/k) G(0,z),
\eea
where $\widetilde Q$ are the mode functions~\eqref{Qnormalized} corresponding to each region.

Consistently with the conditions of applicability of our formalism, that we spelled out in \eqref{sec:general-result}, note that we disregard contributions from the \textit{in} region. It is instructive to consider separately the contributions coming from the excited states alone, i.e.~$\Delta^{\rm out}_t$, the one that takes into account the finite time to ``excite" these states, $\Delta^{\rm feat}_t$, and consequently the mixed contribution. 

In order to see why excited states naturally lead to an enhancement in the sourced gravitational-wave background, let us first focus on the contribution coming from the \textit{out} region $\p_t^{\mathrm{out}}$. This is expected to provide the dominant term in the sum \eqref{main-34} since it is there that the excited states have support. This statement will be proved later while computing the contribution from the region of the feature in Sec.~\ref{Region2}. As already mentioned, the model dependence of $\p_t^{\mathrm{out}}$ only lies in the explicit functional form of the Bogoliubov coefficients. In this section we are thus able to derive generic formulas valid for any set of excited states. Furthermore, by considering a scalar power spectrum peaked around a given scale (of which a sharp feature is only a particular case) we provide explicit analytical approximations that capture the main characteristics of the signal in Sec.~\ref{Detailed}. 
\subsection{Integrating over internal time}
\label{int-time}
Let us begin by noticing that the time integral in $\Delta_t^{\rm out}$ given in Eq.~\eqref{Dout}, can be performed analytically. 
Plugging Eq.~\eqref{Region3} and the de Sitter Green's function~\eqref{Gzz} into Eq.~\eqref{time-int}, and relabeling the Bogoliubov coefficient associated with particle production as $\alpha_{\indX i}^+ \equiv \alpha_{\indX i}$ and $\alpha_{\indX i}^- \equiv \beta_{\indX i}$, allows us to write $\p_t^{\mathrm{out}}$ in a compact way as follows:
\bea \label{main-5}
\p_t^{\mathrm{out}}(k) =&& \frac{H^4}{8\pi^4\Mp^4} \int \di x \int \di y  \; \muu(x,y) \; \times 
\nn\\&&
\sum_{i,j} \Bigg| \sum_{X;\,{\rm s_{1,2}=\pm}}  \!\!\!\!{ \alpha}_{Xi}^{\rm s_1}(xk) { \alpha}_{Xj}^{\rm s_2} (yk)  {\cal G}({\rm s_1} x,{\rm s_2} y,\zf)\Bigg|^2.
\eea
This generic formula for the tensor power spectrum in presence of excited states is one of the main results of this work. The time integral has been factored out as 
\begin{align}
 \label{G-1}
{\cal G}(x, y, \zf)\equiv \int_{\zf}^0 \zeta(xz) \zeta(yz) G(0,z)  =  \int_{\zf}^0 \frac{\di z}{{z}^2} \; \zeta(xz) \zeta(yz) \frac{\zeta(z) - \zeta^*(z)}{2i}.
\end{align}
Note that from this definition it follows that
% %
\begin{align} \label{Gs}
{\cal G}(x,y,\zf) &= {\cal G}(y,x,\zf), 
 \nn\\
{\cal G}(-x,-y,\zf) &= {\cal G}^*(x,y,\zf),
 \nn\\
{\cal G}(x,-y,\zf) &= {\cal G}^*(-x,y,\zf).
\end{align}
% %
%

The integral~\eqref{G-1} can be computed explicitly:\footnote{Note that for a fixed external momentum $k$, $\cG$ diverges in the $x,y \rightarrow \infty $ limits. However, as discussed in Sec. \ref{sec:general-result}, a finite cutoff for the internal scalar momenta $x k$ and $y k$ has to be imposed when computing $\p_t$.  
In addition, $x\rightarrow \infty$ but $x k = \mathrm{const}$ corresponds to $k \rightarrow 0$, this long wavelength limit sends to zero $\mathcal{G}$ of Eq.~\eqref{G-1} via its third argument $\zf=k\tauf$, i.e.~the integration domain in Eq.~\eqref{G-1} collapses to a point.}
\be \label{G-I}
{\cal G}(x,y,\zf) = {\cal K}(x,y) -  {\cal F}(x,y,\zf) - {\cal F}^*(-x,-y,\zf),
\ee
with
\be \label{Kxy}
{\cal K}(x,y) = 
\frac{1-2xy-(x+y)^2}{\left( 1-(x+y)^2 \right)^2},
\ee
and
\begin{align} \label{Fxy}
{\cal F}(x,y,z) &= \frac{e^{-i(1+x+y)z}}{2(1+x+y)^2}  \; \times
\\ &
\left(i\frac{(1+x+y)^2}{z} -ixy(1+x+y)z - x-y-(x+y)^2-xy(2+x+y) \right).\nn
\end{align}
For simplicity, let us focus only on one component of the spectrum corresponding to a single quantum degree of freedom. To that end, we will drop the indices from $\alpha_{\indX i}^\pm$ by implicitly fixing them to, say, $\indX=\zeta$ and $i=1$, e.g.~$\alpha_{\zeta 1}^+\equiv\alpha$ and $\alpha_{\zeta 1}^-\equiv\beta$. The \textit{out}-region spectrum~\eqref{main-5} will contain eight terms and their complex conjugates, which can be written as
\bea\label{complete}
\p^{\rm out}_t(k) &=& \frac{H^4}{8\pi^4\Mp^4} \int \di x\int \di y \; \muu(x,y) \; \left| {\alpha} (xk) \right|^2|{\alpha} (yk)|^2 \times \nn
\\ 
&& \Big(  \left(1+|\ra(xk)|^2 |\ra(yk)|^2 \right) \left| \mathcal{G}(x,y) \right|^2 + 2 {\rm Re} \left[ \ra^*(xk) \ra^*(yk) \mathcal{G}^2(x,y) 
 \right]  \nn\\
&+& 2 {\rm Re} \left[ \left(1+|\ra(xk)|^2\right)\ra^*(yk) \mathcal{G}(x,y) \mathcal{G}(-x,y) +    \left( 1+|\ra(yk)|^2 \right) \ra^*(xk) \mathcal{G}(x,y) \mathcal{G}(x,-y) \right] 
\nn\\
&+& \left(|\ra(xk)|^2+ |\ra(yk)|^2\right) \left| \mathcal{G}(x,-y)\right|^2 + 2 {\rm Re} \left[ \ra^*(xk) \ra(yk) \mathcal{G}^2(x,-y) \right] 
\Big),
\eea
where we have defined 
\be
\rho \equiv \frac{\beta}{\alpha},
\ee
which keeps track of terms related to particle production induced by the excited state.
Note that we have also omitted the third argument in $\mathcal{G}$.

\subsection{Enhancement of the tensor spectrum from excited scalar states}\label{enhancement}
The prefactor outside the integral in Eq.~\eqref{complete} can be rewritten as $H^4/(8\pi^4\Mp^4) = 8\epsilon^2\p_0^2$,
which, together with Eq. \eqref{eq:Pzeta-strong-feature} for $|\alpha|^2$, confirms the expectation that the tensor power spectrum is proportional to $\epsilon ^2 \mathcal{P}_\zeta ^2$. The sum in parenthesis then determines the enhancement of the tensor power spectrum with respect to this naive expectation.  
Note that there are six different combinations of $\mathcal{G}$'s weighing different components of the spectrum:
\be \label{G-order}
\Big|{\cal G}(x,y)\Big|^2, \Big|{\cal G}(x,-y)\Big|^2, {\cal G}(x,y){\cal G}(-x,y), {\cal G}(x,y){\cal G}(x,-y), {\cal G}^2(x,-y), {\cal G}^2(x,y).
\ee

To better appreciate the peculiarity of excited states, let us write the time integral explicitly:
\be\label{cG-2}
\mathcal{G}(x,\pm y,\zf) =  \int_{\zf}^0 \frac{\di z}{z^2} \; e^{-i(x\pm y)z}( 1 + ixz)(1 \pm iyz) \mathrm{Im}[\zeta(z)].
\ee
When the momenta of the scalar modes running in the loop are deep inside the horizon, i.e.~when $x z \gg 1$ and $y z \gg 1$, the highly oscillating exponential damps the value of the integral for non-excited states where only the $(+)$ contribution is present. In contrast, excited states add contributions with a $(-)$ sign. These are terms in which the constructive interference between positive and negative frequency modes enhances the integral in the large argument regime, i.e.~$|\mathcal{G}(x,-y)| \gg |\mathcal{G}(x,y)|$ for large $x$ and $y$, and close to the resonant point $x \simeq y$ where the frequency of the oscillating piece is small. This effect of constructive interferences is well known from studies of primordial non-Gaussianities in the presence of excited states, resulting in an enhancement of the bispectrum near flattened configurations~\cite{Chen:2006nt,Holman:2007na,Meerburg:2009ys,Agarwal:2012mq,Ganc:2011dy,Flauger:2013hra,Aravind:2013lra}. In fact, the argument of the square in Eq.~\eqref{main-5} contains the tree-level, tensor-scalar-scalar correlator (this could be made explicit via the cutting rules of~\cite{Melville:2021lst,Goodhew:2021oqg,Baumann:2021fxj} applied to the correlator itself).
The similar enhancement of the latter due to excited states leads to the amplification of the tensor power spectrum that we discuss here.

Let us recall that, in general, to have a potentially observable $\OGW$, a significant enhancement of $\p_{\zeta}$ with respect to the single-field, slow-roll result $\p_0$ is required (assuming that the latter corresponds to the CMB pivot value). As reviewed at the end of Sec.~\ref{sec:dyn-ex-state}, in our context, this automatically implies a large amount of particle production, i.e.~$|\alpha| \simeq |\beta| \gg 1 $, and $|\rho| \simeq 1 $. Thus, let us now focus on this arguably most interesting case. Then, for $|\rho| \simeq 1 $, and recalling that $|\mathcal{G}(x,-y)|$ generally dominates over $|\mathcal{G}(x,y)|$ over most of the integration range, it follows that $\p^{\rm out}_t$ is dominated by the last line in \eqref{complete}:
\begin{empheq}[box=\graybox]{align}
\label{complete2}
\p^{\rm out}_t(k) = \frac{H^4}{4\pi^4\Mp^4} 
\int _0^{\infty} \di y \int_{|1-y|}^{1+y} \di x \; \muu(x,y) \; \left| {\alpha} (xk) \right|^2|{\alpha} (yk)|^2 \times \nn\\ 
\Big( \left|\mathcal{G}(x,-y)\right|^2 + {\rm Re}\left[   e^{i(\theta(xk) - \theta(yk))}  \mathcal{G}^2(x,-y) \right]
\Big),
\end{empheq}
where $\theta\equiv \mathrm{Ph}(\beta/\alpha)$. Interestingly, the GW spectrum in this case exhibits many universal features independently of the shape in $k$ of the \bog~coefficients, as can be seen in Fig.~\ref{Enhancement-different}, where we plot $\Omegainf (k)$ for various realisations of an excited state with $|\alpha|^2\simeq |\beta|^2 \gg 1$. 
\begin{figure}[t]
    \centering
    \includegraphics[width=0.9\textwidth]{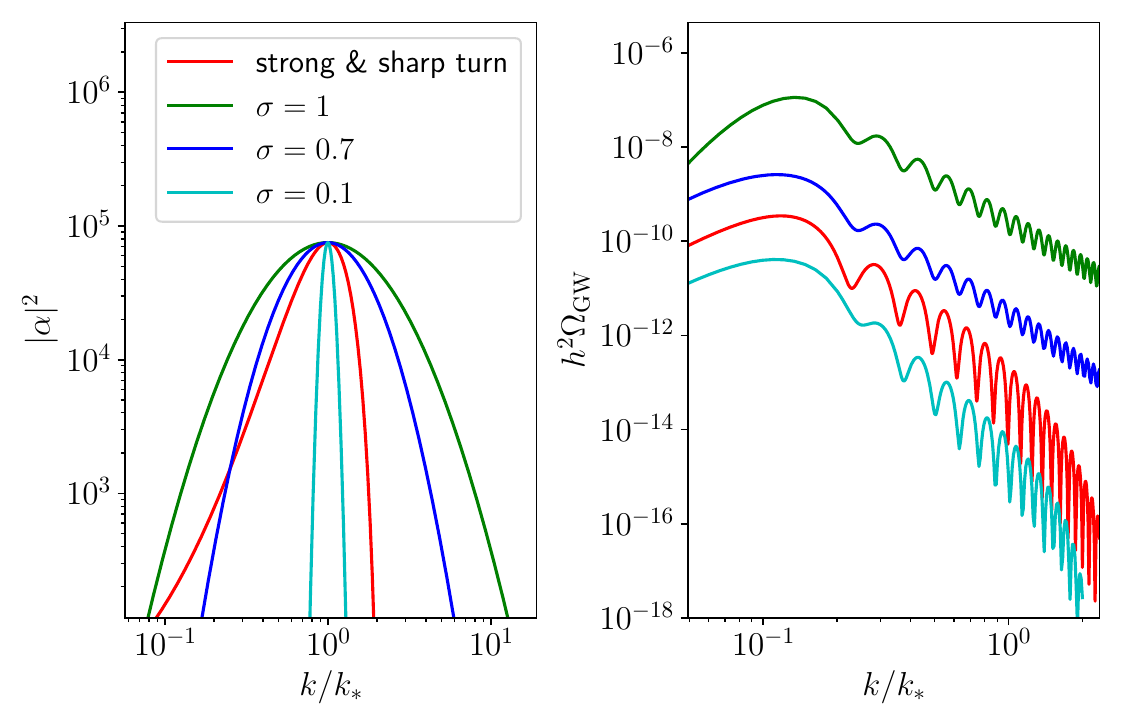}
    \caption{\textit{ Results for $\Omegainf$, computed via \eqref{complete}, as induced by the presence of different excited states, characterised by different profiles of the \bog~coefficients, with $|\alpha|^2\simeq |\beta|^2 \gg 1$. Here we consider $|\alpha(k)|^2 = \overline{\alpha}^2 \exp\left[-\tfrac{1}{2 \sigma^2}\ln^2 (k/k_*)\right]$
    for different values of $\sigma$ and we took $\mathrm{Ph}(\beta/\alpha)=0$ for simplicity, and we also include $|\alpha|^2$ and $\Omegainf$ from the example \ref{Explicit} with $(\etaperp,\delta)=(28,0.25)$ (red). 
    The main finding is that the principal properties of $\Omegainf$ are independent of the precise shape of $|\alpha(k)|^2$, i.e.~$\Omegainf$ exhibits a peak at $k \sim \kf \ll \kstar$ and a UV tail modulated by $\mathcal{O}(1)$ oscillations with frequency $\omega = 2 / \kf$. Although the precise fall-off of the GW spectrum in the UV depends on the shape of the \bog~coefficients, its behaviour is still universal under the (motivated) assumption that the latter are peaked around $k_*$. In fact, for the two examples with the narrowest peaks in $|\alpha (k)|^2$ one can check that the GW spectra become near-identical and only differ by a numerical factor.}} 
    \label{Enhancement-different}
\end{figure}
The GW spectrum exhibits a maximum that can be checked to occur at $k \sim \kf$. Furthermore, there are oscillations on the UV tail with a relative amplitude that quickly approaches unity along the tail. The frequency of the oscillation is the same in all cases and is given by $\omega = 2 / \kf$. 

The appearance of the oscillations on the tail can be understood to arise from the properties of the kernel $\mathcal{G}(x,-y)$. From the explicit expressions in \eqref{G-I}-\eqref{Fxy}, note that the contribution ${\cal F}(x,-y, \zf)$ exhibits an overall prefactor $e^{-i \zf}$ that does not depend on the integration variables $(x,y)$. It is this prefactor that is unaffected by the integral over $(x,y)$ that is responsible for terms of type $\sim \cos (2 \zf) = \cos (2k / \kf )$ in $\Omegainf (k)$, the factor of 2 coming from the fact that the integration kernel contains is quadratic in $\mathcal{G}(x,-y)$.

We will revisit this in more detail when specialising to peaked \bog~coefficients below. The main point of this discussion here and of Fig.~\ref{Enhancement-different} is that many of the properties of  $\Omegainf$ that we will describe in more detail for  peaked \bog~coefficients below, qualitatively also hold for broad profiles for $|\alpha (k)|^2$, like the example with $\sigma=1$ in Fig.~\ref{Enhancement-different}.

We now restrict attention to setups that produce a peak in $\p_\zeta(k)$ at $k=k_*$, or equivalently in $|\alpha(k)|^2$, ---as induced here by the excited state. In this case one can give an approximate expression for the integral in Eq.~\eqref{complete2}, see App.~\ref{Appapprox} for further details. The upshot is that the peaked \bog~coefficients select the preferred locus $kx=ky=k_*$, and thus we can use 
\begin{align}
\label{theapprox}
  \frac{H^4}{8\pi^4\Mp^4}\int \di x \int \di y  \, | \alpha (k x)|^2 | \alpha (k y)|^2 \, F(x,y,k) \approx 2 \epsilon ^2 A^2 \, \kappa^{-1} \, F\big(\kappa^{-1},\kappa^{-1},k \big)  \, , 
\end{align}
for any function $F$. We also used that from Eq.~\eqref{eq:Pzeta-strong-feature}, the amplitude of the scalar power spectrum is given by $\bar{\p}_{\zeta} \simeq 2\p_0|\alpha|^2$.
Omitting the bar from now on, we thus have $|\alpha|^4 H^4/(8\pi^4\Mp^4)  \simeq 2\epsilon^2 \p_\zeta^2$. We then defined
\be\label{Adef}
A \equiv 2 \p_0 \int \di\x \, |\alpha(k)|^2 \approx \p_{\zeta\,*} \, ,
\ee
where $\p_{\zeta\,*}\equiv\p_\zeta(k_*)$ and $\x = k/k_*$. To be complete, note that the approximation in \eqref{theapprox} is valid when the contribution to the integrand apart from the $|\alpha|^2$-terms varies sufficiently slowly.

In the next section, starting from the approximation \eqref{theapprox}, we will be able to derive analytically the main features regarding the shape of the GW spectrum. We stress that we do not need to use the explicit relationship \eqref{simpleBog} between the phases of Bogoliubov coefficients induced by the presence of a sharp feature to draw our main conclusions. In contrast, this phase difference is at the root of the oscillations in $\Omegarad$.

We can also use the analytical form of ${\cal G}(x,y,\zf)$ in \eqref{G-I} to understand how the $(+)$ and $(-)$ contributions individually affect the tensor power spectrum. Let us focus on the locus $x \simeq y$ and distinguish two cases: $k\simeq \kf$ and $k\simeq k_*$ meaning momenta of the GW close to the one of the feature and to the maximum of the scalar power spectrum, respectively. 
In both situations, the momenta of the sourcing scalar modes are $x k \simeq k_*$ so that $x \zf \equiv x k \tauf \simeq k_* \tauf \gg 1$. Thus, for $k\simeq \kf$ ($|\zf|\simeq 1$), $x \gg 1$. In this case, $\mathcal{K}(x,x) \propto 1/x^2$ while $\mathcal{F}(x,x) \propto x$.
In contrast, the same functions with a minus sign in one of the argument -- signature of an excited state -- scale as $\mathcal{K}(x,-x)\propto x^2$ and $\mathcal{F}(x,-x) \propto e^{-i\zf} x^2$. 
In the range $k\simeq k_*$, $x\simeq 1$ and $|\zf| \gg 1$, the function $\mathcal{K}$, which is independent of $\zf$, is not relevant and we have $\mathcal{F}(x,x) \propto e^{-i(1+ {\cal O}(1))\zf} \zf$ and $\mathcal{F}(x,-x) \propto e^{-i\zf} \zf$. Here, the $(+)$ and $(-)$ contributions have similar size and the higher frequency (due to $|\zf|\gg 1$) makes the oscillations visible in the GW signal (see explicit details below). 
We summarise this discussion in Fig.~\ref{Enhancement} from which it is clear that, for a given strength of the scalar source, i.e.~parameterised by $\p_{Q}\sim \epsilon\p_{\zeta}\simeq \epsilon A$, tensors are orders of magnitude enhanced in presence of an excited state.\\[-0.3cm]
\begin{figure}[t]
    \centering
    \includegraphics[width=0.8\textwidth]{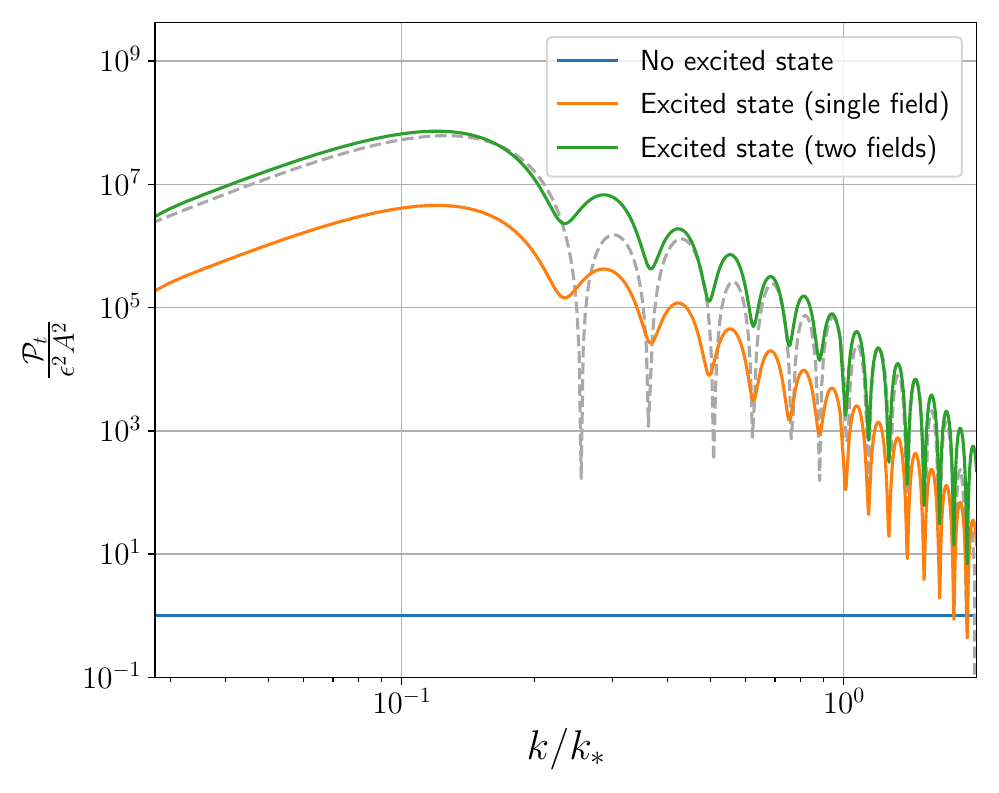}
    \caption{\textit{Enhancement of $\p_t$ boosted by an excited scalar state normalised to the naive expectation $\sim \epsilon^2 A^2$, where $A$ is an estimate of the strength of the primordial scalar power spectrum $\p_{\zeta}$ ---see Eq.~\eqref{Adef}.
    The orange and green curves are computed with the \bog \, coefficients in Sec.~\ref{Explicit} with parameters $(\etaperp,\delta)=(28,0.25)$ for a single field or a two fields scenario respectively. In the latter case, $\p_t$ is enhanced by an extra factor $\mathcal{N}^4\sim 16$ as expected. The grey dashed line labels the analytical approximation in Eq.~\eqref{R3approx}.}}
    \label{Enhancement}
\end{figure}

Having discussed the case of large of particle production, one may entertain the possibility that the scalar power spectrum is enhanced via other mechanisms (such as a change in the Hubble flow) so that $\p_ 0 $ for the scales of interest is already much larger than the power spectrum at CMB scales. In this framework, a non-adiabatic transition generating excited states does not necessarily need to enhance the power spectrum by several orders of magnitude in order to achieve observationally relevant values of $\OGW$. 
Thus, given the generic framework of this section, let us briefly comment on the case where a transient non-adiabatic evolution leads to a small amount of particle production, i.e.~$|\beta|\ll |\alpha|$ (or equivalently $|\rho|\ll 1$). When $\beta = 0$, i.e.~the modes are in a Bunch-Davies state, only the first term in \eqref{complete} contributes and, for a fixed $\p_\zeta$, the tensor power spectrum is damped with respect to the case $|\rho|\simeq 1$. With a small amount of particle production $\rho \ll 1$ (due to the enhancement of scalar modes on sub-Hubble scales), if the enhancement proper to excited states is sufficiently large so that $|\mathcal{G}(-x,y)/ \mathcal{G}(x,y)|> \rho^{-1}$, then the would-be small corrections linear in $\rho$ in Eq.~\eqref{complete} might become dominant, resulting in different shapes of $\p_t$. 
In the limit $x\simeq y$, Eq.~\eqref{complete} becomes
\begin{align}\label{terms}
\p^{\rm out}_t(k) &\simeq   \frac{H^4}{8\pi^4\Mp^4} \int \di x\int \di y  \left| {\alpha} (xk) \right|^2|{\alpha} (yk)|^2 \muu(x,y) \Big( \left(1+|\ra|^4\right) \left| \mathcal{G}(x,x) \right|^2 \\ \nn 
&+  4 |\ra|^2 \left| \mathcal{G}(x,-x)\right|^2 +4 {\rm Re} \left[ \ra^*(1+|\ra|^2) \mathcal{G}(x,x) \right]  \mathcal{G}(-x,x)  +2 {\rm Re} \left[ \ra^* \ra^* \mathcal{G}^2(x,x) 
 \right]
\Big).
\end{align}
It is worth stressing that, in the limit of small $\rho$, the conclusions drawn from the previous formula about the shape of $\p_t$  should be only taken as qualitative. In fact, in this limit, the concept of a preferred time $\tauf$ becomes ill-defined.
That was the key assumption behind the use of the classical Green's function method, see Sec.~\ref{sec:multi} and App.~\ref{app:in-in}, and the possibility to neglect the infinite past contributions therein. 
Thus, the rigorous treatment of this case requires a proper full in-in ``quantum" computation for small $\rho$ that is beyond the scope of the current work.

\subsection{Detailed structure of the spectrum}\label{Detailed}
We now have all the necessary ingredients to extract the overall observable structure of the tensor power spectrum. We will derive analytically the position and the scaling of the peak, which occurs around $k\simeq \kf$, as well as the frequency of the oscillations occurring at smaller scales around $k\simeq k_*$.

\subsubsection{Out-region contribution: maximum and oscillations}
Let us first focus on the \textit{out} region.
By means of the approximation \eqref{theapprox}, Eq.~\eqref{main-5} (in the limit of large particle production) can be easily integrated. The tensor power spectrum becomes

\be \label{R3approx}
\p^{\rm out}_t(k) \simeq 8 \epsilon^2 \mathcal{N}^4 A^2 \x^{-1}\muu(\x^{-1},\x^{-1}) \left| \mathcal{G}(\x^{-1},-\x^{-1},\zf = -\kappa \gam )\right|^2,
\ee
where we have assumed that all fields and quanta contribute equally so that $\mathcal{N}$ is a place holder for the number of fields. 

The explicit expression of the function $\cG$ appearing above simplifies considerably at the point imposed by the approximation \eqref{theapprox}, i.e.~$x\simeq y\simeq 1/\x$, and we stress once more that the minus sign in the second argument of $\cG$ is related to the negative frequency mode present \textit{only} in the case of excited states. After some manipulations, it is useful to rewrite $\cG$ in \eqref{R3approx} as follows: 
\be\label{cG1}
\Big|\cG\left(\x^ {-1},-\x^{-1},-\x \gam \right)\Big| = \gam^2\left[ \frac{\sin(\gam\x)}{\gam\x} -\frac{2\left(1- \cos(\gam\x) \right)}{(\gam\x)^2} -\frac{1}{\gam^2} \left( 1 - \frac{\sin(\gam\x)}{\gam\x}\right)\right],
\ee
where, as before, $k_*$ marks the maximum of the primordial scalar power spectrum $\p_{\zeta}$, while $\gam = - \kstar \tauf = \kstar / \kf $ parameterises how deep inside the horizon was the maximally enhanced mode $\kstar$ at the onset of the \textit{out} region. Recall that we consider $\gam \gg 1$ so that the last term in the previous expression can be safely neglected for all practical purposes. 

By inserting Eq.~\eqref{cG1} into \eqref{R3approx} and \eqref{redshift2}, we obtain an explicit analytical template for $\Omegainf$ that matches remarkably well the numerical results for the full spectrum (see Figs. \ref{fig:Full28}, \ref{AnApprox_05_14} and \ref{Benchmark1LISA}):
\be\label{template1}
\Omegainf = \bar{\Omega}\frac{1}{\x^3}\left(1-\frac{\x^2}{4} \right)^2 \cdot \left(\sin(\gam \x)-2\frac{(1-\cos{\gam \x})}{\gam\x}\right)^2 \, ,
\ee
with $\bar{\Omega}= 8 r_i \epsilon^2 A^2 \mathcal{N}^4
\gam^2$. Further, one may estimate the behaviour of the envelope by sending $\sin(\gam\x) \rightarrow 1$ and $\cos(\gam\x) \rightarrow 1$. This leads to the following simple expression 
\be\label{envelope}
\Omega_{\mathrm{GW}}^{\mathrm{inf-env}}\simeq \bar{\Omega} \frac{1}{\x^3}\left(1-\frac{\x^2}{4}\right)^2 \, .
\ee
As we are going to show in a moment, the signal has its maximum for $\x \ll 1$. Thus, the falloff of the spectrum right after its peak follows approximately the simple power-law behaviour $\x^{-3}$ ---see Fig.~\ref{Benchmark1LISA}. Note that by momentum conservation, the wavenumbers of the tensor modes generated at second order by scalar perturbations peaked at the scale $k_*$ cannot exceed $2 k_*$, i.e.~$\kappa \leq 2$, where the signal indeed vanishes. Furthermore, for small enough momenta, i.e.~for $\gamma \kappa \ll 1$ on the left of the peak of the signal, $\Omegainf$ behaves like $\kappa^3$, resulting in a symmetric envelope of the signal with respect to its peak, which is also well visible in Fig.~\ref{Benchmark1LISA}.

As can be checked a posteriori, the maximum of the spectrum is to a good approximation given by the maximum of the function $\mathcal{G}$, with the prefactors $\x^{-1}\muu(\x^{-1},\x^{-1})$ in~\eqref{R3approx} only leading to small corrections suppressed by $\gamma^{-1}$. Thus, to find the maximum, we set to zero the derivative of~\eqref{cG1} with respect to $\x$ and we look for the solution with the smallest $\x$. 
In the limit $\gam\gg 1$ this leads to
\be
4 -3\gam\x\sin \gam\x+ (\gam^2\x^2 -4)\cos \gam\x = 0 ,
\ee
which depends only on $ \gam\x$.
Thus, one can find a universal first root $\gam \x_{\mathrm{max}} = c$, i.e.
\begin{empheq}[box=\graybox]{equation}
\label{max}
k_{\mathrm{max}} = 
c\,\kf, \quad\text{with}\quad c\simeq3.505,
\end{empheq}
which is valid for any $k_*$ and $\gam \gg 1$.\footnote{Practically, we are neglecting order one terms compared to the one proportional to $\gam^2$, so that in reality $\gam \gtrsim 1$ is enough.} 
It is now easy to estimate how $\p_t^{\rm out}$ scales with $\gam$ at its maximum by inserting~\eqref{max} in Eqs.~\eqref{cG1} and~\eqref{R3approx}, i.e.~by considering the regime $\gam\gg 1$ with $\gam\x_{\rm max} =c$:
\be \label{out-max}
\p_t^{\mathrm{out}}|_{\mathrm{\max}} \simeq 8 \epsilon^2 \mathcal{N}^4 A^2 \frac{\gam}{c} \Big|\cG\left(\x^ {-1},-\x^{-1},-\x \gam \right)\Big|_{\mathrm{max}}^2 \propto \mathcal{N}^2 \epsilon^2 \p_{\zeta}^2 \gam^5 \, , 
\ee
where we used $ A^2 \propto \p_{\zeta}^2/\mathcal{N}^2$.
\begin{figure}[h]
    \centering
    \includegraphics[width=0.65\textwidth]{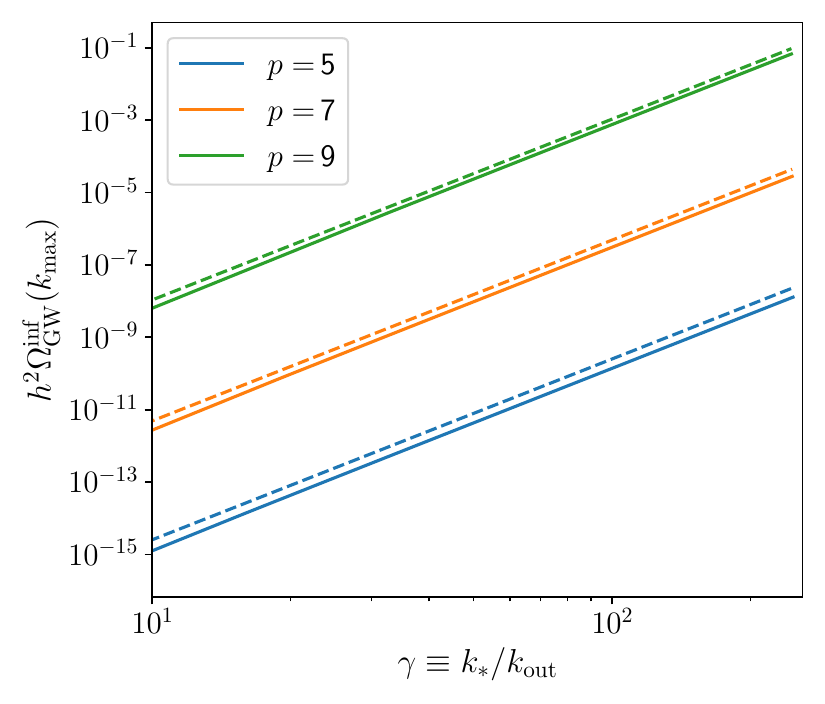}
    \caption{\textit{
    Scaling of the peak in $\Omegainf$ as a function of $\gam$.
    Coloured lines correspond to $h^2\Omegainf$ evaluated at the maximum~\eqref{max}     (computed for the two-field example in Sec.~\ref{Explicit} without the approximation of peaked Bogoliubov coefficients), for different enhancement of the primordial scalar power spectrum $p\equiv 1/2 \ln (\p_{\zeta}/\p_0)$ and fixed $\epsilon = 10^{-2}$. Each dashed line corresponds to $10^{-2} \mathcal{N}^2 \epsilon^2 \gam^5 \Omegarad $ evaluated at $k= 2k_*/\sqrt{3}$, which is the location of the maximum of the scalar-induced GWs sourced during radiation. A given $p$ translates into a fixed $\Omegarad$ maximum amplitude. The figure confirms very accurately the scaling obtained with the simple analytical estimate in Eq.~\eqref{Om-ratio}.}
    }
    \label{gamma5}
\end{figure}
By reintroducing the ``redshift factors" defined in Eqs.~\eqref{redshif1}-\eqref{redshift2}, and putting all together we arrive at
\begin{empheq}[box=\graybox]{equation}
 \label{Om-ratio}
 \frac{\Omegainf|_{k_{\mathrm{max}}}}{\Omegarad|_{2k_*/\sqrt{3}}} = {\cal O}(1) 10^{-2}\mathcal{N}^2\epsilon^2 \gam^5 ,
\end{empheq}
relating the amplitude of the two peaks, i.e.~the one in $\Omegainf$ and the one in $\Omegarad$ respectively, to the slow-roll parameter $\epsilon$ and $\gam$. 
The factor $10^{-2}$ comes from the ratio of the redshift factors $r_i/r_r \simeq 4 \cdot 10^{-2}$. 
Remarkably, despite the expected $\epsilon^2$ suppression, the inflationary contribution is boosted by a $\gam^5$ factor, which we confirm numerically in Fig.~\ref{gamma5}.
There, the ratio between each solid line ($\Omegainf$) and the corresponding dashed line ($\mathcal{N}^2\epsilon^210^{-2}\gam^5\Omegarad$) is indeed always an order one number (taking values between 1 and 2).
The relation \eqref{Om-ratio} underlines that, for a given $\p_\zeta$, the amplitude of $\Omegainf$ can easily be comparable or larger than the one of $\Omegarad$ at their respective maxima. Further, since the different maxima are located at different scales, i.e.~$k^{\mathrm{inf}}_{\mathrm{max}}\simeq \kf$ while $k^{\mathrm{rad}}_{\mathrm{max}}\simeq k_{*}=\gam \kf$, the two contributions might be of the same order and still be distinguishable providing a unique pattern for the total $\Omega_{\mathrm{GW}}$ ---see for instance Fig.~\ref{fig:Full28}.

Let us now study the shape of the spectrum around $k=k_*$. Here, we can approximate Eq.~\eqref{cG1} by going to the limit  $\gam\x \gg 1$. The first term in \eqref{cG1} dominates and $\p_t^{\mathrm{out}}$ oscillates with a constant frequency:
\be\label{oscill}
\p_t^{\mathrm{out}}\simeq   8 \epsilon^2 \mathcal{N}^4 A^2 \muu(\x^{-1},\x^{-1})\gam^2 \frac{\sin^2(\gam\x)}{\x^3}.
\ee
Although in this region the signal is suppressed compared to its value at the peak $k\simeq\kf$, it exhibits order one oscillations with a frequency $\omega$ in $k$-space given by 
\begin{empheq}[box=\graybox]{equation}
\label{freq}
\Delta k = \pi\kf \implies \omega \equiv \frac{ 2 \pi}{\Delta k} =\frac{2} {\kf},
\end{empheq}
which is the same frequency that also controls the modulations in the primordial scalar power spectrum, see Eq.~\eqref{eq:Pzeta-strong-feature}, i.e.~similar to scalar fluctuations, tensor modes enhanced during inflation select the same preferred scale $\kf$ determining the onset of the \textit{out} region. 
As reviewed in Sec.~\ref{sec:Omega-rad-review}, the same mechanism leads to oscillations in the post-inflationary scalar-induced GWs, but with a larger frequency, i.e.~$\omega_{\mathrm{rad}} = \sqrt{3} \omega $, where the numerical factor stems from the fact that scalar perturbations only source tensor modes once they re-enter their sound horizon.\footnote{The frequency of the oscillations of the post-inflationary stochastic background depends on the universe's cosmic expansion at the time of horizon re-entry for the relevant enhanced modes, i.e.~$\omega_{\mathrm{post-inf}} = c_s^{-1}\omega$, with $c_s$ the propagation speed of the scalar fluctuations ($c_s^2 = w$ for a perfect fluid). See \cite{Witkowski:2021raz} for a detailed discussion of the interplay between primordial features and the expansion history of the universe.} Note that the periodic peak-structure in $\Omegarad$ is due to a resonance mechanism that ``averages" order-one oscillations in $\p_\zeta$ to $\sim 10 \%$ oscillations in $\Omegarad$.  In contrast, the oscillations in $\Omegainf$ from Eqs.~\eqref{template1} and \eqref{oscill} are genuinely of order one, offering 
even better prospects of detection in future GWs observatories.

GWs signals exhibiting oscillations with the same frequency have been shown to arise for instantaneous sources active during inflation in~\cite{An:2020fff}. The mechanism presented in the current work is based on a dynamically emergent excited state, and not only provides an explicit realisation for the appearance of such oscillations but it also yields a richer built-in structure for the GWs spectrum; for instance, the enhancement of the peak due to the constructive interference between positive and negative frequency modes as described above, the particular shape of $\Omegainf$ as given by Eq.~\eqref{R3approx} and the possible oscillatory counterpart in the post-inflationary $\Omegarad$. The same line of thought applies to \cite{Peng:2021zon,Cai:2021wzd} where visually similar shapes have been numerically computed for models with resonances. It is likely that our formalism and results are applicable there.

Finally, let us stress once more that all the main results of this section hold generally for excited states sourcing GWs during
inflation; these are the three equations in the squared boxes regarding the position of the maximum \eqref{max}, the enhancement of $\Omegainf$ at its peak \eqref{Om-ratio} and the frequency of the oscillations around $k\simeq k_*$ \eqref{freq}. In fact, these outcomes rely only on the functional shape of $\cG$ that comes from the time integral over the three de Sitter mode functions, together with the assumption that the Bogoliubov coefficients are peaked around a given scale. The latter assumption is guaranteed when the event (the feature) creating the excited state is sharp. 

\begin{figure}[t]
    \centering
    \includegraphics[scale=.77]{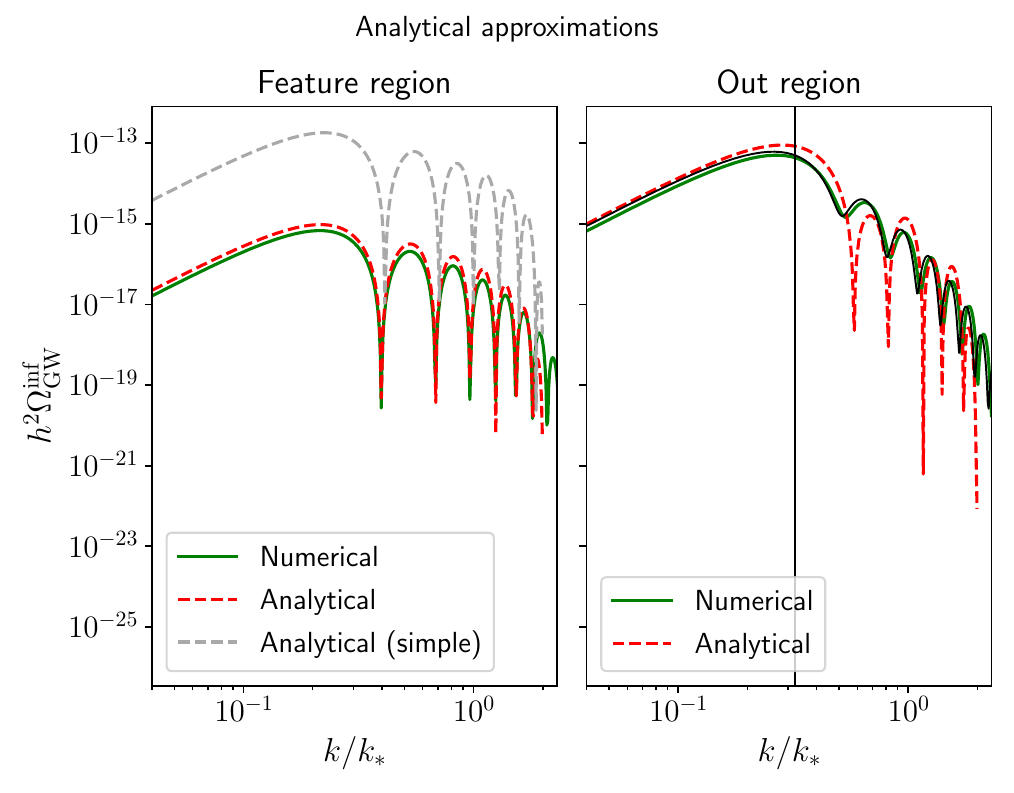}
    \caption{\textit{Comparison between analytical approximations and full numerical results for different contributions to $\Omegainf$ in the explicit scenario of Sec.~\ref{Explicit} with parameters $(\de,\etaperp)=(0.5, 14)$. On the left-hand side the contribution from the region of the feature (green), versus the approximations in Eq.~\eqref{appR2a} (dashed red) and in Eq.~\eqref{appR2b} (dashed gray). On the right-hand side the dominant contribution from the $out$ region (green) versus the analytical approximation in Eq.~\eqref{template1} (dashed red) which, quite remarkably, provides a very good estimate of the overall shape, amplitude and frequency of the full $\Omegainf$ signal (black). Eventually, the vertical black line highlights the position of the global maximum found analytically in Eq.~\eqref{max}.}}
    \label{AnApprox_05_14}
\end{figure}

\subsubsection{Feature-region contribution}\label{Region2}
Until now we have presented results for the total tensor power spectrum $\p_t$ as if the only relevant contribution was the one from the \textit{out} region, i.e.~$\p_t \simeq \p_t ^{\mathrm{out}}$. Thus, let us end this section by showing explicitly that the contribution from the feature region, i.e.~the one coming from the time dependence of the mode functions in the region of the feature, is indeed small. 
In particular, as we are going to show, the contribution is suppressed by the small parameter $\delta$ describing the finite duration (in $e$-folds) of the feature.

Let us start by considering the term inside the modulus in Eq.~\eqref{Dfeat}, using Eq.~\eqref{R2}: 
\begin{align}\label{featt}
I^{\mathrm{feat}}_{ij}=\left(\sum _X \widetilde{Q}^{\mathrm{out}}_{Xi}(xk ,\tauf)\widetilde{Q}^{\mathrm{out}}_{Xj}(yk,\tauf) \right) \mathcal{I}(g,\de,\zf),
\end{align}
with 
\begin{align}\label{R2timein}
  \mathcal{I}\left( \g,\de,\zf \right)=  \int_{e^{\delta} \zf}^{\zf} \di z \left(\frac{\zf}{z}\right)^{g} G(0, z),
\end{align}
where $\g \equiv \g(xk)+\g(yk)$ comes from the model-dependent exponent $\g(k)$ defined in Eq.~\eqref{R2}, and we recall that $G(0, z)$ is given in Eq.~\eqref{Gzz0}. The time-integral~\eqref{R2timein} can be performed analytically:
\begin{align}\label{II}
\mathcal{I}(\g,\de,\zf) = & \frac{g}{g+1}\mathrm{Re}\Big[ (i \zf)^g \Gamma[-g,i e^\de \zf] - (i\zf)^g \Gamma[-g,i\zf]  \Big]
\nn\\ &  + \frac{1}{g+1} \frac{\sin(\zf) - e^{-(g+1)\de } \sin\left( e^\de \zf\right)}{\zf},
\end{align}
where $\Gamma$ is the incomplete gamma function $ \Gamma[a,x] = \int _{x}^{\infty}\di t \, t^{a-1} e^{-t} $.
Expanding for $\de\to0$, we obtain
\be \label{I2-de0-1}
\mathcal{I}(\g,\de,\zf) =  -\zf G(0,\zf) \de  + {\cal O} \left( \de^2 \right),
\ee
which can simply be obtained by neglecting the variation of $G(0,z)$ in the integrand in \eqref{R2timein} compared to the exponentially growing factor. Note that to first order in $\de$ the model-dependent parameter $\g$ drops out. Let us now also consider the prefactor in parenthesis in Eq.~\eqref{featt} and, for simplicity, focus again on a single mode (that is, disregarding the sum over fields by fixing the quanta indices to $i=j=1$).
Taking the modulus square of \eqref{featt},
we obtain 
\begin{align}
\left|I^{\mathrm{feat}}\right|^2 &
\simeq  \left|\widetilde{Q}^{\mathrm{out}}(k x,\tauf)\right|^2 \left|\widetilde{Q}^{\mathrm{out}}(k y,\tauf)\right|^2 \mathcal{I}^2(g,\delta,\zf),
\label{Ifeat}
\end{align}
where 
\begin{align}
\left|\widetilde{Q}^{\mathrm{out}}(p,\tauf)\right|^2 \simeq\,& (|\alpha(p)|^2+|\beta(p)|^2)  |\zeta(p\tauf)|^2 + 2\mathrm{Re}\left[\alpha(p)\beta^*(p) \zeta^2(p\tauf)\right]\nn\\[1mm]
\simeq\,& 2|\alpha(p)|^2 \left(1 + p^2\tauf^2 + (1- p^2 \tauf ^2)\cos\varphi +2 p \tauf   \sin\varphi\right); \label{QQ}
\end{align}
in the second step we have used Eq.~\eqref{simpleBog}, i.e.~$\beta(k)  = \alpha(k) e^{-2ik\tauf + i \varphi}$.
Putting everything together, namely inserting the expression \eqref{QQ} into Eqs.~\eqref{Ifeat},~\eqref{Dfeat} and \eqref{constituent}, and then using the narrow-peak
approximation~\eqref{theapprox}, we arrive at
 \begin{align}\label{appR2a}
 \mathcal{P}_t^{\mathrm{feat}}\simeq\,& 8\epsilon ^2 A^2 \mathcal{N}^4 \x^{-1}\muu(\x^{-1},\x^{-1})  \mathcal{I}^2(g(k_*),\delta,-\gam \x)
 \gam^4 (1-\cos\varphi)^2\\[1mm]
 \simeq\,& 8\epsilon^2 A^2 {\cal N}^{4} \x^{-1}\muu(\x^{-1},\x^{-1})  |\gam\x G(0,-\gam\x)|^2  \de^2 
\gam^4 (1-\cos\varphi)^2\,.
\label{appR2b}
 \end{align}
In the first expression we have used the definition $\gam= -\kstar\tauf$, together with $\gam \gg 1$. 
We have assumed that all mode functions in the multifield sum give the same contribution. This brings a factor ${\cal N}^4$ multiplying the single-field result.
Equation~\eqref{appR2b} is valid under the additional approximation on the integral $\mathcal{I}$ in Eq.~\eqref{I2-de0-1}. This overestimates the overall amplitude of $\p_t^{\mathrm{feat}}$ ---see Fig.~\ref{AnApprox_05_14}, unless $\de \ll 10^{-2}$.\footnote{For not too small $\delta$ the amplitude is overestimated by the approximation \eqref{I2-de0-1}, although we have checked numerically that one still has a good match for the frequency unless $|k \tauf| \gtrsim 20$.} In any case, from Eq.~\eqref{appR2b} we learn that the contribution from the feature region is suppressed at least by a factor of $\de^2$. Moreover, this expression allows us to understand the oscillatory pattern arising from this contribution. In fact, $| G(0,-\gamma \kappa=k/\kf )|^2$ provides oscillations with the same frequency as given in \eqref{freq}. 
Overall, from Eq.~\eqref{appR2b} we can infer how the contribution from the feature region scales with the various parameters:
\be
\mathcal{P}_t^{\mathrm{feat}}\propto \epsilon^2 {\cal N}^2 \gam^4 \delta^2 \p_{\zeta}^2,
\ee
where again we used $A^2  \simeq \p_{\zeta}^2/{\cal N}^2$. 

To summarise, the feature-region contribution is enhanced by a factor $\gam^4$ and suppressed by, at least, a factor $\de^2$. In contrast, the \textit{out} region ---c.f. Eq.~\eqref{out-max}--- scales as $\gam^5$, and thus, in general, dominates the spectrum.

\subsection{Explicit example: turning in field space}\label{Explicit}
We provide here a concrete example in which an excited state emerges as the result of a sharp feature along the inflationary dynamics. The setup illustrated in this section relies on a model-independent, multifield mechanism first proposed in \cite{Palma:2020ejf,Fumagalli:2020adf} to seed primordial black holes. All numerical plots in the current paper, e.g.~Figs.~\ref{AnApprox_05_14}-\ref{Benchmark2LISA}, are produced using this mechanism as a benchmark, which serves as an explicit illustration for our results. In this example, the sharp feature triggering the excited state corresponds to a sudden and strong turn of the trajectory in field space. Let us start with the Lagrangian for the general class of non-linear sigma models whose target-space geometry is given by the metric $G_{IJ}$: 
\begin{equation} \label{L-background}
    \frac{\mathcal{L}}{ \sqrt{-g}}
    = -\frac{1}{2}G_{IJ}(\boldsymbol{\phi})\partial^\mu \phi ^I \partial_\mu \phi ^J - V(\boldsymbol{\phi}),
\end{equation}
where $\boldsymbol{\phi}=\left(\phi^1,\cdots,\phi^{\cal N}\right)$ and $V(\boldsymbol{\phi})$ is a generic multifield potential. Let us restrict for simplicity to two fields (${\cal N}=2$). A convenient way to organise perturbation theory is the adiabatic-entropic basis defined by the unit vectors $T^I = \dot{\phi}^I /(G_{JK}\dot{\phi}^J\dot{\phi}^K)^{1/2}$ (tangent to the background trajectory) and $N^I$ (orthogonal to the trajectory) with the pair $\left(T^I,N^I\right)$ selecting a definite orientation. Here,  $\dot{\phantom{a}}\equiv\di/\di t.$ Expressing the field fluctuations as 
$\delta \phi^I= Q_\phi T^I + Q_\psi N^I$ and fixing the comoving gauge, i.e.~$Q_\phi=0$ \& $g_{ij}=a^2 e^{2 \zeta} \delta_{ij}$, we may proceed to obtain the dynamics of the two degrees of freedom $Q_\psi, Q_\zeta \equiv \Mp\sqrt{2\epsilon}\zeta$.

The effective action at second order in these linear fluctuations around a given
homogeneous background reads\footnote{Note that the isocurvature self-interactions neglected here would be relevant for the discussion below Eq.~\eqref{eq:h-h} about non-Gaussian corrections to the tensor power spectrum.}~\cite{Sasaki:1995aw,GrootNibbelink:2001qt,Langlois:2008mn}
\begin{align}
\frac{\mathcal L}{a^3} =&  \Mp^2 \epsilon\left( \dot{\zeta}^2 -   \frac{(\partial_i \zeta)^2}{a^{2}} \right)   +  \frac12\left(\dot{Q}_\psi^2  
- \frac{(\partial_i Q_\psi)^2}{a^2} 
- m_s^2 Q_\psi^2 \right)+ 2 \sqrt{2 \epsilon} H \Mp \eta_\perp \dot{\zeta} Q_\psi , \label{Lagrangian-zeta-psi}
\end{align}
where the last term describes the coupling between the two types of perturbations. Here, $\etaperp \equiv \frac1H N_I D_t T^I$ is a dimensionless parameter measuring the turning rate of the trajectory, with $D_t A^I \equiv \dot{A}^I+\Gamma^I_{JK} \dot{\phi}^J A^K$ the covariant time derivative along the background trajectory, which deviates from a geodesic in field-space when $\etaperp \neq 0 $. 
The resulting equations of motion read
\bea
\dot{\Pi}_{\zeta}  + 3 H \Pi_\zeta + \frac{k^2}{a^2} Q_\zeta =0  , \label{eom-1} \\
\ddot{Q}_\psi + 3 H \dot{Q}_\psi + \left( m_s^2 + \frac{k^2}{a ^2} \right) Q_\psi -2 H\eta_\perp \dot{Q}_\zeta = 0 , \label{eom-2} 
\eea
where for simplicity we are considering $H$ constant here and in what follows, and we have used $\Pi_\zeta \equiv \dot{Q}_\zeta  + 2 \eta_\perp H Q_\psi$.

The mass parameter $m_s^2$ turns out to be determined by background quantities as
\be
\label{ms2}
m_s^2=V_{;ss}-H^2\eta_{\perp}^2+\epsilon H^2  \Mp^2 R_{\rm fs}\,,
\ee
with $V_{;ss}=e_s^I e_s^J V_{;IJ}$ the projection of the covariant Hessian of the potential along the entropic direction, and $R_{{\rm fs}}$ the field-space scalar curvature. 
An alternative and also useful notion of mass is that of the entropy mass given by
\be
\mu^2 = m_s^2 + 4 H^2 \eta_\perp^2 . \label{entropy-mass-mu}
\ee
It can be shown that $\mu$ corresponds to the rest-energy of the massive particle state of the spectrum on  sub-Hubble scales}~\cite{Achucarro:2012yr,Castillo:2013sfa}. In addition, on super-Hubble scales, one can integrate Eq.~\eqref{eom-1} once. Upon doing so, Eq.~\eqref{eom-2} reduces to $\ddot Q_\psi + 3 H \dot Q_\psi + \mu^2 Q_\psi = 0$, which shows that $\mu$ also plays the role of $Q_\psi$'s mass there.

From the expression \eqref{ms2}, one sees that 
if $\etaperp^2\gg 1$, so that at a given time the bending parameter is large enough to overcome the other two contributions, then the entropic field experiences a transient tachyonic instability for $k^2/a^2 \lesssim |m_s^2|$, first noticed in \cite{Cremonini:2010ua}. As a series of recent works has shown~\cite{Renaux-Petel:2015mga,Garcia-Saenz:2018ifx, Garcia-Saenz:2018vqf, Fumagalli:2019noh, Bjorkmo:2019qno, Ferreira:2020qkf}, this does not lead to a background instability but rather to a  transient exponential growth of fluctuations until (effective sound) horizon crossing. Through the derivative coupling between the two perturbations,
whose strength is determined by the same parameter $\etaperp$, this growth also affects the curvature perturbation. Thus, both adiabatic and entropic perturbations are subject to the same exponential growth (on sub-Hubble scales) compared to standard setups (see e.g.~\cite{Chakraborty:2019dfh,Aragam:2020uqi,Anguelova:2020nzl,Aragam:2021scu,Renaux-Petel:2021yxh} for recent discussions about this regime of strongly non-geodesic motion).

Considering a brief period in which the bending parameter is large provides an explicit example of a transient non-adiabatic evolution leading to the dynamical appearance of an excited state. Using the language of Sec.~\ref{sec:dyn-ex-state}, a large bending parameter determines a feature region, corresponding to the time interval in which $\etaperp \gg 1$, followed by an \textit{out} region where the bending is negligible and the system is in an excited state.

\subsubsection*{Mode functions for strong sharp turns}

Equations (\ref{eom-1}) and (\ref{eom-2}) can be re-expressed as equations of motion for Bogoliubov coefficients keeping track of the production of excited states during inflation. We may define time-dependent Bogoliubov coefficients as %
\bea
\hat{Q}_{\zeta} (\bk,\tau) &=& \frac{H}{\sqrt{2 k^3}} \sum_{i=1,2} \left[ \alpha_{{\zeta} i} (\tau) \zeta(k \tau) + \beta_{{\zeta} i}(\tau) \zeta^*(k \tau) \right] a_i (\bk) + {\rm h.c.} (-{\bf k}) ,  \label{sol-intro-1} \\
\hat{Q}_{\psi} (\bk,\tau) &=& \frac{H}{\sqrt{2 k^3}} \sum_{i=1,2} \left[ \alpha_{{\psi} i} (\tau) \psi(k \tau) + \beta_{{\psi} i}(\tau) \psi^*(k \tau) \right] a_i (\bk) + {\rm h.c.} (-{\bf k}) ,  \label{sol-intro-2}
\eea
where $\zeta( k \tau ) = (1 + i k \tau) e^{-i k \tau}$ is the massless dS mode function given by Eq.~\eqref{zpm}, whereas $\psi (k \tau) = i \sqrt{\frac{\pi}{2}} (- k \tau)^{3/2} H_{\nu}^{(1)} (- k \tau)$, with $\nu = \sqrt{9/4 - \mu^2/H^2}$, is the dS mode function for a massive field of mass $\mu$. In fact, $\mu$ here coincides with the entropy mass defined in Eq.~\eqref{entropy-mass-mu}. Both mode functions respect Bunch-Davies initial conditions at $k\tau=-\infty$. 

Then, it can be shown that $\alpha_{\zeta i} (\tau)$, $\alpha_{\psi i} (\tau)$, $\beta_{\zeta i} (\tau)$ and $\beta_{\psi i} (\tau)$ obey the following first-order equations of motion
\bea
  \frac{\di}{\di \tau}\left(\begin{array}{c} {  \alpha_{\zeta i}}  \\ {\beta_{\zeta i}}
  \end{array}\right) =  \mathcal{A} (k, \tau)  \left( \begin{array}{c} \alpha_{\psi i} \\  \beta_{\psi i} \end{array}\right) , \qquad
 \frac{\di}{\di \tau}  \left(\begin{array}{c}    {\alpha_{\psi i}}  \\  {\beta_{\psi i}} \end{array}\right) =  \mathcal{B} (k, \tau)  \left(\begin{array}{c} \alpha_{\zeta i} \\ \beta_{\zeta i} \end{array}\right) , \label{prime-M-new}
\eea
where the coefficient matrices, satisfying $\det{\cal A}=\det{\cal B}=0$ and $\mathcal{A}\mathcal{B}=\mathcal{B}\mathcal{A}=0$, are given by
\bea
\mathcal{A} (k, \tau) =  - i \frac{ \etaperp (\tau)}{ k^3 \tau^3} \left(\begin{array}{cc} \zeta^{*'} \psi  &  \zeta^{*'} \psi^{*}  \\  -  \zeta^{'} \psi &   - \zeta^{'} \psi^{*}  \end{array}\right)  , \quad \mathcal{B}(k, \tau) = i \frac{ \etaperp (\tau)}{  k^3 \tau^3} \left(\begin{array}{cc} -\zeta^{'} \psi^{*} & -\zeta^{*'} \psi^{*} \\ \zeta^{'} \psi  & \zeta^{*'} \psi    \end{array}\right).  
\eea

Noteworthily, the first-order, coupled differential equations~\eqref{prime-M-new} are valid for any time-dependent $\eta_{\perp}(\tau)$ and preserve the relations \eqref{bog-constraint-1} and \eqref{bog-constraint-2} for the time-dependent Bogoliubov coefficients written in Eqs.~\eqref{sol-intro-1} and~\eqref{sol-intro-2}. This can be shown by taking the time derivative of Eqs.~\eqref{bog-constraint-1},~\eqref{bog-constraint-2} and using Eq.~\eqref{prime-M-new} to show that they stay invariant as long as the Bogoliubov coefficients satisfy them at a given initial time. Given that $\zeta(k \tau)$ and $\psi(k \tau)$ coincide with the standard single-field solutions for massless and massive fields respectively, Eq.~\eqref{prime-M-new} shows explicitly that turns in field space, parametrised by $\eta_\perp$, induce the excitation of modes through the mixing of Bogoliubov coefficients. 
More to the point, starting from the Bunch-Davies initial conditions \eqref{Region1} in the \textit{in} region where $\eta_\perp$ vanishes, these equations make manifest that the bending in the feature region feeds non-vanishing $\beta$ coefficients, which become time-independent once the bending is over in the \textit{out} region, to coincide with the ones defined in Eq.~\eqref{Region3}.

Although using a different method, i.e.~directly solving for the mode functions $Q_{Xi}$, analytical solutions were found explicitly in \cite{Palma:2020ejf}, with results generalised in~\cite{Fumagalli:2020nvq}, by considering a top-hat profile for the time dependence of the bending parameter\footnote{See Sec.~2.1 of~\cite{Fumagalli:2020nvq} for an extensive discussion about similarities and differences when using a different time-dependent profile for $\etaperp(N)$ or different parameterisations for $m_s^2$.} $\etaperp(N)$, with height and width given by the constants $\etaperp$ and $\delta$, and in the regime of a sharp turn $\delta\lesssim 1$. 
In particular, by setting Bunch-Davies initial conditions, the mode functions in the \textit{out} and feature region take the simple forms~\eqref{Region3} and~\eqref{R2}, respectively. For the sake of completeness, we write here the corresponding functions in a convenient form. In the current setup, the model-dependent function $g$ in the exponent of Eq.~\eqref{R2} is given by:
\be
g = \etaperp S(k),
\ee
while the \bog\,  coefficients in Eq.~\eqref{Region3} can be written as
\begin{align}
\alpha_{\zeta\, 1}&=-\frac{e^{\deltaeta\, S}}{4 S \sqrt{1+\X}},\qquad
\alpha_{\zeta \,2}=\frac{i e^{ \deltaeta\, S}}{4 S(1+\X+\sqrt{\X(1+\X)}},\nn\\[1mm]
\alpha_{\psi\, i}&=i\frac{ S^2+\x^2}{2\x} \alpha_{\zeta i},\qquad \mathrm{for}\,\,i=1,2\,, \\[1mm]
\beta_{\zeta \,i} &= -e^{i \theta_k}\alpha_{\zeta\, i },\qquad \beta_{\psi \,i} = e^{i \theta_k}\alpha_{\psi\, i },\qquad \mathrm{for}\,\,i=1,2 \,,
\end{align}
with 
\begin{align}
\theta_k&=2 e^{-\delta/2} \x \etaperp+2 \arctan(\x/S)\,, \label{phase} \\[1mm]
\X &= \frac{{(3+\xi)}^2}{16 \x^2},\quad S(k)=\sqrt{ \sqrt{4 \x^2 +\frac{(3+\xi)^2}{4}}-\left(\x^2 +\frac{(3+\xi)}{2} \right)
}\,,
\end{align}
and 
\be
\label{def-kappa-X}
\x = \frac{k}{\kstar} = \frac{k}{\kf \, \etaperp e^{-\delta/2}}.
\ee
Here, $\xi$ parametrises the contributions from the potential and field-space curvature to the entropic mass \eqref{ms2}, i.e.
\be
m_s^2 = (\xi-1) \etaperp^2(N) H^2 \, .
\ee
In this notation, $\xi < 1$ and $\etaperp \neq 0 $ correspond to the appearance of the transient tachyonic instability. These simple expressions are valid for scales $k\sim k_*$, where modes are exponentially amplified and one finds $|\alpha_{X\,i}|\simeq |\beta_{X\,i}|\gg 1$. Furthermore, from Eqs.~\eqref{phase} and~\eqref{def-kappa-X} we observe the typical pattern of oscillations proper to a sharp feature, i.e.~with a frequency given by $\omega\sim 2/\kf$ as described around Eq.~\ref{simpleBog}. Here, the slowly varying $k$-dependent phase factor is given explicitly by $\varphi_k \simeq 2\arctan(\kappa/S)$. 

For the numerical results shown throughout the paper we have considered the particular case $\xi=-3$~\cite{Palma:2020ejf} corresponding to the case in which the entropy mass \eqref{entropy-mass-mu} vanishes ($\mu=0$)~\cite{Achucarro:2016fby}. Apart from being a convenient choice, this case can be well-justified from a holographic perspective, wherein one can construct multifield models of inflation with the help of ``fake" super-potentials~\cite{Achucarro:2018ngj}. As shown in~\cite{Palma:2020ejf}, in this type of models a shift symmetry of the super-potential (with respect to one of the fields) enforces the value $\xi=-3$ throughout the full evolution of the multifield system, before, during and after the turn. 
Thus, we have normalised $k$ in the previous expressions to the maximum of the primordial power spectrum for $\xi = -3$.
 
Moreover, in this example, the parameter defined in Eq.~\eqref{gamma}, quantifying how deep inside the horizon the maximally enhanced mode lies
at the beginning of the \textit{out} region, is given by
\be \label{gamma-eta-p}
\gam=\eta_\perp e^{-\delta/2} \simeq \eta_\perp,
\ee 
while the maximal enhancement of the primordial power spectrum at $k=k_*$, reads
\be \label{p-example}
\frac{\p_\zeta(k_*)}{\p_0} = \frac{e^{2\etaperp \delta}}{2}.
\ee

\section{Theoretical bounds and observational prospects}
\label{sec:obs}
\subsection{Backreaction and perturbativity}
It is well-known that a sharp feature along the inflationary dynamics (a transient period of non-adiabatic evolution) can lead to strong coupling of fluctuations at a given scale, jeopardising perturbative unitary and thus leading to a loss of theoretical control.
Furthermore, the quanta copiously produced during this epoch can backreact and compete with the underlying background evolution rendering unsatisfactory the standard treatment of perturbation theory, see e.g.~\cite{Holman:2007na,Bartolo:2013exa,Adshead:2014sga,Cannone:2014qna}.

In this section, as a proof of principle, we show how bounds coming from the requirements that unitarity and backreaction be under control can constrain a significant part of the parameter space of a given scenario that would otherwise lead to a signal above the threshold of observability in GWs detectors. \begin{figure}[h]
    \centering
    \includegraphics[width=0.75\textwidth]{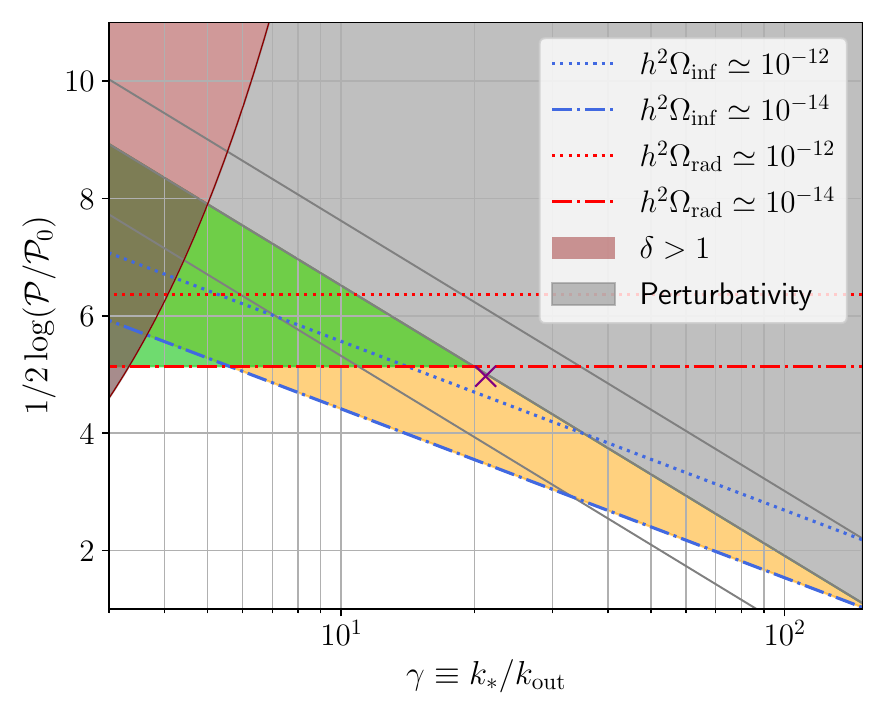}
    \caption{\textit{Perturbativity constraints versus amplitudes of $\OGW$. We highlight two values of constant $\Omegarad$ (red lines) and two values of constant $\Omegainf$ (blue lines) for a fixed $\epsilon = 0.05$. Their behaviour can be understood from Eqs.~\eqref{infsca}-\eqref{radsca}. The grey lines correspond to the bound \eqref{boundpert} with three different fudge factors $C=1,10,100$, i.e.~larger $C$ translates into decreasing constraining bounds. The grey area corresponds to the region of parameter space excluded by perturbativity constraints if $C =10$. Assuming that a signal is being potentially detectable if $h^2\OGW \gtrsim 10^{-14}$, then the green area corresponds to the region of parameter space where $\Omegarad$ has observational relevance while the theoretical framework is under control. This allowed and relevant region of parameter space is extended by an additional strip (in yellow) once the $\Omegainf$ signal is also considered. For $\delta>1$ (red region) the feature described in Sec.~\ref{Explicit} is not sharp and the solutions for the Bogoliubov coefficients does not hold.
    }}
    \label{figPert}
\end{figure}
In \cite{Fumagalli:2020nvq} we have derived tentative bounds applicable to the example of Sec.~\ref{Explicit}: 
perturbative control requires
\be\label{pert}
\gam^4 \p_{\zeta}\lesssim C,
\ee
while the constraint for backreaction to be under control is
\be\label{back}
\epsilon\gam ^4\p_{\zeta} \lesssim C,
\ee
where the $C$ on the right-hand side in both inequalities has to be thought of as a fudge factor taking values ${\cal O}(1)-{\cal O}(100)$ and which can be estimated in full computations. Note that, interestingly, the perturbativity bound~\eqref{pert} is in agreement with the constraint arising from energy conservation discussed in~\cite{Inomata:2021zel}. Let us characterise the enhancement of the scalar power spectrum at its maximum by the parameter
\be
 p \equiv \frac{1}{2}\ln\left(\frac{\p_{\zeta}}{\p_0}\right)\,,
\ee
where for definiteness, $\p_0 = 2.4 \times 10^{-9}$ is considered in this section.
\begin{figure}[h]
    \centering
    \includegraphics[width=0.75\textwidth]{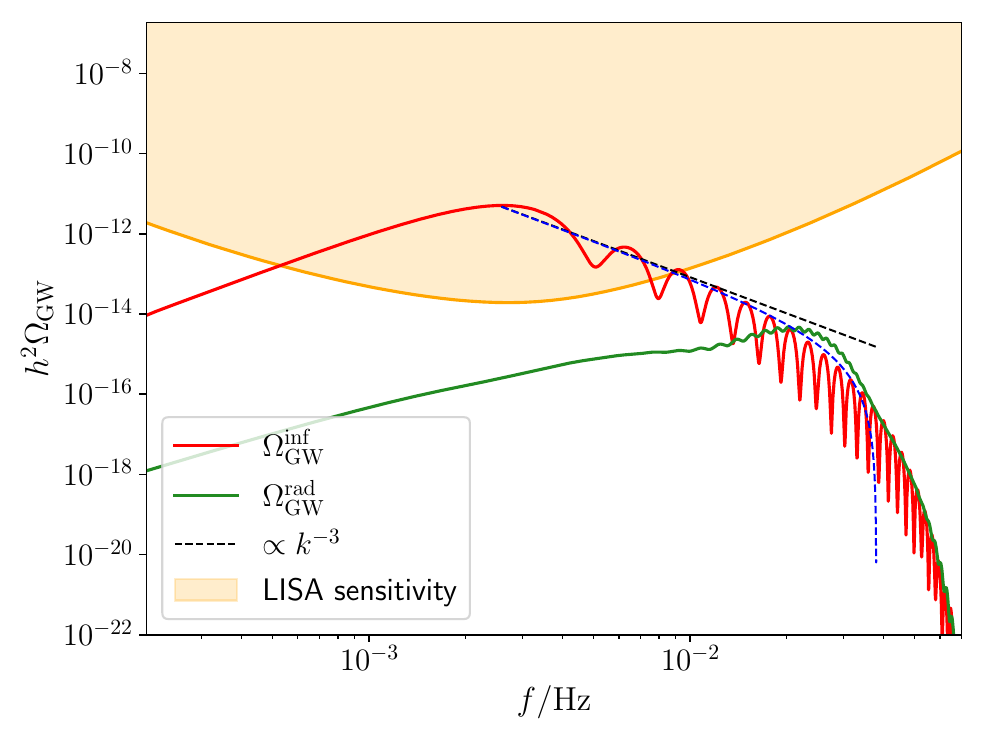}
    \caption{\textit{$\Omegarad$ and $\Omegainf$ computed for the parameter choice space highlighted with a cross in Fig.~\ref{figPert}, i.e.~$(\gam, p )= (21.2,5)$ or $(\eta_{\perp},\delta)= (23.7,0.225)$. For illustrative purposes the time of the feature has been adjusted so that the signal peaks in the LISA frequency band. We also show the approximate envelope of $\Omegainf$, cf.~Eq.~\eqref{envelope} (dashed blue), and highlighted the power law $f^{-3}$ falloff after the peak (dashed black). 
   }}
    \label{Benchmark1LISA}
\end{figure}
The bounds \eqref{pert} and \eqref{back} can be rewritten in terms of $p$ and $\gamma$ as
\begin{align}
\label{boundpert}
   \mathrm{Perturbativity}:\qquad p &\lesssim \frac{1}{2}\log \left(C\p_0^{-1}\gamma^{-4}\right),\\[1mm]
   \label{boundback}
    \mathrm{Backreaction}:\qquad p &\lesssim \frac{1}{2}\log \left(C\epsilon^{-1}\p_0^{-1}\gamma^{-4}\right).
\end{align}
\subsection{Constraining the parameter space}
Let us now understand how the bounds from the previous section constrain part of the parameter space that would naively lead to a pronounced signal in $\Omega_{\rm GW}$.
As discussed in the previous section, the amplitude of the gravitational-wave energy density generated during inflation and in the post-inflationary evolution will depend on $\gam$ and on the enhancement $p$ of the primordial scalar power spectrum. In particular, we have shown that $\Omegainf\propto \epsilon^2 \p^2_{\zeta} \gamma^5$ at its global maximum \eqref{max}, so that we can rewrite it as
\be\label{infsca}
h^2\Omegainf(\gamma, p) = h^2\Omegainf(\tilde{\gamma},\tilde{p},\epsilon)\left(\frac{\gamma}{\tilde{\gamma}}\right)^5\left(\frac{\p_{\zeta}}{\tilde{\p_{\zeta}}}\right)^2= r_i c_i \epsilon^2 \p_0 ^2\gamma^5 e^{4p},
\ee
where quantities with a tilde are meant to be evaluated at a given pivot point $(\tilde{\gam},\tilde{p})$, $r_i$ is the redshift factor defined in \eqref{redshift2} and
\be 
c_i(\tilde{\gam},\tilde{p}) = \frac{h^2\Omegainf(\tilde{\gam},\tilde{p},\epsilon)}{r_i\tilde{\gam}^5\epsilon^2\tilde{\p}_{\zeta}^2} = \frac{\p_t^{\mathrm{out}}(\tilde{\gam},\tilde{p},\epsilon)}{\tilde{\gam}^5\epsilon^2\tilde{\p}_{\zeta}^2},
\ee
is an order-one number, almost independent of the pivot scale chosen.\footnote{For instance, in our two-field example of Sec.~\ref{Explicit}, $c_i(\eta_{\perp}\delta= 7 - 15) \simeq 0.2 - 0.13 $ and $c_r$ in Eq.~\eqref{radsca}, satisfies $c_r(\eta_{\perp}\delta =7 - 15)/r_r = 0.42 - 0.15$.} The amplitude of $\Omegarad$ (at its maximum) depends instead only on the enhancement of the scalar power spectrum and can be written as 
\be\label{radsca}
h^2\Omegarad(p) = r_r c_r \p_0^2 e^{4p},
\ee
with $r_r$ the redshift factor defined in \eqref{redshift2} and $c_r$ again an order-one number.

A certain enhancement $p$ of the primordial scalar power spectrum $\p_\zeta$ fixes the amplitude of $\Omegarad$. In contrast, the amplitude of $\Omegainf$, for the same value of $p$, is still allowed to grow as $\gam^5$. Thus, one can imagine scenarios (part of the parameter space) characterised by a small enhancement of the scalar fluctuations compared to CMB scales, which lead to an unobservable amount of $\Omegarad$ but for which $\Omegainf$ is still large in amplitude. Naturally, in such a case, the backreaction and perturbativity constraints~\eqref{pert}-\eqref{back} are more easily satisfied. Furthermore, let us recall that the maxima of the two contributions $\Omegainf$ and $\Omegarad$ are located at different frequencies: from Eq.~\eqref{max} we have $k^{\mathrm{inf}}_{\mathrm{max}}\simeq 3.5 \kf$ and
 $k^{\mathrm{rad}}_{\mathrm{max}}= \left(2/\sqrt{3}\right) \gamma\kf$ (recall that $\gam\gg1$). Thus, in the region of parameter space where the two peaks have similar amplitude, one can still hope to see the imprint of both in the full $\OGW$. 
 
 For a given example of a sharp feature, the underlying parameters defining the mechanism are directly related to $\gam$ and $p$. In the case of a strong turn in field-space, these are given by Eqs.~\eqref{gamma-eta-p} and \eqref{p-example}, i.e.~$p = \delta\eta_{\perp} -\ln \sqrt{2}$ and $
\gamma = \eta_{\perp}e^{-\delta/2}$.
By fixing the product $\etaperp\delta$ we fix $p$, and $\gam$ is then roughly given by $p/\de$, where $\de$ is the parameter controlling the duration of the feature. 

In this context, Fig.~\ref{figPert} can be thought of as a roadmap in parameter space for sharp features generating excited states and sourcing GWs.
On the y-axis we have the enhancement $p$ of the power spectrum. From Eq.~\eqref{radsca} it follows that lines of constant $p$ coincide with lines of constant $\Omegarad$ and that the larger the value of $p$, the larger $\Omegarad$. Lines of constant $\Omegainf$ can be understood by looking at Eq.~\eqref{infsca}: increasing $p$ and $\gam$ leads to an enhancement of $\Omegainf$ at its peak.  Thus, if we take a specific value of $\Omega_{\mathrm{GW}}=\bar{\Omega}$ as a proxy for the threshold of detection in a given GWs observatory, all the parameter space above the corresponding lines of $\Omegarad$ and $\Omegainf$ would correspond, a priori, to a ``positive detection" for that particular experiment. However, and this is the purpose of this section, perturbativity bounds could forbid a huge part of this parameter space. 
\begin{figure}[h]
    \centering
    \includegraphics[width=0.75\textwidth]{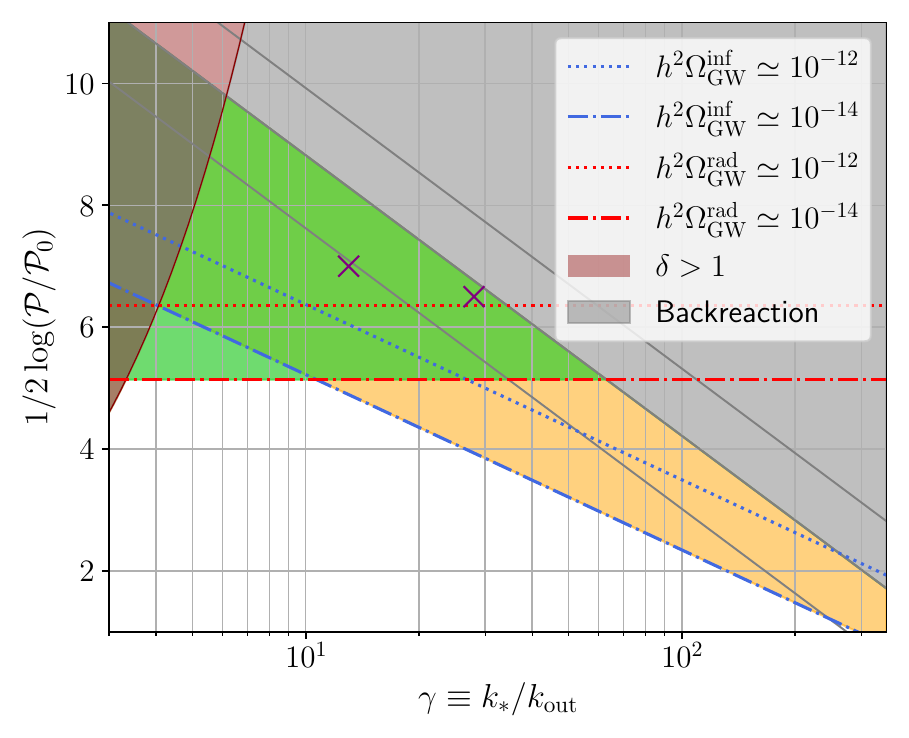}
    \caption{\textit{Backreaction constraints versus amplitudes of $\OGW$. The colour code is the same as in Fig.~\ref{figPert}. Different grey lines correspond now to the bound in Eq.~\eqref{boundback}.  
    For given $(\gam,p)$, a change in the slow-roll parameter (here fixed to $\epsilon=0.01$) would mildly influence the position of the backreaction constraints ---see text.}}
    \label{figBack}
\end{figure}
As an example, let us focus on the value $\bar{\Omega}\simeq 10^{-14}$ that in the mHz band can be seen as the optimistic threshold of detectability for LISA. If we consider first the contribution from $\Omegarad$, any parameter corresponding to the the points above the dash-dotted red line in Fig.~\ref{figPert} would give a signal in the LISA sensitivity band. Even so, by taking $C\simeq 10$ in Eq.~\eqref{boundpert} the region highlighted in grey would correspond to points where perturbativity is lost and the theory is not under theoretical control, leaving us with the highlighted green triangular region as the de facto available region of parameter space. Considering the contribution from $\Omegainf$ adds a strip in parameter space, highlighted in yellow, were $\Omegainf >\bar{\Omega}$ and still perturbativity is roughly under control. Note that in our specific set-up, the pair $(p,\gam)$ also determines the region in which $\delta < 1$, cf.~Eqs.~\eqref{gamma-eta-p},~\eqref{p-example}. The complementary region is marked in red as our analytical solutions to the dynamics of fluctuations is not accurate there.  
\begin{figure}[h]
    \centering
    \includegraphics[width=0.75\textwidth]{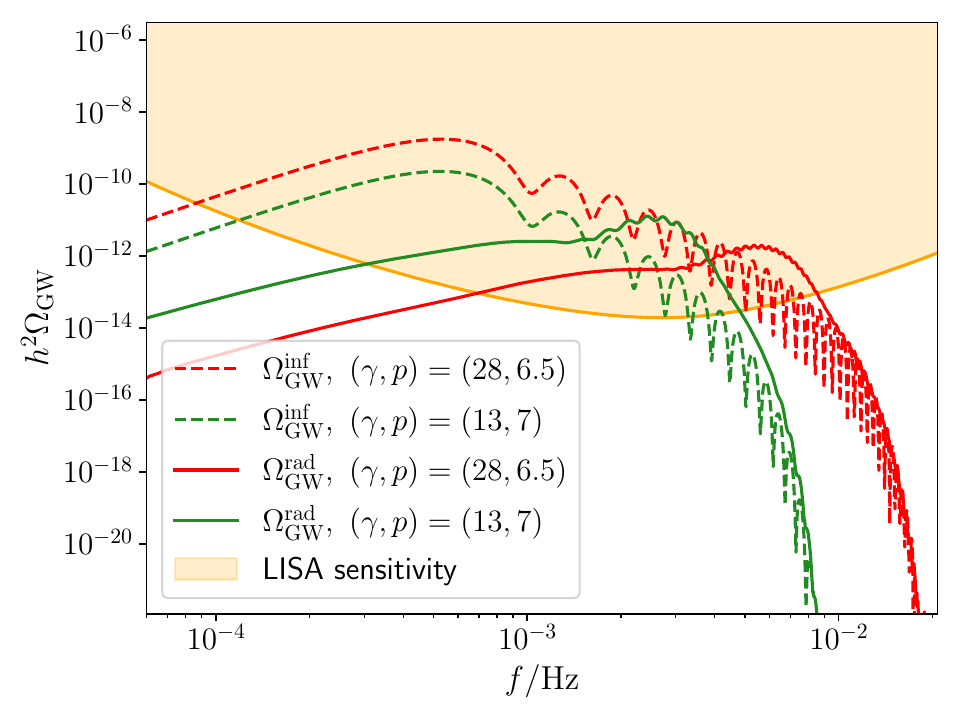}
    \caption{\textit{$\Omegarad$ and $\Omegainf$ corresponding  to  the  points  in  parameter  space highlighted with crosses in Fig.~\ref{figBack}. For illustrative purpose we choose the same time of the feature $N_{\mathrm{out}}=28.5$ for both scenarios. The position of the peak of $\Omegainf$ is then fixed ---see Eq.~\eqref{max}--- and larger $\gam$ implies a higher and wider signal. Conversely, increasing $\gam$ shifts the peak of $\Omegarad$ to higher frequencies, while its overall amplitude depends only on $p$, i.e.~the enhancement of the power spectrum of the curvature perturbation.}} 
    \label{Benchmark2LISA}
\end{figure}

This discussion applies equally to the bound~\eqref{boundback} from backreaction. The resulting Fig.~\ref{figBack} would be the analogous roadmap in parameter space for this case. The only small difference is what happens once a given value of the slow-roll parameter $\epsilon$ is fixed. While in Fig.~\ref{figPert}, $\epsilon$ is controlling only the size of $\Omegainf\propto \epsilon^2$, in Fig.~\ref{figBack}, $\epsilon$ also determines the position of the backreaction constraints through Eq.~\eqref{back}. Thus, one should keep in mind that a smaller $\epsilon$ would ameliorate backreaction but at the same time would lift all lines of constant $\Omegainf$.

Summarising, by looking at Fig.~\ref{figPert} (Fig.~\ref{figBack}) we can select specific points in parameter space for which perturbativity (backreaction) bounds are satisfied and $\Omegainf$ and/or $\Omegarad$ are potentially observable in a given experiment. For instance, the gravitational wave signal generated during inflation corresponding to the point in parameter space marked with a cross in Fig.~\ref{figPert} is shown in Fig.~\ref{Benchmark1LISA} together with the LISA sensitivity band. The small value of $p$ implies a subdominant $\Omegarad \simeq 10^{-5}\cdot\p_0^2 e^{4\cdot 5}\simeq 5\cdot 10^{-15}$ compared to $\Omegainf$. 
In Fig.~\ref{Benchmark2LISA} we show $\Omegarad$ and $\Omegainf$ for the two points in parameter space marked in Fig.~\ref{figBack}.

To conclude, scenarios in which a sharp feature leads to a significant enhancement of the power spectrum of the scalar fluctuations should be studied carefully, and at the same time, from a theoretical and phenomenological point of view. In particular, the significant constraining power coming from the --- often neglected --- requirement of remaining in a regime of theoretical control, demands a detailed determination of the corresponding bounds \eqref{pert},~\eqref{back} once a particular set-up is proposed.

\section{Summary of results}
\label{sec:conc}

We have derived analytical predictions for the \textit{enhanced} SGWB sourced during inflation as induced by the presence of a scalar excited state. Excited states with large occupation numbers associated with small scales can dynamically emerge as the result of a sudden transition along the inflationary trajectory. We illustrate this concept by means of an explicit example in a multifield setup but analogous dynamics can emerge in other contexts. Allowing non-trivial dynamics during inflation not only provides an unconstrained and exciting phenomenological playground at scales smaller than the CMB; given the theoretical difficulties in building a consistent high-energy embedding for models displaying a prolonged period of standard slow-roll evolution, it is reasonable to assume that the few e-folds probed by current experiments might have been, in reality, rather exceptional.

Let us schematically summarise our main findings. First, and independently of the current application to the case of an excited state:
\begin{itemize}
  \item[$\star$]
  We have derived a generic formula ---Eq.~\eqref{Master}--- for the tensor power spectrum sourced at second order by a generic multi-scalar system during inflation.
\end{itemize}
The non-trivial sum arising from properly quantising the multifield system leads to a $\mathcal{N}^4$ enhancement of the tensor power spectrum, with $\mathcal{N}$ the number of scalar degrees of freedom, when all fields equally contribute.
In particular, the former is given in terms of an integral over the scalar-field momenta and the time evolution. 
At the time when the excited state has been generated, the relevant enhanced modes are, by definition, deep inside the Hubble radius. Assuming, from that time onwards, an almost de Sitter like evolution of fluctuations, the time evolution can be explicitly integrated. 
\begin{itemize}
    \item[$\star$] Equation~\eqref{complete2} provides one with a kernel that can be applied to compute the corresponding GWs background for any excited state with sufficiently large occupation numbers, i.e.~whenever the \bog \, coefficients satisfy $|\alpha |\simeq |\beta|\gg 1$.
\end{itemize}
The spectral shape of the SGWB associated with a dynamically generated excited state shows clear recognisable features that we have characterised in detail: the signal has a principal peak at $k^{\mathrm{inf}}_{\mathrm{max}}\simeq 3.5 \kf$, followed by a series of order one oscillations with frequency $\omega=2/\kf$, where $\kf$ denotes the scale associated with the emergence of the excited state. The occurrence of these oscillations is the counterpart for the tensor power spectrum of the well-known oscillatory pattern of the scalar power spectrum characteristic of sharp features, with $\kf$ the scale of the feature \cite{Slosar:2019gvt}. In our context, these oscillations modulate an envelope that is decreasing as $k^{-3}$ in the case where only a narrow range of scales are put into an excited state. 
This brings us to the third main result of the paper.
\begin{itemize}
    \item[$\star$] The inflationary-era contribution to the GW spectrum due to an excited state during inflation is well-captured by the following analytical template:
\be\label{template}
\Omegainf(k) = \bar{\Omega}^{\mathrm{inf}}_{\mathrm{GW}} \frac{1}{(\omega k)^3}\left(1-\frac{(\omega k)^2}{16\gam^2}\right)^2\cdot\left( \sin(\omega k/2) -4\frac{(1-\cos(\omega k/2))}{\omega k}\right)^2 \, ,
\ee
which depends on the three parameters  $(\bar{\Omega}_{\mathrm{GW}}^{\mathrm{inf}},\gam, \omega)$, with $\gamma \gg 1$. Here, $\bar{\Omega}_{\mathrm{GW}}^{\mathrm{inf}} = 4 r_i\, \mathcal{N}^4 \gamma^5 H^4/(\pi\Mp)^4$, with $r_i$ a redshift factor~\eqref{redshift2} and $\mathcal{N}$ the number of scalars affected by the excited state.
The expression~\eqref{template} holds
for generic excited states with large occupation numbers that peak around a given scale $k_*=2 \gamma/\omega$, and is valid in the relevant range $k \leq 2 k_*$.
\end{itemize}
The latter condition is automatically satisfied if the excited state is triggered by a sharp feature along the inflationary trajectory. In this case, a non-trivial scalar-induced contribution to the stochastic background sourced after inflation is also present~\cite{Fumagalli:2020nvq}. 
The maximum of the post-inflationary contribution $\Omegarad$, here assuming a radiation-dominated universe at the time of horizon re-entry of the relevant scales, is located at higher momenta with respect to $k^{\mathrm{inf}}_{\mathrm{max}}$, i.e.~$k^{\mathrm{rad}}_{\mathrm{max}}= \left(2/\sqrt{3}\right) \kstar =  \left(2/\sqrt{3}\right) \gamma\kf$. In addition, the peak in $\Omegarad$ is also modulated by oscillations of amplitude that can be of order $20\%$ of the envelope and frequency equal to $\sqrt{3}\,\omega$. Thus, one may consider the interesting possibility of detecting both contributions through a signal that can hardly be mimicked by other phenomena. Further, the features in $\Omegarad$, in contrast to $\Omegainf$, non-trivially encode information about the cosmic expansion of the universe at the time the post-inflationary GWs are sourced~\cite{Witkowski:2021raz} (see \cite{Fumagalli:2021dtd} for a preliminary study of the detection prospects of these oscillatory signals with LISA).
Note that, in general, quantities such as the scalar induced GWs sourced after inflation as well as the primordial black holes abundance are sensitive to the statistical properties of the curvature fluctuation at the end of inflation. In contrast, the scalar induced GWs sourced during inflation are sensitive to all (scalar) degrees of freedom entering in the energy-momentum tensor and to their time evolution during inflation. In the case of an excited state, this manifests itself as an enhancement of the tensor power spectrum proportional to $\gamma^5$, where $\gamma$ measures how deep inside the horizon was the maximally enhanced mode at the time the excited state emerges. 
This enhancement can be understood from the constructive interference between positive and negative frequency modes, similar to the mechanism resulting in the amplification of the bispectrum near flattened configurations in presence of an excited state~\cite{Chen:2006nt,Holman:2007na,Meerburg:2009ys,Agarwal:2012mq,Ganc:2011dy,Flauger:2013hra,Aravind:2013lra}.
Thus, despite a suppression factor $\epsilon^2$ of the inflationary-era GW spectrum compared to the post-inflationary one, we find that $\Omega_\textrm{GW}^{\textrm{inf}}$ can easily be comparable to or larger than $\Omega_\textrm{GW}^{\textrm{rad}}$.

All in all, our results reinforce the importance of GW cosmology in reconstructing the history of the very early universe. Our work leaves several interesting open questions to pursue in the future. On the phenomenological side, for instance, it is possible to obtain more accurate templates to reconstruct the GW signal by taking into account the subdominant component of the mixed contribution in Eq.~\eqref{omdeco}, as well as consider more complicated dynamics once excited states have emerged. In addition, although we have estimated the size of backreaction effects, it is necessary to have a more rigorous understanding of how they affect the evolution of the system.
On the theoretical side, it would be interesting to study cases where the particle production is not large. There, the classical treatment we have used in this work is not accurate enough and one has to resort to the full \textit{in-in} formalism.

\section*{Acknowledgements}
We are grateful to Guillermo Ballesteros, Sadra Jazayeri, Mauro Pieroni and Lucas Pinol for useful discussions, as well as to the participants of the workshop Gravitational-Wave Primordial Cosmology where preliminary results of this work were presented. J.F, S.RP, and L.T.W are supported by the European Research Council under the European Union's Horizon 2020 research and innovation programme (grant agreement No 758792, project GEODESI). 
J.F is currently supported by a Contrato de Atracción de Talento (Modalidad 1) de la Comunidad de Madrid (Spain), number 2017-T1/TIC-5520 and the IFT Centro de Excelencia Severo Ochoa Grant SEV-2. G.A.P and C.Z. are supported by the Fondecyt Regular Project No.~1210876 (ANID). S.S is supported by the ``CUniverse'' research promotion project by Chulalongkorn University (grant reference CUAASC).
%%%%%%%%
\appendix
%%%%%%%%
\section{In-in, one-loop tensor power spectrum} 
\label{app:in-in}
In Sec.~\ref{ssec:int-inin} we argued that the Green's function solution~\eqref{S-to-I} corresponds to a truncation of the field operator to first order in the interaction-picture field. Here, we further show that the power spectrum obtained in this way is approximately equal to the one obtained from the in-in, one-loop, two-point correlation function: 
\be \label{hh-in-in}
\left\langle \hat{h}^\lambda_{\bk}(\tau) \hat{h}^\mu_{{\bk}'}(\tau) \right\rangle = \Bigg\langle \left[ \bar{\rm T} e^{i \int^\tau_{-\infty_+} \!\! \di\tau_1 \; \hat{\cal H}_{\rm int}(\tau_1)} \right] \hat{h}^\lambda_{\bk} (\tau)\hat{h}^\mu_{\bk}(\tau) \left[ {\rm T} e^{-i \int^\tau_{-\infty_-} \!\! \di \tau_2 \; \hat{\cal H}_{\rm int}(\tau_2)} \right] \Bigg\rangle,
\ee
where $\hat{\cal H}_{\rm int}$ is the interaction picture Hamiltonian~\eqref{Hint}. Our task is thus to compare \eqref{hh-in-in} to \eqref{eq:h-h}, which we rewrite here for convenience:
\begin{align} 
\left\langle \hat{h}^\lambda_{\bk}(\tau) \hat{h}^\mu_{{\bk}'}(\tau) \right\rangle
&=
 \int^\tau \mathrm{d}\tau_1 \int^\tau \mathrm{d}\tau_2 \; g_k(\tau,\tau_1) g_k(\tau,\tau_2) \left\langle \hat{S}^\lambda_{{\bk}}(\tau_1) \hat{S}^\mu_{{\bk}'}(\tau_2) \right\rangle.
\label{eq:h-h-app}
\end{align}
In this work we have been interested in cases where there is a preferred scale $\tauf$, when the scalars are excited away from the Bunch-Davies vacuum state, sourcing GWs. Since these excited states have support from $\tauf$ onward until the end of inflation at $\tau=0$ (\textit{out} region), in order to avoid the complications discussed in Sec.~\ref{ssec:int-inin}, coming from the $i\epsilon$ prescription, we may as well focus on the spectrum therein.   
Let us thus use the notation of Eq.~\eqref{main-34} and label $\p_t^{\rm out}$ the part of the spectrum corresponding to the \textit{out} region. Furthermore, let us ignore the momentum conserving delta function and also write $\p^{\rm out}_t(k)\equiv\lim_{k\tau\to 0}\p^{\rm out}_t(k,\tau)$ as in Sec.~\ref{sec:GW-ExSt}. 

The in-in formalism yields two one-loop contributions for the tensor power spectrum: 
\be \label{1lA}
\p^{\rm out}_{t;A}(k) = \frac{k^3}{2\pi^2 \Mp^4} \int^0_{\tauf} \di\tau_2 \int^0_{\tauf} \di \tau_1 \; \left\langle  \hat{\cal H}_{\rm int}(\tau_1) \hat{h}_{\bk} (\tau)\hat{h}_{\bk}(\tau)  \hat{\cal H}_{\rm int}(\tau_2) \right\rangle,
\ee
and
\be \label{1lB}
\p^{\rm out}_{t;B}(k) =  -\frac{k^3}{\pi^2 \Mp^4} {\rm Re}\, \int^0_{\tauf} \di\tau_2 \int^{\tau_2}_{\tauf} \di \tau_1 \; \left\langle  \hat{\cal H}_{\rm int}(\tau_1) \hat{\cal H}_{\rm int}(\tau_2) \hat{h}_{\bk} (\tau)\hat{h}_{\bk}(\tau)  \right\rangle.
\ee

A crucial observation is that the integral over the $(\tau_1,\tau_2)$ plane in~\eqref{eq:h-h-app} picks up the real part of the integrand. 
Moreover, since the power spectra are factorisable in time ---cf.~\eqref{eq:P-without-renormlaisation}--- the integrand, i.e.~the source correlator~\eqref{eq:S-S}, and consequently the graviton two-point function~\eqref{eq:h-h-app}, will inherit this structure. To catalyse the comparison, let us make these properties manifest by 
rewriting the \textit{out}-region tensor power spectrum~\eqref{Master} as
\bea \label{P-cl}
\!\!\p^{\rm out}_{t}(k) &=& \sum_\alpha \int \mathrm{d}^3  \bp\!\!
\int^0_{\tauf} \! \!\di \tau_1 \;
g_k(0,\tau_1) f^\alpha_{p,|\bk-\bp|}(\tau_1) \int^0_{\tauf} \!\! \di\tau_2 \; g_k(0,\tau_2) f^\alpha_{p,|\bk-\bp|}(\tau_2),
\eea
where the Green's function satisfies $g_k(0,\tau) \propto {\rm Im} \, \zeta(k\tau)$ ---c.f.~\eqref{eq:green-function}, while $f^\alpha_{p,q}(\tau)$ are real functions defined via the relation 
\be \label{S=SS}
\frac{k^3}{4\pi^5 \Mp^4} p^4 \sin^4\theta \; {\rm Re}  \sum_{X,Y} P_{XY}(\tau_1,\tau_2;p) P_{XY}\left( \tau_1,\tau_2;q \right) =  \sum_\alpha f^\alpha_{p,q}(\tau_1)  f^\alpha_{p,q}(\tau_2).
\ee
This is nothing but rewriting the modulus squared in the second line of Eq.~\eqref{Master} (recall that $g_k$ is real) as the sum of the squared\footnote{Recall that these are integrands and thus $\tau_{1,2}$ are dummy variables. ``Squared" here means the product of the function with itself evaluated once at $\tau_1$ and once at $\tau_2$.} real and imaginary parts of the quantity $\sum_{X}Q_{Xi}(p,\tau_1) Q_{Xj}\left( q,\tau_1 \right)$. The index $\alpha$ in the right-hand side just reorganises the summation over $ij$; it takes $2{\cal N}^2$ values (for ${\cal N}$ scalar fields). 

Next, let us note that since the scalar source enters in the same way in both methods (the only difference lies in how the external gravitons connect to the internal vertices), we may use this notation to also reorganise the one-loop diagrams as
\bea \label{P-1l-A}
\p^{\rm out}_{t;A}(k) &=&2\sum_\alpha
\int \mathrm{d}^3  \bp \int^0_{\tauf}  \!\!\!  \di \tau_1 \; \zeta(k\tau_1) f^\alpha_{p,|\bk-\bp|}(\tau_1) \int^0_{\tauf}  \!\!\! \di\tau_2 \;\zeta^*(k\tau_2) f^\alpha_{p,|\bk-\bp|}(\tau_2),
\eea
and
\bea \label{P-1l-B}
\p^{\rm out}_{t;B}(k)  &=& \! -4\!  \sum_\alpha{\rm Re}\!\int \!\!\mathrm{d}^3  \bp \! \int^0_{\tauf} \!\! \di\tau_2 \!\!\int^{\tau_2}_{\tauf}  \!\!\!  \di \tau_1  \zeta(k\tau_1) f^\alpha_{p,|\bk-\bp|}(\tau_1)  \zeta(k\tau_2) f^\alpha_{p,|\bk-\bp|}(\tau_2),
\eea
where we have taken into account the combinatorial factor of $2$ coming from the distinct $h$-contractions. Note that in this way, we may compare \eqref{P-cl}, \eqref{P-1l-A} and \eqref{P-1l-B} term by term and then resum.

Finally, by noting that due to the reality condition on $f^\alpha$, the integrand of~\eqref{P-1l-B} is symmetric under $\tau_1\leftrightarrow \tau_2$, we may disentangle the integrals in $\p^{{\rm 1};B}_t$ and rewrite it as~\cite{Adshead:2009cb}
\bea \label{B-decoupled}
\p^{\rm out}_{t;B}(k)  &=&  -2\sum_\alpha {\rm Re} \! \!\int \!\mathrm{d}^3  \bp \! \! \int^{0}_{\tauf} \! \! \di \tau_1 \zeta(k\tau_1) f^\alpha_{p,|\bk-\bp|}(\tau_1) \!\!\int^0_{\tauf} \!\!\di\tau_2
\zeta(k\tau_2) f^\alpha_{p,|\bk-\bp|}(\tau_2).
\eea
Upon doing so, we obtain 
\be \label{c-1-eq}
\p^{\rm out}_{t;A}(k) + \p^{\rm out}_{t;B}(k) = \p^{\rm out}_{t}(k),
\ee
which is the desired result.
\section{Approximate expression for the GW spectrum}\label{Appapprox}

For simplicity we suppress all labels on the Bogoliubov coefficients and proceed as if there was only a single scalar mode. From Eq.~\eqref{complete2}, the spectrum of induced GWs takes the form
\begin{align}
\label{eq:GWind-simple-ansatz}
\p^{\rm out}_t(k) = \int_1^\infty \di s \int_0^1 \di d \, \Big| \alpha \Big(\tfrac{k}{2} (s+d) \Big) \Big|^2 \Big| \alpha \Big(\tfrac{k}{2} (s-d) \Big) \Big|^2 \, F(x,y,k) \, ,
\end{align} 
where we find it convenient to perform the integration using the variables $d,s$ defined in Eq.~\eqref{sd}, and the kernel $F(x,y,k)$ can be read off from Eq.~\eqref{complete2}. 

We also restrict our attention to situations where the excited state induces a peak in the scalar power spectrum, as this is the case most relevant for phenomenology. In terms of the Bogoliubov coefficients, this implies that $|\alpha(k)|^2$ will exhibit a peak, see Eq.~\eqref{eq:Pzeta-excited}. For a first analysis, we consider the extreme case $|\alpha(k)|^2 = C \, \delta[\log(k/\kstar)]$, i.e., where $|\alpha(k)|^2$ is given by a single spike. 
Inserting this profile into Eq.~\eqref{eq:GWind-simple-ansatz} one finds
\begin{align}
\label{eq:GWind-simple-ansatz-delta}
\p^{\textrm{out}, \, \delta}_t(k) = C^2 \, \frac{\kstar^2}{k^2} \, F\bigg(\frac{\kstar}{k},\frac{\kstar}{k}, k \bigg) \, \Theta \bigg( \frac{2 \kstar}{k} -1\bigg) \, .
\end{align}
In practice, however, the peak in $|\alpha(k)|^2$ will have finite extent. If it is sufficiently narrow, the above single-spike 
result is expected to still give a good approximation to $\p^{\rm out}_t(k)$, especially for $k \sim \kstar$. However, if the peak is not narrow enough or for sufficiently small $k$ this is not the case anymore.

To illustrate this point, for a peak of finite width $\Delta k$ we write
\be\label{tophat}
|\alpha(k)|^2 =C \frac{\kstar}{\Delta k} \, \Theta(k-k_1)\Theta(k_2-k),\quad \textrm{with} \quad \Delta k \equiv k_2 - k_1 \, ,
\ee
where the Heaviside theta functions enforce the finite extension of the peak, with $k_1 < \kstar < k_2$. Inserting Eq.~\eqref{tophat} into Eq.~\eqref{eq:GWind-simple-ansatz} the effect of the Heaviside theta functions is to restrict the integration domain to
\begin{align}
\label{eq:sd-limits-top-hat}
    \frac{2k_1}{k} < s-d < \frac{2k_2}{k} \, , \quad \frac{2k_1}{k} < s+d < \frac{2k_2}{k} \, .
\end{align}
We can then confirm that for a sufficiently narrow peak with $\Delta k \ll \kstar$ one recovers the single-spike result. In this case the two instances of $\mathcal{P}$ in Eq.~\eqref{eq:GWind-simple-ansatz} only have significant overlap when the two arguments are close to each other, i.e.~in the vicinity of $d=0$. To approximate this, we can thus set $d=0$ and integrate over the region of overlap as given in Eq.~\eqref{eq:sd-limits-top-hat}. For the $d$-integral this implies integrating up to $d_\textrm{max}= \Delta k / k$, which gives a factor of $\Delta k / k$. This implicitly assumes that we are considering values $k \sim \kstar$, so that $\Delta k / k$ is small. The $s$-integration is to be performed over $s \in [2 k_1 / k, \, 2k_2 / k]$, i.e.~over an interval of width $\Delta k / k$ enclosing the value $2\kstar / k$. For sufficiently small $\Delta k / k$ the integrand can be taken as constant over this interval. The integral can thus be approximated by evaluating the integrand at $s=2\kstar / k$ and multiplying by the width $2\Delta k / k$. Upon including a factor of $1/2$ that accounts for the triangular shape of the $(d,s)$-integration domain, one recovers the single-spike result~\eqref{eq:GWind-simple-ansatz-delta}.

We now examine how this analysis is changed if the peak in $|\alpha(k)|^2$ is not narrow, or if we are interested in small values of $k$. In both these cases the interval $\Delta k / k$ is not necessarily small, which will affect how the $d$- and $s$-integrals can be performed. In particular, consider $\Delta k / k > 1$. In this case $d_\textrm{max}= \Delta k / k > 1$ and the constraint on the $d$-integral due to the finite size of the peak is void, as the range of integration in Eq.~\eqref{eq:GWind-simple-ansatz} is $d \in [0,1]$. As long as the integrand only varies slowly with $d$, we can proceed as in the narrow peak case, set $d=0$ and perform the integral. The difference however is that this does not give a factor of $\Delta k / k$ anymore. The integral over $s$ is again over an interval of width $2\Delta k / k$, which is no longer small. Still, if the integrand only varies slowly over this range, we can proceed as before. Then, for a peak in $|\alpha(k)|^2$ and for $\Delta k / k > 1$ the spectrum of induced GWs can be written as in the single-spike case, albeit with one less factor of $\Delta k / k$, i.e.
\begin{align}
\label{eq:GWind-approx-result}
\p^{\textrm{out}}_t(k) \approx C^2 \, \frac{\kstar}{\Delta k} \kappa^{-1} \, F\big(\kappa^{-1},\kappa^{-1},k \big) \, ,
\end{align}
with $\kappa = k / \kstar$. Recall that an assumption for arriving at this result is that $F(x,y,k)$ in Eq.~\eqref{eq:GWind-simple-ansatz} varies sufficiently slowly. Here we did not include the Heaviside theta function from Eq.~\eqref{eq:GWind-simple-ansatz-delta}, as this piece is also affected by the finite width of the peak in $|\alpha(k)|^2$. The correct factor is not very important here, as this mainly affects the UV part of the spectrum, where this approximate expression is not expected to hold.

In case of sharp features, it is reasonable to assume that $k_*/\Delta k \simeq 1$. In the main text we have also redefined the constant $C$ of this appendix such that $C = A(2 \p_0)^{-1}$. In this way $A$ gives a direct estimate of the peak of the primordial scalar power spectrum ---see Eq.~\eqref{Adef}. Thus, we finally arrive at the approximation in Eq.~\eqref{theapprox} used throughout the paper.
This, when appropriately applied to the various contributions of the GW spectrum, will provide a very good match to numerical results for $k \lesssim \kstar$. As we explain in the main text, this includes the regime of maximal amplitude of the GW spectrum.

\bibliographystyle{JHEP}
\bibliography{Biblio-2020}

\providecommand{\href}[2]{#2}\begingroup\raggedright\begin{thebibliography}{100}

\bibitem{Abbott:2016blz}
{\scshape LIGO Scientific, Virgo} collaboration, \emph{{Observation of
  Gravitational Waves from a Binary Black Hole Merger}},
  \href{https://doi.org/10.1103/PhysRevLett.116.061102}{\emph{Phys. Rev. Lett.}
  {\bfseries 116} (2016) 061102}
  [\href{https://arxiv.org/abs/1602.03837}{{\ttfamily 1602.03837}}].

\bibitem{Guth:1980zm}
A.H.~Guth, \emph{{The Inflationary Universe: A Possible Solution to the Horizon
  and Flatness Problems}},
  \href{https://doi.org/10.1103/PhysRevD.23.347}{\emph{Phys. Rev. D} {\bfseries
  23} (1981) 347}.

\bibitem{Starobinsky:1980te}
A.A.~Starobinsky, \emph{{A New Type of Isotropic Cosmological Models Without
  Singularity}},
  \href{https://doi.org/10.1016/0370-2693(80)90670-X}{\emph{Phys. Lett.}
  {\bfseries B91} (1980) 99}.

\bibitem{Linde:1981mu}
A.D.~Linde, \emph{{A New Inflationary Universe Scenario: A Possible Solution of
  the Horizon, Flatness, Homogeneity, Isotropy and Primordial Monopole
  Problems}}, \href{https://doi.org/10.1016/0370-2693(82)91219-9}{\emph{Phys.
  Lett.} {\bfseries B108} (1982) 389}.

\bibitem{Albrecht:1982wi}
A.J.~Albrecht and P.J.~Steinhardt, \emph{{Cosmology for Grand Unified Theories
  with Radiatively Induced Symmetry Breaking}},
  \href{https://doi.org/10.1103/PhysRevLett.48.1220}{\emph{Phys. Rev. Lett.}
  {\bfseries 48} (1982) 1220}.

\bibitem{Mukhanov:1981xt}
V.F.~Mukhanov and G.V.~Chibisov, \emph{{Quantum Fluctuation and Nonsingular
  Universe. (In Russian)}}, {\emph{JETP Lett.} {\bfseries 33} (1981) 532}.

\bibitem{Starobinsky:1979ty}
A.A.~Starobinsky, \emph{{Spectrum of relict gravitational radiation and the
  early state of the universe}}, {\emph{JETP Lett.} {\bfseries 30} (1979) 682}.

\bibitem{Rubakov:1982df}
V.A.~Rubakov, M.V.~Sazhin and A.V.~Veryaskin, \emph{{Graviton Creation in the
  Inflationary Universe and the Grand Unification Scale}},
  \href{https://doi.org/10.1016/0370-2693(82)90641-4}{\emph{Phys. Lett. B}
  {\bfseries 115} (1982) 189}.

\bibitem{Fabbri:1983us}
R.~Fabbri and M.d.~Pollock, \emph{{The Effect of Primordially Produced
  Gravitons upon the Anisotropy of the Cosmological Microwave Background
  Radiation}}, \href{https://doi.org/10.1016/0370-2693(83)91322-9}{\emph{Phys.
  Lett. B} {\bfseries 125} (1983) 445}.

\bibitem{Abbott:1984fp}
L.F.~Abbott and M.B.~Wise, \emph{{Constraints on Generalized Inflationary
  Cosmologies}},
  \href{https://doi.org/10.1016/0550-3213(84)90329-8}{\emph{Nucl. Phys. B}
  {\bfseries 244} (1984) 541}.

\bibitem{LISA:web}
\emph{https://www.elisascience.org}.

\bibitem{SKA:web}
\emph{https://www.skatelescope.org}.

\bibitem{IPTA:web}
\emph{https://www.ipta4gw.org}.

\bibitem{BICEP:2021xfz}
{\scshape BICEP, Keck} collaboration, \emph{{Improved Constraints on Primordial
  Gravitational Waves using Planck, WMAP, and BICEP/Keck Observations through
  the 2018 Observing Season}},
  \href{https://doi.org/10.1103/PhysRevLett.127.151301}{\emph{Phys. Rev. Lett.}
  {\bfseries 127} (2021) 151301}
  [\href{https://arxiv.org/abs/2110.00483}{{\ttfamily 2110.00483}}].

\bibitem{Baumann:2014nda}
D.~Baumann and L.~McAllister, \emph{{Inflation and String Theory}}, Cambridge
  University Press (2015), [\href{https://arxiv.org/abs/1404.2601}{{\ttfamily
  1404.2601}}].

\bibitem{Clesse:2017bsw}
S.~Clesse and J.~Garc\'\i{}a-Bellido, \emph{{Seven Hints for Primordial Black
  Hole Dark Matter}},
  \href{https://doi.org/10.1016/j.dark.2018.08.004}{\emph{Phys. Dark Univ.}
  {\bfseries 22} (2018) 137}
  [\href{https://arxiv.org/abs/1711.10458}{{\ttfamily 1711.10458}}].

\bibitem{Garcia-Bellido:2020pwq}
J.~Garc\'\i{}a-Bellido, J.F.~Nu\~no Siles and E.~Ruiz~Morales, \emph{{Bayesian
  analysis of the spin distribution of LIGO/Virgo black holes}},
  \href{https://doi.org/10.1016/j.dark.2021.100791}{\emph{Phys. Dark Univ.}
  {\bfseries 31} (2021) 100791}
  [\href{https://arxiv.org/abs/2010.13811}{{\ttfamily 2010.13811}}].

\bibitem{Franciolini:2021tla}
G.~Franciolini, V.~Baibhav, V.~De~Luca, K.K.Y.~Ng, K.W.K.~Wong, E.~Berti
  et~al., \emph{{Quantifying the evidence for primordial black holes in
  LIGO/Virgo gravitational-wave data}},
  \href{https://arxiv.org/abs/2105.03349}{{\ttfamily 2105.03349}}.

\bibitem{Novikov-pbh}
Y.B.~{Zel'dovich} and I.D.~{Novikov}, \emph{{The Hypothesis of Cores Retarded
  during Expansion and the Hot Cosmological Model}}, {\emph{Soviet Astronomy}
  {\bfseries 10} (1967) 602}.

\bibitem{Hawking:1971ei}
S.~Hawking, \emph{{Gravitationally collapsed objects of very low mass}},
  {\emph{Mon. Not. Roy. Astron. Soc.} {\bfseries 152} (1971) 75}.

\bibitem{Fumagalli:2020nvq}
J.~Fumagalli, S.~Renaux-Petel and L.T.~Witkowski, \emph{{Oscillations in the
  stochastic gravitational wave background from sharp features and particle
  production during inflation}},
  \href{https://doi.org/10.1088/1475-7516/2021/08/030}{\emph{JCAP} {\bfseries
  08} (2021) 030} [\href{https://arxiv.org/abs/2012.02761}{{\ttfamily
  2012.02761}}].

\bibitem{Adshead:2020bji}
P.~Adshead, N.~Afshordi, E.~Dimastrogiovanni, M.~Fasiello, E.A.~Lim and
  G.~Tasinato, \emph{{Multimessenger cosmology: Correlating cosmic microwave
  background and stochastic gravitational wave background measurements}},
  \href{https://doi.org/10.1103/PhysRevD.103.023532}{\emph{Phys. Rev. D}
  {\bfseries 103} (2021) 023532}
  [\href{https://arxiv.org/abs/2004.06619}{{\ttfamily 2004.06619}}].

\bibitem{Unal:2020mts}
C.~\"Unal, E.D.~Kovetz and S.P.~Patil, \emph{{Multimessenger probes of
  inflationary fluctuations and primordial black holes}},
  \href{https://doi.org/10.1103/PhysRevD.103.063519}{\emph{Phys. Rev. D}
  {\bfseries 103} (2021) 063519}
  [\href{https://arxiv.org/abs/2008.11184}{{\ttfamily 2008.11184}}].

\bibitem{Malhotra:2020ket}
A.~Malhotra, E.~Dimastrogiovanni, M.~Fasiello and M.~Shiraishi,
  \emph{{Cross-correlations as a Diagnostic Tool for Primordial Gravitational
  Waves}}, \href{https://doi.org/10.1088/1475-7516/2021/03/088}{\emph{JCAP}
  {\bfseries 03} (2021) 088}
  [\href{https://arxiv.org/abs/2012.03498}{{\ttfamily 2012.03498}}].

\bibitem{Ricciardone:2021kel}
A.~Ricciardone, L.V.~Dall'Armi, N.~Bartolo, D.~Bertacca, M.~Liguori and
  S.~Matarrese, \emph{{Cross-correlating Astrophysical and Cosmological
  Gravitational Wave Backgrounds with the Cosmic Microwave Background}},
  \href{https://arxiv.org/abs/2106.02591}{{\ttfamily 2106.02591}}.

\bibitem{Braglia:2021fxn}
M.~Braglia and S.~Kuroyanagi, \emph{{Probing pre-Recombination Physics by the
  Cross-Correlation of Stochastic Gravitational Waves and CMB Anisotropies}},
  \href{https://arxiv.org/abs/2106.03786}{{\ttfamily 2106.03786}}.

\bibitem{Dimastrogiovanni:2021mfs}
E.~Dimastrogiovanni, M.~Fasiello, A.~Malhotra, P.D.~Meerburg and G.~Orlando,
  \emph{{Testing the Early Universe with Anisotropies of the Gravitational Wave
  Background}},  \href{https://arxiv.org/abs/2109.03077}{{\ttfamily
  2109.03077}}.

\bibitem{Starobinsky:1992ts}
A.A.~Starobinsky, \emph{{Spectrum of adiabatic perturbations in the universe
  when there are singularities in the inflation potential}}, {\emph{JETP Lett.}
  {\bfseries 55} (1992) 489}.

\bibitem{Kaloper:2003nv}
N.~Kaloper and M.~Kaplinghat, \emph{{Primeval corrections to the CMB
  anisotropies}}, \href{https://doi.org/10.1103/PhysRevD.68.123522}{\emph{Phys.
  Rev. D} {\bfseries 68} (2003) 123522}
  [\href{https://arxiv.org/abs/hep-th/0307016}{{\ttfamily hep-th/0307016}}].

\bibitem{Ashoorioon:2006wc}
A.~Ashoorioon and A.~Krause, \emph{{Power Spectrum and Signatures for Cascade
  Inflation}},  \href{https://arxiv.org/abs/hep-th/0607001}{{\ttfamily
  hep-th/0607001}}.

\bibitem{Bean:2008na}
R.~Bean, X.~Chen, G.~Hailu, S.H.H.~Tye and J.~Xu, \emph{{Duality Cascade in
  Brane Inflation}},
  \href{https://doi.org/10.1088/1475-7516/2008/03/026}{\emph{JCAP} {\bfseries
  03} (2008) 026} [\href{https://arxiv.org/abs/0802.0491}{{\ttfamily
  0802.0491}}].

\bibitem{Ashoorioon:2017toq}
A.~Ashoorioon, R.~Casadio, G.~Geshnizjani and H.J.~Kim, \emph{{Getting
  Super-Excited with Modified Dispersion Relations}},
  \href{https://doi.org/10.1088/1475-7516/2017/09/008}{\emph{JCAP} {\bfseries
  09} (2017) 008} [\href{https://arxiv.org/abs/1702.06101}{{\ttfamily
  1702.06101}}].

\bibitem{Ashoorioon:2018uey}
A.~Ashoorioon, R.~Casadio, M.~Cicoli, G.~Geshnizjani and H.J.~Kim,
  \emph{{Extended Effective Field Theory of Inflation}},
  \href{https://doi.org/10.1007/JHEP02(2018)172}{\emph{JHEP} {\bfseries 02}
  (2018) 172} [\href{https://arxiv.org/abs/1802.03040}{{\ttfamily
  1802.03040}}].

\bibitem{Ballesteros:2018wlw}
G.~Ballesteros, J.~Beltran~Jimenez and M.~Pieroni, \emph{{Black hole formation
  from a general quadratic action for inflationary primordial fluctuations}},
  \href{https://doi.org/10.1088/1475-7516/2019/06/016}{\emph{JCAP} {\bfseries
  06} (2019) 016} [\href{https://arxiv.org/abs/1811.03065}{{\ttfamily
  1811.03065}}].

\bibitem{Ballesteros:2021fsp}
G.~Ballesteros, S.~C\'espedes and L.~Santoni, \emph{{Large power spectrum and
  primordial black holes in the effective theory of inflation}},
  \href{https://arxiv.org/abs/2109.00567}{{\ttfamily 2109.00567}}.

\bibitem{Tasinato:2020vdk}
G.~Tasinato, \emph{{An analytic approach to non-slow-roll inflation}},
  \href{https://doi.org/10.1103/PhysRevD.103.023535}{\emph{Phys. Rev. D}
  {\bfseries 103} (2021) 023535}
  [\href{https://arxiv.org/abs/2012.02518}{{\ttfamily 2012.02518}}].

\bibitem{Dalianis:2021iig}
I.~Dalianis, G.P.~Kodaxis, I.D.~Stamou, N.~Tetradis and A.~Tsigkas-Kouvelis,
  \emph{{Spectrum oscillations from features in the potential of single-field
  inflation}}, \href{https://doi.org/10.1103/PhysRevD.104.103510}{\emph{Phys.
  Rev. D} {\bfseries 104} (2021) 103510}
  [\href{https://arxiv.org/abs/2106.02467}{{\ttfamily 2106.02467}}].

\bibitem{Inomata:2021tpx}
K.~Inomata, E.~McDonough and W.~Hu, \emph{{Amplification of Primordial
  Perturbations from the Rise or Fall of the Inflaton}},
  \href{https://arxiv.org/abs/2110.14641}{{\ttfamily 2110.14641}}.

\bibitem{Chung:1999ve}
D.J.H.~Chung, E.W.~Kolb, A.~Riotto and I.I.~Tkachev, \emph{{Probing Planckian
  physics: Resonant production of particles during inflation and features in
  the primordial power spectrum}},
  \href{https://doi.org/10.1103/PhysRevD.62.043508}{\emph{Phys. Rev. D}
  {\bfseries 62} (2000) 043508}
  [\href{https://arxiv.org/abs/hep-ph/9910437}{{\ttfamily hep-ph/9910437}}].

\bibitem{Barnaby:2009dd}
N.~Barnaby and Z.~Huang, \emph{{Particle Production During Inflation:
  Observational Constraints and Signatures}},
  \href{https://doi.org/10.1103/PhysRevD.80.126018}{\emph{Phys. Rev. D}
  {\bfseries 80} (2009) 126018}
  [\href{https://arxiv.org/abs/0909.0751}{{\ttfamily 0909.0751}}].

\bibitem{Cook:2011hg}
J.L.~Cook and L.~Sorbo, \emph{{Particle production during inflation and
  gravitational waves detectable by ground-based interferometers}},
  \href{https://doi.org/10.1103/PhysRevD.85.023534}{\emph{Phys. Rev. D}
  {\bfseries 85} (2012) 023534}
  [\href{https://arxiv.org/abs/1109.0022}{{\ttfamily 1109.0022}}].

\bibitem{Carney:2012pk}
D.~Carney, W.~Fischler, E.D.~Kovetz, D.~Lorshbough and S.~Paban, \emph{{Rapid
  field excursions and the inflationary tensor spectrum}},
  \href{https://doi.org/10.1007/JHEP11(2012)042}{\emph{JHEP} {\bfseries 11}
  (2012) 042} [\href{https://arxiv.org/abs/1209.3848}{{\ttfamily 1209.3848}}].

\bibitem{Achucarro:2010da}
A.~Achucarro, J.-O.~Gong, S.~Hardeman, G.A.~Palma and S.P.~Patil,
  \emph{{Features of heavy physics in the CMB power spectrum}},
  \href{https://doi.org/10.1088/1475-7516/2011/01/030}{\emph{JCAP} {\bfseries
  1101} (2011) 030} [\href{https://arxiv.org/abs/1010.3693}{{\ttfamily
  1010.3693}}].

\bibitem{Palma:2020ejf}
G.A.~Palma, S.~Sypsas and C.~Zenteno, \emph{{Seeding primordial black holes in
  multifield inflation}},
  \href{https://doi.org/10.1103/PhysRevLett.125.121301}{\emph{Phys. Rev. Lett.}
  {\bfseries 125} (2020) 121301}
  [\href{https://arxiv.org/abs/2004.06106}{{\ttfamily 2004.06106}}].

\bibitem{Fumagalli:2020adf}
J.~Fumagalli, S.~Renaux-Petel, J.W.~Ronayne and L.T.~Witkowski, \emph{{Turning
  in the landscape: a new mechanism for generating Primordial Black Holes}},
  \href{https://arxiv.org/abs/2004.08369}{{\ttfamily 2004.08369}}.

\bibitem{Braglia:2020taf}
M.~Braglia, X.~Chen and D.K.~Hazra, \emph{{Probing Primordial Features with the
  Stochastic Gravitational Wave Background}},
  \href{https://doi.org/10.1088/1475-7516/2021/03/005}{\emph{JCAP} {\bfseries
  03} (2021) 005} [\href{https://arxiv.org/abs/2012.05821}{{\ttfamily
  2012.05821}}].

\bibitem{Iacconi:2021ltm}
L.~Iacconi, H.~Assadullahi, M.~Fasiello and D.~Wands, \emph{{Revisiting
  small-scale fluctuations in $\alpha$-attractor models of inflation}},
  \href{https://arxiv.org/abs/2112.05092}{{\ttfamily 2112.05092}}.

\bibitem{Polarski:1992dq}
D.~Polarski and A.A.~Starobinsky, \emph{{Spectra of perturbations produced by
  double inflation with an intermediate matter dominated stage}},
  \href{https://doi.org/10.1016/0550-3213(92)90062-G}{\emph{Nucl. Phys. B}
  {\bfseries 385} (1992) 623}.

\bibitem{Adams:1997de}
J.A.~Adams, G.G.~Ross and S.~Sarkar, \emph{{Multiple inflation}},
  \href{https://doi.org/10.1016/S0550-3213(97)00431-8}{\emph{Nucl. Phys. B}
  {\bfseries 503} (1997) 405}
  [\href{https://arxiv.org/abs/hep-ph/9704286}{{\ttfamily hep-ph/9704286}}].

\bibitem{Pi:2017gih}
S.~Pi, Y.-l.~Zhang, Q.-G.~Huang and M.~Sasaki, \emph{{Scalaron from
  $R^2$-gravity as a heavy field}},
  \href{https://doi.org/10.1088/1475-7516/2018/05/042}{\emph{JCAP} {\bfseries
  05} (2018) 042} [\href{https://arxiv.org/abs/1712.09896}{{\ttfamily
  1712.09896}}].

\bibitem{Pi:2019ihn}
S.~Pi, M.~Sasaki and Y.-l.~Zhang, \emph{{Primordial Tensor Perturbation in
  Double Inflationary Scenario with a Break}},
  \href{https://doi.org/10.1088/1475-7516/2019/06/049}{\emph{JCAP} {\bfseries
  06} (2019) 049} [\href{https://arxiv.org/abs/1904.06304}{{\ttfamily
  1904.06304}}].

\bibitem{DAmico:2020euu}
G.~D'Amico and N.~Kaloper, \emph{{Rollercoaster cosmology}},
  \href{https://doi.org/10.1088/1475-7516/2021/08/058}{\emph{JCAP} {\bfseries
  08} (2021) 058} [\href{https://arxiv.org/abs/2011.09489}{{\ttfamily
  2011.09489}}].

\bibitem{DAmico:2021vka}
G.~D'Amico, N.~Kaloper and A.~Westphal, \emph{{Double Monodromy Inflation: A
  Gravity Waves Factory for CMB-S4, LiteBIRD and LISA}},
  \href{https://arxiv.org/abs/2101.05861}{{\ttfamily 2101.05861}}.

\bibitem{Ragavendra:2020vud}
H.V.~Ragavendra, L.~Sriramkumar and J.~Silk, \emph{{Could PBHs and secondary
  GWs have originated from squeezed initial states?}},
  \href{https://doi.org/10.1088/1475-7516/2021/05/010}{\emph{JCAP} {\bfseries
  05} (2021) 010} [\href{https://arxiv.org/abs/2011.09938}{{\ttfamily
  2011.09938}}].

\bibitem{Domenech:2021ztg}
G.~Dom\`enech, \emph{{Scalar Induced Gravitational Waves Review}},
  \href{https://doi.org/10.3390/universe7110398}{\emph{Universe} {\bfseries 7}
  (2021) 398} [\href{https://arxiv.org/abs/2109.01398}{{\ttfamily
  2109.01398}}].

\bibitem{Thrane:2013oya}
E.~Thrane and J.D.~Romano, \emph{{Sensitivity curves for searches for
  gravitational-wave backgrounds}},
  \href{https://doi.org/10.1103/PhysRevD.88.124032}{\emph{Phys. Rev. D}
  {\bfseries 88} (2013) 124032}
  [\href{https://arxiv.org/abs/1310.5300}{{\ttfamily 1310.5300}}].

\bibitem{Maggiore:2007ulw}
M.~Maggiore, \emph{{Gravitational Waves. Vol. 1: Theory and Experiments}},
  Oxford Master Series in Physics, Oxford University Press (2007).

\bibitem{Caprini:2018mtu}
C.~Caprini and D.G.~Figueroa, \emph{{Cosmological Backgrounds of Gravitational
  Waves}}, \href{https://doi.org/10.1088/1361-6382/aac608}{\emph{Class. Quant.
  Grav.} {\bfseries 35} (2018) 163001}
  [\href{https://arxiv.org/abs/1801.04268}{{\ttfamily 1801.04268}}].

\bibitem{Acquaviva:2002ud}
V.~Acquaviva, N.~Bartolo, S.~Matarrese and A.~Riotto, \emph{{Second order
  cosmological perturbations from inflation}},
  \href{https://doi.org/10.1016/S0550-3213(03)00550-9}{\emph{Nucl. Phys.}
  {\bfseries B667} (2003) 119}
  [\href{https://arxiv.org/abs/astro-ph/0209156}{{\ttfamily
  astro-ph/0209156}}].

\bibitem{Mollerach:2003nq}
S.~Mollerach, D.~Harari and S.~Matarrese, \emph{{CMB polarization from
  secondary vector and tensor modes}},
  \href{https://doi.org/10.1103/PhysRevD.69.063002}{\emph{Phys. Rev. D}
  {\bfseries 69} (2004) 063002}
  [\href{https://arxiv.org/abs/astro-ph/0310711}{{\ttfamily
  astro-ph/0310711}}].

\bibitem{Ananda:2006af}
K.N.~Ananda, C.~Clarkson and D.~Wands, \emph{{The Cosmological gravitational
  wave background from primordial density perturbations}},
  \href{https://doi.org/10.1103/PhysRevD.75.123518}{\emph{Phys. Rev. D}
  {\bfseries 75} (2007) 123518}
  [\href{https://arxiv.org/abs/gr-qc/0612013}{{\ttfamily gr-qc/0612013}}].

\bibitem{Baumann:2007zm}
D.~Baumann, P.J.~Steinhardt, K.~Takahashi and K.~Ichiki, \emph{{Gravitational
  Wave Spectrum Induced by Primordial Scalar Perturbations}},
  \href{https://doi.org/10.1103/PhysRevD.76.084019}{\emph{Phys. Rev. D}
  {\bfseries 76} (2007) 084019}
  [\href{https://arxiv.org/abs/hep-th/0703290}{{\ttfamily hep-th/0703290}}].

\bibitem{Biagetti:2013kwa}
M.~Biagetti, M.~Fasiello and A.~Riotto, \emph{{Enhancing Inflationary Tensor
  Modes through Spectator Fields}},
  \href{https://doi.org/10.1103/PhysRevD.88.103518}{\emph{Phys. Rev. D}
  {\bfseries 88} (2013) 103518}
  [\href{https://arxiv.org/abs/1305.7241}{{\ttfamily 1305.7241}}].

\bibitem{Turner:1993vb}
M.S.~Turner, M.J.~White and J.E.~Lidsey, \emph{{Tensor perturbations in
  inflationary models as a probe of cosmology}},
  \href{https://doi.org/10.1103/PhysRevD.48.4613}{\emph{Phys. Rev. D}
  {\bfseries 48} (1993) 4613}
  [\href{https://arxiv.org/abs/astro-ph/9306029}{{\ttfamily
  astro-ph/9306029}}].

\bibitem{Liu:2015psa}
X.-J.~Liu, W.~Zhao, Y.~Zhang and Z.-H.~Zhu, \emph{{Detecting Relic
  Gravitational Waves by Pulsar Timing Arrays: Effects of Cosmic Phase
  Transitions and Relativistic Free-Streaming Gases}},
  \href{https://doi.org/10.1103/PhysRevD.93.024031}{\emph{Phys. Rev. D}
  {\bfseries 93} (2016) 024031}
  [\href{https://arxiv.org/abs/1509.03524}{{\ttfamily 1509.03524}}].

\bibitem{Salopek:1988qh}
D.S.~Salopek, J.R.~Bond and J.M.~Bardeen, \emph{{Designing Density Fluctuation
  Spectra in Inflation}},
  \href{https://doi.org/10.1103/PhysRevD.40.1753}{\emph{Phys. Rev.} {\bfseries
  D40} (1989) 1753}.

\bibitem{GrootNibbelink:2001qt}
S.~Groot~Nibbelink and B.J.W.~van Tent, \emph{{Scalar perturbations during
  multiple field slow-roll inflation}},
  \href{https://doi.org/10.1088/0264-9381/19/4/302}{\emph{Class. Quant. Grav.}
  {\bfseries 19} (2002) 613}
  [\href{https://arxiv.org/abs/hep-ph/0107272}{{\ttfamily hep-ph/0107272}}].

\bibitem{Tsujikawa:2002qx}
S.~Tsujikawa, D.~Parkinson and B.A.~Bassett, \emph{{Correlation-consistency
  cartography of the double inflation landscape}},
  \href{https://doi.org/10.1103/PhysRevD.67.083516}{\emph{Phys. Rev.}
  {\bfseries D67} (2003) 083516}
  [\href{https://arxiv.org/abs/astro-ph/0210322}{{\ttfamily
  astro-ph/0210322}}].

\bibitem{Weinberg:2008zzc}
S.~Weinberg, \emph{{Cosmology}}, Oxford Univ. Press (2008).

\bibitem{Pinol:2020cdp}
L.~Pinol, S.~Renaux-Petel and Y.~Tada, \emph{{A manifestly covariant theory of
  multifield stochastic inflation in phase space: solving the discretisation
  ambiguity in stochastic inflation}},
  \href{https://doi.org/10.1088/1475-7516/2021/04/048}{\emph{JCAP} {\bfseries
  04} (2021) 048} [\href{https://arxiv.org/abs/2008.07497}{{\ttfamily
  2008.07497}}].

\bibitem{Weinberg:2005vy}
S.~Weinberg, \emph{{Quantum contributions to cosmological correlations}},
  \href{https://doi.org/10.1103/PhysRevD.72.043514}{\emph{Phys. Rev.}
  {\bfseries D72} (2005) 043514}
  [\href{https://arxiv.org/abs/hep-th/0506236}{{\ttfamily hep-th/0506236}}].

\bibitem{Musso:2006pt}
M.~Musso, \emph{{A new diagrammatic representation for correlation functions in
  the in-in formalism}},
  \href{https://doi.org/10.1007/JHEP11(2013)184}{\emph{JHEP} {\bfseries 11}
  (2013) 184} [\href{https://arxiv.org/abs/hep-th/0611258}{{\ttfamily
  hep-th/0611258}}].

\bibitem{Seery:2008qj}
D.~Seery, K.A.~Malik and D.H.~Lyth, \emph{{Non-gaussianity of inflationary
  field perturbations from the field equation}},
  \href{https://doi.org/10.1088/1475-7516/2008/03/014}{\emph{JCAP} {\bfseries
  0803} (2008) 014} [\href{https://arxiv.org/abs/0802.0588}{{\ttfamily
  0802.0588}}].

\bibitem{Senatore:2009cf}
L.~Senatore and M.~Zaldarriaga, \emph{{On Loops in Inflation}},
  \href{https://doi.org/10.1007/JHEP12(2010)008}{\emph{JHEP} {\bfseries 12}
  (2010) 008} [\href{https://arxiv.org/abs/0912.2734}{{\ttfamily 0912.2734}}].

\bibitem{Adshead:2009cb}
P.~Adshead, R.~Easther and E.A.~Lim, \emph{{The 'in-in' Formalism and
  Cosmological Perturbations}},
  \href{https://doi.org/10.1103/PhysRevD.80.083521}{\emph{Phys. Rev. D}
  {\bfseries 80} (2009) 083521}
  [\href{https://arxiv.org/abs/0904.4207}{{\ttfamily 0904.4207}}].

\bibitem{Baumgart:2020oby}
M.~Baumgart and R.~Sundrum, \emph{{Manifestly Causal In-In Perturbation Theory
  about the Interacting Vacuum}},
  \href{https://doi.org/10.1007/JHEP03(2021)080}{\emph{JHEP} {\bfseries 03}
  (2021) 080} [\href{https://arxiv.org/abs/2010.10785}{{\ttfamily
  2010.10785}}].

\bibitem{Komatsu:2001rj}
E.~Komatsu and D.N.~Spergel, \emph{{Acoustic signatures in the primary
  microwave background bispectrum}},
  \href{https://doi.org/10.1103/PhysRevD.63.063002}{\emph{Phys. Rev.}
  {\bfseries D63} (2001) 063002}
  [\href{https://arxiv.org/abs/astro-ph/0005036}{{\ttfamily
  astro-ph/0005036}}].

\bibitem{Palma:2019lpt}
G.A.~Palma, B.~Scheihing~Hitschfeld and S.~Sypsas, \emph{{Non-Gaussian CMB and
  LSS statistics beyond polyspectra}},
  \href{https://doi.org/10.1088/1475-7516/2020/02/027}{\emph{JCAP} {\bfseries
  02} (2020) 027} [\href{https://arxiv.org/abs/1907.05332}{{\ttfamily
  1907.05332}}].

\bibitem{Chen:2006nt}
X.~Chen, M.-x.~Huang, S.~Kachru and G.~Shiu, \emph{{Observational signatures
  and non-Gaussianities of general single field inflation}}, {\emph{JCAP}
  {\bfseries 0701} (2007) 002}
  [\href{https://arxiv.org/abs/hep-th/0605045}{{\ttfamily hep-th/0605045}}].

\bibitem{Holman:2007na}
R.~Holman and A.J.~Tolley, \emph{{Enhanced Non-Gaussianity from Excited Initial
  States}}, \href{https://doi.org/10.1088/1475-7516/2008/05/001}{\emph{JCAP}
  {\bfseries 0805} (2008) 001}
  [\href{https://arxiv.org/abs/0710.1302}{{\ttfamily 0710.1302}}].

\bibitem{Meerburg:2009ys}
P.D.~Meerburg, J.P.~van~der Schaar and P.S.~Corasaniti, \emph{{Signatures of
  Initial State Modifications on Bispectrum Statistics}},
  \href{https://doi.org/10.1088/1475-7516/2009/05/018}{\emph{JCAP} {\bfseries
  0905} (2009) 018} [\href{https://arxiv.org/abs/0901.4044}{{\ttfamily
  0901.4044}}].

\bibitem{Agarwal:2012mq}
N.~Agarwal, R.~Holman, A.J.~Tolley and J.~Lin, \emph{{Effective field theory
  and non-Gaussianity from general inflationary states}},
  \href{https://doi.org/10.1007/JHEP05(2013)085}{\emph{JHEP} {\bfseries 05}
  (2013) 085} [\href{https://arxiv.org/abs/1212.1172}{{\ttfamily 1212.1172}}].

\bibitem{Ganc:2011dy}
J.~Ganc, \emph{{Calculating the local-type fNL for slow-roll inflation with a
  non-vacuum initial state}},
  \href{https://doi.org/10.1103/PhysRevD.84.063514}{\emph{Phys. Rev. D}
  {\bfseries 84} (2011) 063514}
  [\href{https://arxiv.org/abs/1104.0244}{{\ttfamily 1104.0244}}].

\bibitem{Flauger:2013hra}
R.~Flauger, D.~Green and R.A.~Porto, \emph{{On squeezed limits in single-field
  inflation. Part I}},
  \href{https://doi.org/10.1088/1475-7516/2013/08/032}{\emph{JCAP} {\bfseries
  08} (2013) 032} [\href{https://arxiv.org/abs/1303.1430}{{\ttfamily
  1303.1430}}].

\bibitem{Aravind:2013lra}
A.~Aravind, D.~Lorshbough and S.~Paban, \emph{{Non-Gaussianity from Excited
  Initial Inflationary States}},
  \href{https://doi.org/10.1007/JHEP07(2013)076}{\emph{JHEP} {\bfseries 07}
  (2013) 076} [\href{https://arxiv.org/abs/1303.1440}{{\ttfamily 1303.1440}}].

\bibitem{Maldacena:2002vr}
J.M.~Maldacena, \emph{{Non-Gaussian features of primordial fluctuations in
  single field inflationary models}},
  \href{https://doi.org/10.1088/1126-6708/2003/05/013}{\emph{JHEP} {\bfseries
  05} (2003) 013} [\href{https://arxiv.org/abs/astro-ph/0210603}{{\ttfamily
  astro-ph/0210603}}].

\bibitem{Barnaby:2011qe}
N.~Barnaby, E.~Pajer and M.~Peloso, \emph{{Gauge Field Production in Axion
  Inflation: Consequences for Monodromy, non-Gaussianity in the CMB, and
  Gravitational Waves at Interferometers}},
  \href{https://doi.org/10.1103/PhysRevD.85.023525}{\emph{Phys. Rev. D}
  {\bfseries 85} (2012) 023525}
  [\href{https://arxiv.org/abs/1110.3327}{{\ttfamily 1110.3327}}].

\bibitem{delRio:2018vrj}
A.~del Rio, R.~Durrer and S.P.~Patil, \emph{{Tensor Bounds on the Hidden
  Universe}}, \href{https://doi.org/10.1007/JHEP12(2018)094}{\emph{JHEP}
  {\bfseries 12} (2018) 094}
  [\href{https://arxiv.org/abs/1808.09282}{{\ttfamily 1808.09282}}].

\bibitem{Nilles:2001fg}
H.P.~Nilles, M.~Peloso and L.~Sorbo, \emph{{Coupled fields in external
  background with application to nonthermal production of gravitinos}},
  \href{https://doi.org/10.1088/1126-6708/2001/04/004}{\emph{JHEP} {\bfseries
  04} (2001) 004} [\href{https://arxiv.org/abs/hep-th/0103202}{{\ttfamily
  hep-th/0103202}}].

\bibitem{Domenech:2021and}
G.~Dom\`enech, S.~Passaglia and S.~Renaux-Petel, \emph{{Gravitational waves
  from dark matter isocurvature}},
  \href{https://arxiv.org/abs/2112.10163}{{\ttfamily 2112.10163}}.

\bibitem{Garcia-Bellido:2017aan}
J.~Garcia-Bellido, M.~Peloso and C.~Unal, \emph{{Gravitational Wave signatures
  of inflationary models from Primordial Black Hole Dark Matter}},
  \href{https://doi.org/10.1088/1475-7516/2017/09/013}{\emph{JCAP} {\bfseries
  09} (2017) 013} [\href{https://arxiv.org/abs/1707.02441}{{\ttfamily
  1707.02441}}].

\bibitem{Unal:2018yaa}
C.~Unal, \emph{{Imprints of Primordial Non-Gaussianity on Gravitational Wave
  Spectrum}}, \href{https://doi.org/10.1103/PhysRevD.99.041301}{\emph{Phys.
  Rev. D} {\bfseries 99} (2019) 041301}
  [\href{https://arxiv.org/abs/1811.09151}{{\ttfamily 1811.09151}}].

\bibitem{Cai:2018dig}
R.-g.~Cai, S.~Pi and M.~Sasaki, \emph{{Gravitational Waves Induced by
  non-Gaussian Scalar Perturbations}},
  \href{https://doi.org/10.1103/PhysRevLett.122.201101}{\emph{Phys. Rev. Lett.}
  {\bfseries 122} (2019) 201101}
  [\href{https://arxiv.org/abs/1810.11000}{{\ttfamily 1810.11000}}].

\bibitem{Atal:2021jyo}
V.~Atal and G.~Dom\`enech, \emph{{Probing non-Gaussianities with the high
  frequency tail of induced gravitational waves}},
  \href{https://doi.org/10.1088/1475-7516/2021/06/001}{\emph{JCAP} {\bfseries
  06} (2021) 001} [\href{https://arxiv.org/abs/2103.01056}{{\ttfamily
  2103.01056}}].

\bibitem{Adshead:2021hnm}
P.~Adshead, K.D.~Lozanov and Z.J.~Weiner, \emph{{Non-Gaussianity and the
  induced gravitational wave background}},
  \href{https://doi.org/10.1088/1475-7516/2021/10/080}{\emph{JCAP} {\bfseries
  10} (2021) 080} [\href{https://arxiv.org/abs/2105.01659}{{\ttfamily
  2105.01659}}].

\bibitem{Fumagalli:2021cel}
J.~Fumagalli, S.~Renaux-Petel and L.T.~Witkowski, \emph{{Resonant features in
  the stochastic gravitational wave background}},
  \href{https://doi.org/10.1088/1475-7516/2021/08/059}{\emph{JCAP} {\bfseries
  08} (2021) 059} [\href{https://arxiv.org/abs/2105.06481}{{\ttfamily
  2105.06481}}].

\bibitem{Espinosa:2018eve}
J.R.~Espinosa, D.~Racco and A.~Riotto, \emph{{A Cosmological Signature of the
  SM Higgs Instability: Gravitational Waves}},
  \href{https://doi.org/10.1088/1475-7516/2018/09/012}{\emph{JCAP} {\bfseries
  09} (2018) 012} [\href{https://arxiv.org/abs/1804.07732}{{\ttfamily
  1804.07732}}].

\bibitem{Kohri:2018awv}
K.~Kohri and T.~Terada, \emph{{Semianalytic calculation of gravitational wave
  spectrum nonlinearly induced from primordial curvature perturbations}},
  \href{https://doi.org/10.1103/PhysRevD.97.123532}{\emph{Phys. Rev. D}
  {\bfseries 97} (2018) 123532}
  [\href{https://arxiv.org/abs/1804.08577}{{\ttfamily 1804.08577}}].

\bibitem{Inomata:2019zqy}
K.~Inomata, K.~Kohri, T.~Nakama and T.~Terada, \emph{{Gravitational Waves
  Induced by Scalar Perturbations during a Gradual Transition from an Early
  Matter Era to the Radiation Era}},
  \href{https://doi.org/10.1088/1475-7516/2019/10/071}{\emph{JCAP} {\bfseries
  10} (2019) 071} [\href{https://arxiv.org/abs/1904.12878}{{\ttfamily
  1904.12878}}].

\bibitem{Inomata:2019ivs}
K.~Inomata, K.~Kohri, T.~Nakama and T.~Terada, \emph{{Enhancement of
  Gravitational Waves Induced by Scalar Perturbations due to a Sudden
  Transition from an Early Matter Era to the Radiation Era}},
  \href{https://doi.org/10.1103/PhysRevD.100.043532}{\emph{Phys. Rev. D}
  {\bfseries 100} (2019) 043532}
  [\href{https://arxiv.org/abs/1904.12879}{{\ttfamily 1904.12879}}].

\bibitem{Domenech:2019quo}
G.~Dom\`enech, \emph{{Induced gravitational waves in a general cosmological
  background}}, \href{https://doi.org/10.1142/S0218271820500285}{\emph{Int. J.
  Mod. Phys. D} {\bfseries 29} (2020) 2050028}
  [\href{https://arxiv.org/abs/1912.05583}{{\ttfamily 1912.05583}}].

\bibitem{Domenech:2020kqm}
G.~Dom\`enech, S.~Pi and M.~Sasaki, \emph{{Induced gravitational waves as a
  probe of thermal history of the universe}},
  \href{https://doi.org/10.1088/1475-7516/2020/08/017}{\emph{JCAP} {\bfseries
  08} (2020) 017} [\href{https://arxiv.org/abs/2005.12314}{{\ttfamily
  2005.12314}}].

\bibitem{Witkowski:2021raz}
L.T.~Witkowski, G.~Dom\`enech, J.~Fumagalli and S.~Renaux-Petel,
  \emph{{Expansion history-dependent oscillations in the scalar-induced
  gravitational wave background}},
  \href{https://arxiv.org/abs/2110.09480}{{\ttfamily 2110.09480}}.

\bibitem{Cai:2019amo}
R.-G.~Cai, S.~Pi, S.-J.~Wang and X.-Y.~Yang, \emph{{Resonant multiple peaks in
  the induced gravitational waves}},
  \href{https://doi.org/10.1088/1475-7516/2019/05/013}{\emph{JCAP} {\bfseries
  05} (2019) 013} [\href{https://arxiv.org/abs/1901.10152}{{\ttfamily
  1901.10152}}].

\bibitem{Melville:2021lst}
S.~Melville and E.~Pajer, \emph{{Cosmological Cutting Rules}},
  \href{https://doi.org/10.1007/JHEP05(2021)249}{\emph{JHEP} {\bfseries 05}
  (2021) 249} [\href{https://arxiv.org/abs/2103.09832}{{\ttfamily
  2103.09832}}].

\bibitem{Goodhew:2021oqg}
H.~Goodhew, S.~Jazayeri, M.H.~Gordon~Lee and E.~Pajer, \emph{{Cutting
  cosmological correlators}},
  \href{https://doi.org/10.1088/1475-7516/2021/08/003}{\emph{JCAP} {\bfseries
  08} (2021) 003} [\href{https://arxiv.org/abs/2104.06587}{{\ttfamily
  2104.06587}}].

\bibitem{Baumann:2021fxj}
D.~Baumann, W.-M.~Chen, C.~Duaso~Pueyo, A.~Joyce, H.~Lee and G.L.~Pimentel,
  \emph{{Linking the Singularities of Cosmological Correlators}},
  \href{https://arxiv.org/abs/2106.05294}{{\ttfamily 2106.05294}}.

\bibitem{An:2020fff}
H.~An, K.-F.~Lyu, L.-T.~Wang and S.~Zhou, \emph{{A unique gravitational wave
  signal from phase transition during inflation}},
  \href{https://arxiv.org/abs/2009.12381}{{\ttfamily 2009.12381}}.

\bibitem{Peng:2021zon}
Z.-Z.~Peng, C.~Fu, J.~Liu, Z.-K.~Guo and R.-G.~Cai, \emph{{Gravitational waves
  from resonant amplification of curvature perturbations during inflation}},
  \href{https://doi.org/10.1088/1475-7516/2021/10/050}{\emph{JCAP} {\bfseries
  10} (2021) 050} [\href{https://arxiv.org/abs/2106.11816}{{\ttfamily
  2106.11816}}].

\bibitem{Cai:2021wzd}
R.-G.~Cai, C.~Chen and C.~Fu, \emph{{Primordial black holes and stochastic
  gravitational wave background from inflation with a noncanonical spectator
  field}}, \href{https://doi.org/10.1103/PhysRevD.104.083537}{\emph{Phys. Rev.
  D} {\bfseries 104} (2021) 083537}
  [\href{https://arxiv.org/abs/2108.03422}{{\ttfamily 2108.03422}}].

\bibitem{Sasaki:1995aw}
M.~Sasaki and E.D.~Stewart, \emph{{A General analytic formula for the spectral
  index of the density perturbations produced during inflation}},
  \href{https://doi.org/10.1143/PTP.95.71}{\emph{Prog. Theor. Phys.} {\bfseries
  95} (1996) 71} [\href{https://arxiv.org/abs/astro-ph/9507001}{{\ttfamily
  astro-ph/9507001}}].

\bibitem{Langlois:2008mn}
D.~Langlois and S.~Renaux-Petel, \emph{{Perturbations in generalized
  multi-field inflation}},
  \href{https://doi.org/10.1088/1475-7516/2008/04/017}{\emph{JCAP} {\bfseries
  0804} (2008) 017} [\href{https://arxiv.org/abs/0801.1085}{{\ttfamily
  0801.1085}}].

\bibitem{Achucarro:2012yr}
A.~Achucarro, V.~Atal, S.~Cespedes, J.-O.~Gong, G.A.~Palma and S.P.~Patil,
  \emph{{Heavy fields, reduced speeds of sound and decoupling during
  inflation}}, \href{https://doi.org/10.1103/PhysRevD.86.121301}{\emph{Phys.
  Rev.} {\bfseries D86} (2012) 121301}
  [\href{https://arxiv.org/abs/1205.0710}{{\ttfamily 1205.0710}}].

\bibitem{Castillo:2013sfa}
E.~Castillo, B.~Koch and G.~Palma, \emph{{On the integration of fields and
  quanta in time dependent backgrounds}},
  \href{https://doi.org/10.1007/JHEP05(2014)111}{\emph{JHEP} {\bfseries 05}
  (2014) 111} [\href{https://arxiv.org/abs/1312.3338}{{\ttfamily 1312.3338}}].

\bibitem{Cremonini:2010ua}
S.~Cremonini, Z.~Lalak and K.~Turzynski, \emph{{Strongly Coupled Perturbations
  in Two-Field Inflationary Models}},
  \href{https://doi.org/10.1088/1475-7516/2011/03/016}{\emph{JCAP} {\bfseries
  1103} (2011) 016} [\href{https://arxiv.org/abs/1010.3021}{{\ttfamily
  1010.3021}}].

\bibitem{Renaux-Petel:2015mga}
S.~Renaux-Petel and K.~Turzy{\'n}ski, \emph{{Geometrical Destabilization of
  Inflation}},
  \href{https://doi.org/10.1103/PhysRevLett.117.141301}{\emph{Phys. Rev. Lett.}
  {\bfseries 117} (2016) 141301}
  [\href{https://arxiv.org/abs/1510.01281}{{\ttfamily 1510.01281}}].

\bibitem{Garcia-Saenz:2018ifx}
S.~Garcia-Saenz, S.~Renaux-Petel and J.~Ronayne, \emph{{Primordial fluctuations
  and non-Gaussianities in sidetracked inflation}},
  \href{https://doi.org/10.1088/1475-7516/2018/07/057}{\emph{JCAP} {\bfseries
  1807} (2018) 057} [\href{https://arxiv.org/abs/1804.11279}{{\ttfamily
  1804.11279}}].

\bibitem{Garcia-Saenz:2018vqf}
S.~Garcia-Saenz and S.~Renaux-Petel, \emph{{Flattened non-Gaussianities from
  the effective field theory of inflation with imaginary speed of sound}},
  \href{https://doi.org/10.1088/1475-7516/2018/11/005}{\emph{JCAP} {\bfseries
  1811} (2018) 005} [\href{https://arxiv.org/abs/1805.12563}{{\ttfamily
  1805.12563}}].

\bibitem{Fumagalli:2019noh}
J.~Fumagalli, S.~Garcia-Saenz, L.~Pinol, S.~Renaux-Petel and J.~Ronayne,
  \emph{{Hyper-Non-Gaussianities in Inflation with Strongly Nongeodesic
  Motion}}, \href{https://doi.org/10.1103/PhysRevLett.123.201302}{\emph{Phys.
  Rev. Lett.} {\bfseries 123} (2019) 201302}
  [\href{https://arxiv.org/abs/1902.03221}{{\ttfamily 1902.03221}}].

\bibitem{Bjorkmo:2019qno}
T.~Bjorkmo, R.Z.~Ferreira and M.D.~Marsh, \emph{{Mild Non-Gaussianities under
  Perturbative Control from Rapid-Turn Inflation Models}},
  \href{https://doi.org/10.1088/1475-7516/2019/12/036}{\emph{JCAP} {\bfseries
  12} (2019) 036} [\href{https://arxiv.org/abs/1908.11316}{{\ttfamily
  1908.11316}}].

\bibitem{Ferreira:2020qkf}
R.Z.~Ferreira, \emph{{Non-Gaussianities in models of inflation with large and
  negative entropic masses}},
  \href{https://doi.org/10.1088/1475-7516/2020/08/034}{\emph{JCAP} {\bfseries
  08} (2020) 034} [\href{https://arxiv.org/abs/2003.13410}{{\ttfamily
  2003.13410}}].

\bibitem{Chakraborty:2019dfh}
D.~Chakraborty, R.~Chiovoloni, O.~Loaiza-Brito, G.~Niz and I.~Zavala,
  \emph{{Fat inflatons, large turns and the $\eta$-problem}},
  \href{https://doi.org/10.1088/1475-7516/2020/01/020}{\emph{JCAP} {\bfseries
  01} (2020) 020} [\href{https://arxiv.org/abs/1908.09797}{{\ttfamily
  1908.09797}}].

\bibitem{Aragam:2020uqi}
V.~Aragam, S.~Paban and R.~Rosati, \emph{{The Multi-Field, Rapid-Turn
  Inflationary Solution}},
  \href{https://doi.org/10.1007/JHEP03(2021)009}{\emph{JHEP} {\bfseries 03}
  (2021) 009} [\href{https://arxiv.org/abs/2010.15933}{{\ttfamily
  2010.15933}}].

\bibitem{Anguelova:2020nzl}
L.~Anguelova, \emph{{On Primordial Black Holes from Rapid Turns in Two-field
  Models}}, \href{https://doi.org/10.1088/1475-7516/2021/06/004}{\emph{JCAP}
  {\bfseries 06} (2021) 004}
  [\href{https://arxiv.org/abs/2012.03705}{{\ttfamily 2012.03705}}].

\bibitem{Aragam:2021scu}
V.~Aragam, R.~Chiovoloni, S.~Paban, R.~Rosati and I.~Zavala, \emph{{Rapid-turn
  inflation in supergravity is rare and tachyonic}},
  \href{https://arxiv.org/abs/2110.05516}{{\ttfamily 2110.05516}}.

\bibitem{Renaux-Petel:2021yxh}
S.~Renaux-Petel, \emph{{Inflation with strongly non-geodesic motion:
  theoretical motivations and observational imprints}},  in \emph{{European
  Physical Society Conference on High Energy Physics 2021}}, 11, 2021
  [\href{https://arxiv.org/abs/2111.00989}{{\ttfamily 2111.00989}}].

\bibitem{Achucarro:2016fby}
A.~Ach\'ucarro, V.~Atal, C.~Germani and G.A.~Palma, \emph{{Cumulative effects
  in inflation with ultra-light entropy modes}},
  \href{https://doi.org/10.1088/1475-7516/2017/02/013}{\emph{JCAP} {\bfseries
  02} (2017) 013} [\href{https://arxiv.org/abs/1607.08609}{{\ttfamily
  1607.08609}}].

\bibitem{Achucarro:2018ngj}
A.~Ach{\'u}carro, S.~C{\'e}spedes, A.-C.~Davis and G.A.~Palma,
  \emph{{Constraints on Holographic Multifield Inflation and Models Based on
  the Hamilton-Jacobi Formalism}},
  \href{https://doi.org/10.1103/PhysRevLett.122.191301}{\emph{Phys. Rev. Lett.}
  {\bfseries 122} (2019) 191301}
  [\href{https://arxiv.org/abs/1809.05341}{{\ttfamily 1809.05341}}].

\bibitem{Bartolo:2013exa}
N.~Bartolo, D.~Cannone and S.~Matarrese, \emph{{The Effective Field Theory of
  Inflation Models with Sharp Features}},
  \href{https://doi.org/10.1088/1475-7516/2013/10/038}{\emph{JCAP} {\bfseries
  10} (2013) 038} [\href{https://arxiv.org/abs/1307.3483}{{\ttfamily
  1307.3483}}].

\bibitem{Adshead:2014sga}
P.~Adshead and W.~Hu, \emph{{Bounds on nonadiabatic evolution in single-field
  inflation}}, \href{https://doi.org/10.1103/PhysRevD.89.083531}{\emph{Phys.
  Rev. D} {\bfseries 89} (2014) 083531}
  [\href{https://arxiv.org/abs/1402.1677}{{\ttfamily 1402.1677}}].

\bibitem{Cannone:2014qna}
D.~Cannone, N.~Bartolo and S.~Matarrese, \emph{{Perturbative Unitarity of
  Inflationary Models with Features}},
  \href{https://doi.org/10.1103/PhysRevD.89.127301}{\emph{Phys. Rev. D}
  {\bfseries 89} (2014) 127301}
  [\href{https://arxiv.org/abs/1402.2258}{{\ttfamily 1402.2258}}].

\bibitem{Inomata:2021zel}
K.~Inomata, \emph{{Bound on induced gravitational waves during inflation era}},
  \href{https://doi.org/10.1103/PhysRevD.104.123525}{\emph{Phys. Rev. D}
  {\bfseries 104} (2021) 123525}
  [\href{https://arxiv.org/abs/2109.06192}{{\ttfamily 2109.06192}}].

\bibitem{Slosar:2019gvt}
A.~Slosar et~al., \emph{{Scratches from the Past: Inflationary Archaeology
  through Features in the Power Spectrum of Primordial Fluctuations}},
  \href{https://arxiv.org/abs/1903.09883}{{\ttfamily 1903.09883}}.

\bibitem{Fumagalli:2021dtd}
J.~Fumagalli, M.~Pieroni, S.~Renaux-Petel and L.T.~Witkowski, \emph{{Detecting
  primordial features with LISA}},
  \href{https://arxiv.org/abs/2112.06903}{{\ttfamily 2112.06903}}.

\end{thebibliography}\endgroup

\end{document}